\documentclass[11pt]{article}
\usepackage[left=0.8in,top=0.8in,right=0.8in,bottom=0.8in]{geometry}

\usepackage{graphicx}
\usepackage{amsmath}
\usepackage{amssymb}
\usepackage{amsthm}
\usepackage{latexsym}
\usepackage{bm}
\usepackage{array,supertabular}
\usepackage{color}
\usepackage{framed}
\usepackage{setspace}
\usepackage{fancyhdr}
\usepackage{stmaryrd}
\usepackage{mathrsfs}
\usepackage{mdframed}
\usepackage{url}
\usepackage{multirow}
\usepackage{algorithm}
\usepackage{algpseudocode}
\newcolumntype{P}[1]{>{\centering\arraybackslash}p{#1}}

\SetSymbolFont{stmry}{bold}{U}{stmry}{m}{n}

\newtheorem{remark}{Remark}

\numberwithin{equation}{section}

\newenvironment{myenv}[1]
  {\mdfsetup{
    frametitle={\colorbox{white}{\space#1\space}},
    innertopmargin=10pt,
    frametitleaboveskip=-\ht\strutbox,
    frametitlealignment=\center
    }
  \begin{mdframed}
  }
  {\end{mdframed}}

\begin{document}
\title{Vascular fluid-structure interaction: unified continuum formulation, image-based mesh generation pipeline, and scalable fully implicit solver technology}
\author{Ju Liu$^{\textup{ a,b,*}}$, Jiayi Huang$^{\textup{ a}}$, Qingshuang Lu$^{\textup{ a}}$, Yujie Sun$^{\textup{ a}}$ \\
$^a$ \textit{\small Department of Mechanics and Aerospace Engineering,}\\
\textit{\small Southern University of Science and Technology,}\\
\textit{\small 1088 Xueyuan Avenue, Shenzhen, Guangdong 518055, China}\\
$^b$ \textit{\small Guangdong-Hong Kong-Macao Joint Laboratory for Data-Driven Fluid Mechanics}\\
\textit{\small and Engineering Applications,}\\
\textit{\small Southern University of Science and Technology}\\
\textit{\small 1088 Xueyuan Avenue, Shenzhen, Guangdong 518055, China}\\
$^{*}$ \small \textit{E-mail address:} liuj36@sustech.edu.cn, liujuy@gmail.com
}
\date{}
\maketitle

\section*{Abstract}
We propose a computational framework for vascular fluid-structure interaction (FSI), focusing on biomechanical modeling, geometric modeling, and solver technology. The biomechanical model is constructed based on the unified continuum formulation recently proposed in \cite{Liu2018}, which is well-behaved in both compressible and incompressible regimes for fluids and solids. The variational multiscale formulation is chosen as the spatial discretization approach to provide a subgrid turbulence model in the fluid subproblem and to circumvent the inf-sup condition when invoking equal-order interpolations for incompressible materials. The first-order generalized-$\alpha$ scheme is applied uniformly for the fluid and solid subproblems. We highlight that the chosen time integration scheme differs from existing implicit FSI integration methods in that it is \textit{indeed} second-order accurate, does not suffer from the overshoot phenomenon, and optimally dissipates high-frequency modes in both subproblems. In particular, the previously numerically observed no-overshoot property is theoretically justified in this work. For anatomically realistic geometric modeling, we propose a pipeline for generating subject-specific meshes for FSI analysis. Different from most existing methodologies that operate directly on the wall surface mesh, our pipeline starts from the image segmentation stage. This allows using existing algorithms for the surface fairing to remove undesired model artifacts. This strategy can flexibly assign variable wall thickness, which will be propitious in studying pathological cases. With high-quality surface meshes obtained, the volumetric meshes are then generated, guaranteeing a boundary-layered mesh in the fluid subdomain and a matching mesh across the fluid-solid interface. These features are crucial for ensuring a proper coupling between the two subproblems and obtaining accurate simulation results. In the last, we propose a combined suite of nonlinear and linear solver technologies. Invoking a segregated algorithm within the Newton-Raphson iteration, the problem reduces to solving two linear systems in the multi-corrector stage. The first linear system is associated with the mesh motion and can be straightforwardly addressed by the algebraic multigrid (AMG) method. Thanks to the unified continuum formulation, the matrix related to the balance equations presents a two-by-two block structure in both subproblems. Using the Schur complement reduction (SCR) technique reduces the problem to solving matrices of smaller sizes of the elliptic type, and the AMG method again becomes a natural preconditioner candidate. Thus, the SCR combined with AMG serves as an effective preconditioning technique for the second matrix problem. The benefit of the unified formulation is demonstrated in parallelizing the solution algorithms as the number of unknowns matches in both subdomains. It is thus straightforward to achieve load balancing when parallelizing the algorithm via a domain decomposition approach. We use the Greenshields-Weller benchmark as well as a patient-specific vascular model to demonstrate the robustness, efficiency, and scalability of the overall FSI solver technology.

\vspace{5mm}

\noindent \textbf{Keywords:} Unified continuum model, Fluid-structure interaction, Image-based modeling, Iterative methods, Generalized-$\alpha$ method, Vascular biomechanics

\section{Introduction}
Over the years, computational fluid-structure interaction (FSI) serves as a valuable tool for the investigation of a variety of engineering and biomedical problems \cite{Bazilevs2012}. In particular, vascular biomechanics has become a major playground for developing novel FSI techniques because modeling vasculature is challenging in several aspects. From the perspective of coupling strategy, vascular FSI is an exemplary hydroelastic problem, where the densities of the blood and vessel wall have comparable magnitudes. This engenders the well-known added-mass effect, rendering the partitioned approach quite non-robust. An immense amount of research has been performed to design proper transmission conditions to enhance the algorithmic robustness of the partitioned strategy. Nevertheless, collected numerical experiences indicate that, within the partitioned coupling framework, methods for handling the added-mass effect often result in the sacrifice of other critical numerical properties \cite{Causin2005,Kadapa2021}. Thus, although the monolithic coupling strategy demands an extra amount of work for algorithm and code development, it becomes appealing and, sometimes, necessary.

The description of the moving fluid-solid interface is another factor affecting FSI solver designs. The strategies invoked are similar to those in multiphase flows. The immersed-boundary \cite{Peskin1972} and fictitious domain \cite{Baaijens2001} methods, as examples utilizing the interface-capturing idea, are often capable of handling problems with topological changes, with a representative example being the cardiac valve dynamics. In the meantime, similar to interface-capturing methods in multiphase flow modeling, those methods exhibit poor resolutions of the fluid-solid interface if the mesh cannot be adaptively refined and coarsened. This can become rather critical in cardiovascular biomechanics, as interfacial quantities, such as the wall shear stress, oscillatory shear index, etc., are biomarkers related to a variety of diseases \cite{Berger2000,Ku1985}. On the other hand, the Arbitrary Lagrangian-Eulerian (ALE) formulation \cite{Hughes1981} enables an interface-tracking, boundary-fitted strategy in FSI modeling. It enables resolving fluid-solid interfaces as well as fluid boundary layers with the mesh, and it satisfies the coupling conditions by construction. For problems undergoing topological changes in subdomains, mesh distortion inevitably necessitates expensive remeshing and repartitioning procedures in ALE methods \cite{Johnson1999}, which may be viewed as a major drawback of this method.

In our recent work, a unified continuum framework was proposed in order to unify the description of fluid and solid subproblems using the pressure primitive variables. It has a thermomechanical basis rooted in the adoption of the Gibbs free energy as the potential \cite{Liu2018,Liu2021}. It has been elucidated in prior computational fluid dynamics (CFD) studies that using the pressure primitive variables results in a well-behaved formulation in both compressible and incompressible regimes \cite{Hauke1994,Hauke1998}, which partly motivates the development of the unified continuum theory. Within this framework, important material models such as fluids and elastic solids were first recovered \cite{Liu2018}, and it was later extended to account for inelastic materials \cite{Liu2021}. A unified mathematical structure for both subproblems bridges the gap between CFD and computational structural dynamics. The well-studied inf-sup stable mixed element technology for the Stokes problem has been applied to incompressible hyperelasticity \cite{Liu2019} and viscoelasticity \cite{Liu2021}, which delivers provably nonlinear stability of the discretization method. It is worth pointing out that, although the mixed element technology had been applied to study incompressible finite elasticity before \cite{Auricchio2005,Auricchio2013,Schroeder2017}, the scheme based on the unified framework is the only one that delivers nonlinear stability so far, to the authors' knowledge. In addition to stable methods, stabilized methods, in particular the ones based on the variational multiscale (VMS) formulation, have also been leveraged for investigating FSI \cite{Liu2018,Liu2020b}, hyperelasticity \cite{Liu2019a,Scovazzi2016}, and viscoelasticity \cite{Zeng2017}. The geometrical flexibility of low-order simplex elements and the implementational convenience of equal-order interpolations have made this approach attractive to practitioners. In addition to the aforementioned advances, there are multiple other issues related to \textit{vascular} FSI modeling, which we aim to address in this work.

\subsection{Image-based modeling and mesh generation}
The first challenge comes from the generation of high-quality FSI meshes based on clinical images. There have been mature mesh generation strategies developed for vascular lumina \cite{Antiga2002,Arthurs2021,Updegrove2017}. Typically, the luminal surface is first delineated on a stack of two-dimensional images, which is known as the segmentation step \cite{Radaelli2010}. Through a lofting operation, the representation of the luminal wall can be obtained in either the computer-aided designed (CAD) model or polygonal surface meshes. After all luminal surfaces are generated, the wall surfaces are joined by boolean operations, and the resulting combined surface often needs to be repaired by removing undesirable modeling artifacts, which is also known as surface fairing. After a satisfactory wall surface is obtained, a volumetric mesh can then be generated from the boundary representation using the Delaunay method \cite{Si2015} or advancing front method \cite{Schoeberl1997}. Oftentimes, a boundary layer mesh is desired for resolving the near-wall shear layers of viscous flows \cite{Antiga2002,Garimella2000,Zhang2007}. 

The generation of a vascular FSI mesh is more involved because one needs to mesh the vascular tissue and maintain the mesh continuity across the fluid-solid interface. The latter requirement is necessary for producing the correct coupling of the FSI system at the discrete level; otherwise, special techniques need to be invoked for non-matching FSI meshes \cite{Bazilevs2012a}. Perhaps the most widely used strategy for generating the wall mesh is based on the existing boundary layer mesh generation algorithm. By changing the layer growth direction to the exterior of the luminal wall, the volumetric wall mesh can be generated. This is often referred to as the ``extrusion procedure" \cite{svfsi-mesh-gen,Marchandise2013,Raut2015,Wu2022}. Despite its simplicity and popularity, this approach is limited due to the arbitrariness of the arterial tree geometry and the lack of guarantee in avoiding self-intersection in the extrusion process. Although there are advanced algorithmic procedures (e.g. smoothing, shrinking, pruning, etc. \cite{Garimella2000}) that can automatically fix this issue, these procedures may engender physiologically non-realistic vascular wall geometry. Moreover, it is indeed allowed to generate variable wall thickness by setting the extrusion thickness to be a fraction of the local vessel radius \cite{Marchandise2013}. Yet, the extrusion procedure is unaware of additional physiological details and is thus difficult to assign detailed or pathological wall properties to the mesh. 

The idea of segmenting interior and exterior wall surfaces has been utilized for studying the correlation between hemodynamics and plaque progression based on black blood magnetic resonance (MR) imaging \cite{Ladak2001,Steinman2002}. Similar modeling approaches have been combined with FSI simulations for investigating atherosclerotic plaques in coronary and carotid arteries, using mutli-contrast MR \cite{Tang2004} or intravascular ultrasound images \cite{Yang2009}. The emphasis of those work was placed on the construction of an arterial wall model with the plaques taken into account, and the geometries considered there were relatively simple. In this work, we further develop an image-based modeling and mesh generation pipeline to generate high-quality patient-specific meshes suitable for FSI analysis. Instead of manipulating the wall surface mesh, we start by segmenting the exterior wall surface on each slice of two-dimensional images. The contour curves can be generated based on images that can resolve the exterior surface. Otherwise, empirical knowledge about the wall thickness can guide the generation of the contours. Manual intervention may be invoked if certain special or pathological shapes of the wall need to be constructed. The outer surface can then be constructed in a similar fashion by lofting, boolean operation, and surface fairing. Existing CAD and polygon mesh processing tools \cite{opencascade-reference,Botsch2010} guarantee a flexible and robust way of model editing.  With both the interior and exterior surface representation obtained, one may determine the boundary representations of the lumen as well as the vascular wall by closing the planar faces at the inlets and outlets. In our pipeline, a high-quality triangular luminal surface mesh is first re-created based on the prescribed mesh size. After that, a boundary layer mesh and a fully unstructured mesh can be generated in the fluid and solid subdomains, respectively. This pipeline allows us to prepare an FSI mesh for the three-dimensional wall model even when the imaging modality cannot resolve the exterior surface. The introduction of a CAD or a polygon mesh allows us to edit the wall model to make sure it is smooth and physiologically realistic. This can be especially beneficial when the arterial tree has complex geometries with multiple branchings.

\subsection{Time integration}
The time integration scheme used in this work is the \textit{first-order} generalized-$\alpha$ scheme, which was originally proposed for investigating the Navier-Stokes equations \cite{Jansen2000}. Recently, it was numerically observed that the velocity error is independent of the time step size when this scheme was applied to integrating structural dynamics \cite{Kadapa2017}. This encouraging finding suggests that this scheme does suffer from overshoot, an issue that has plagued the second-order generalized-$\alpha$ scheme for decades \cite{Chung1993,Hilber1978}. In this study, we perform numerical analysis to further corroborate that observation. Our analysis reveals that the first-order generalized-$\alpha$ method is indeed among the few methods \cite{Kadapa2017,Yu2008} that satisfy all desirable attributes for structural dynamics, as listed by Hilber and Hughes \cite{Hilber1978}. Additionally, we have recently demonstrated that second-order accuracy can be maintained in CFD analysis if the velocity and pressure are concurrently evaluated at the intermediate time step \cite{Liu2021a}. This leads to a minor modification to the predominant approach adopted in the FSI and CFD communities. Furthermore, when used in combination with the unified framework, the first-order generalized-$\alpha$ scheme can be applied uniformly to both subproblems. This again conveniently helps circumvent an issue encountered in conventional FSI schemes, where the fluid and solid subproblems were treated dichotomously by the first- and second-order generalized-$\alpha$ schemes, respectively. The optimal parametrizations of the two schemes are different, and choosing different parametrizations in two subdomains inevitably results in kinematic incompatibility across the interface \cite[p.~120]{Bazilevs2012}. In the unified framework, the whole FSI problem is integrated in time using the same first-order generalized-$\alpha$ scheme, and the compatibility issue does not arise at all.

\subsection{Parallelization}
Parallelization of the conventional FSI formulation has been rather challenging since the computation load in the two subproblems can be quite unbalanced. In specific, the solid subproblem was traditionally written in the pure displacement formulation which has three degrees of freedom (DOFs) per node, while the fluid subproblem was written using the pressure primitive variables (7 DOFs per node). When using the domain decomposition approach for distributed parallel implementation, it is nontrivial to maintain an ideal load balancing in mesh partitioning, and the subsequent computations among processors will be quite uneven. For certain problems, a thin-walled structural model has been invoked for the solid subproblem, which has much fewer DOFs than that in the fluid subproblem. Therefore, it was suggested to assign the structural subproblem to a single processor and distribute the fluid subdomain to the rest processors \cite{Bazilevs2017}. Recently, a performance model was constructed using integer optimization to determine the optimal load balancing strategy in a partitioned FSI solver \cite{Naseri2020,Totounferoush2019}. In any case, it is by no means trivial to partition a mesh with a balanced amount of work per processor and a balanced amount of communications across processors in the conventional FSI formulation. This troubling issue is conveniently circumvented in the unified continuum formulation \cite{Liu2018}, as both subproblems are written in the same pressure primitive variables. One may therefore perform mesh partitioning for the whole continuum body in a straightforward manner and expect balanced load assignment for each processor. One may argue that the load balancing is achieved at the cost of increasing three DOFs per node from three to seven in the solid subproblem. It will be demonstrated that this does not lead to a significant increase in the computational cost by designing a solver strategy that leverages the structure in the discrete problem of the unified FSI formulation.

\subsection{Nonlinear and linear solution method}
As mentioned above, the unified formulation leads to a clearly defined block structure in the resulting discrete problem. This allows us to design a suite of solution methods that exploits this structure. Our overall solver strategy involves two closely related algorithm designs. We first apply the segregated algorithm to develop a predictor multi-corrector algorithm for solving the nonlinear algebraic problem \cite{Liu2018}. With the help of the segregated algorithm, the momentum and mass equations are solved together, and the kinematic field (the solid and fluid mesh displacement) is determined separately. According to Proposition 5 in \cite{Liu2018}, the segregation of the solid displacement update does not induce a loss of consistency as a Newton-Raphson procedure. The nonlinear solver developed here is closely related to the ``quasi-direct coupling" method \cite[Chapter~6]{Bazilevs2012}. Consistent with prior experiences of the quasi-direct coupling, the segregated algorithm has not been observed to fail for non-robust reasons  \cite{Bazilevs2013}. The solution strategy is thus rather appealing because of its convenience in implementation and flexibility in designing iterative solution techniques for the two linear systems.

With the help of the segregated nonlinear solution method, the fully implicit FSI scheme boils down to repeatedly solving two linear problems. Since the mesh motion is typically governed by an elliptic differential operator, the conjugate gradient method preconditioned by the algebraic multigrid (AMG) method is an ideal candidate. In particular, as the harmonic extension algorithm is utilized in this study, this linear system remains invariant over iterations. The construction of the matrix and the setup of its AMG preconditioner only need to be performed once. The mesh motion can be solved rather efficiently and economically within the time stepping. 

The linear system associated with the balance equations takes a two-by-two block structure, and we have good knowledge of their physical origins as well as algebraic structures. After applying a block factorization, one may untangle the coupling between different physical fields and then invokes an appropriately designed preconditioner for them. This strategy is known to be rather effective in addressing saddle-point problems \cite{Benzi2005}, and the crux is to treat the Schur complement arising from the factorization. In our problem, the Schur complement, under certain simplifying assumptions, can be viewed as an algebraic analog of the pressure Laplacian operator \cite{Quarteroni2000a}, and can thus be preconditioned by the AMG method. In the meantime, getting the algebraic representation of the Schur complement is challenging, because it is dense by nature. This fact makes it prohibitively expensive to directly work on it in large-scale computations. In the  Semi-Implicit Method for Pressure-Linked Equations (SIMPLE) method and its variants, a sparse approximation of the Schur complement is made to construct a block preconditioner \cite{Elman2008}.  In certain scenarios (e.g., convection-dominated flows), other preconditioners have been devised to gain a better representation of the Schur complement with the goal of achieving robustness and scalability \cite{Cyr2012,Deparis2014,Elman2006,Elman1999,Kay2002}. In this work, the Schur complement is fitted into the Krylov subspace method in the form of the matrix-vector product, and there is an inner solver embedded in its algorithmic definition. A sparse approximation, such as the one in the SIMPLE preconditioner, is still utilized to accelerate the Krylov iteration. This strategy is known as the SCR procedure \cite{Benzi2005,May2008,Liu2020}. Indeed, in its crudest form, the SCR procedure reduces to a conventional block preconditioner without invoking the inner solver. The introduction of the inner-level solver offers a manner of adjusting and balancing several design factors (i.e., efficiency, robustness, scalability, etc.) for the overall iteration solution technique. The SCR preconditioning technique has been applied for CFD problems within stationary iterative methods \cite{Manguoglu2008,Manguoglu2011}. We recently adopted it to investigate hyperelasticity \cite{Liu2019a,Liu2019}, CFD with geometrically multiscale coupling \cite{Liu2020}, and an FSI problem with the structure modeled as a thin-walled membrane \cite{Lan2022b}, using FGMRES \cite{Saad1993} as the outer Krylov iteration method and AMG method as the preconditioner for the intermediate and inner solvers. In this work, we aim at investigating the efficacy of the solver technology within the vascular FSI problem with hyperelastic wall mechanics, hemodynamics, and reduced-order downstream vascular models all coupled.

In addition to the proposed block preconditioning technique, we review a few existing solver techniques developed for the \textit{conventional} FSI formulation. Due to the physical origins, the algebraic structures in the fluid and solid subdomains are different. Block factorization was performed to segregatedly treat the saddle-point and elasticity problems in the two subdomains. Most studies adopt the AMG method to build the approximate inverse for the block systems \cite{Crosetto2011,Deparis2016,Gee2011,Jodlbauer2019,Langer2016}, while some also considered the algebraic additive Schwarz method \cite{Deparis2016}. In a different approach, the so-called monolithic AMG method was proposed for FSI problems \cite{Gee2011,Verdugo2016}, in which the block preconditioning technique was utilized as the smoothers. In the last, continued success has also been reported with the additive Schwarz domain decomposition preconditioner with incomplete factorization employed to approximate solutions locally \cite{Kong2019}. This strategy has been used with the Jacobian-free approach, constituting a Newton-Krylov-Schwarz framework for FSI problems \cite{Barker2010,Wu2014,Liao2022}. Recently, the scalability of the one-level Schwarz method was further improved by introducing a coarse space and leveraging the multilevel technique \cite{Kong2017}.

\subsection{Structure and content of the article}
The remainder of the article is organized as follows. In Section \ref{sec:unified-continuum-formulation-FSI}, the unified continuum formulation is presented. Following that, the spatiotemporal discretization method is developed for the FSI problem. In Section \ref{sec:vascular-modeling}, we present the specifically designed techniques for vascular modeling, including the geometrical modeling as well as the prestress generation in the tissue. In Section \ref{sec:preconditioning-technique}, we discuss the solver technology for the FSI problem with a focus on the parallel implementation and preconditioner design. In the last, we present two numerical examples, including a patient-specific vascular model, to examine the overall FSI solver technology in Section \ref{sec:numerical_examples}. We present concluding remarks in Section \ref{sec:conclusion}.

\section{Unified continuum formulation for fluid-structure interaction}
\label{sec:unified-continuum-formulation-FSI}
\subsection{Kinematics on moving domains}
\begin{figure}
	\begin{center}
	\begin{tabular}{c}
\includegraphics[angle=0, trim=0 0 430 0, clip=true, scale=0.36]{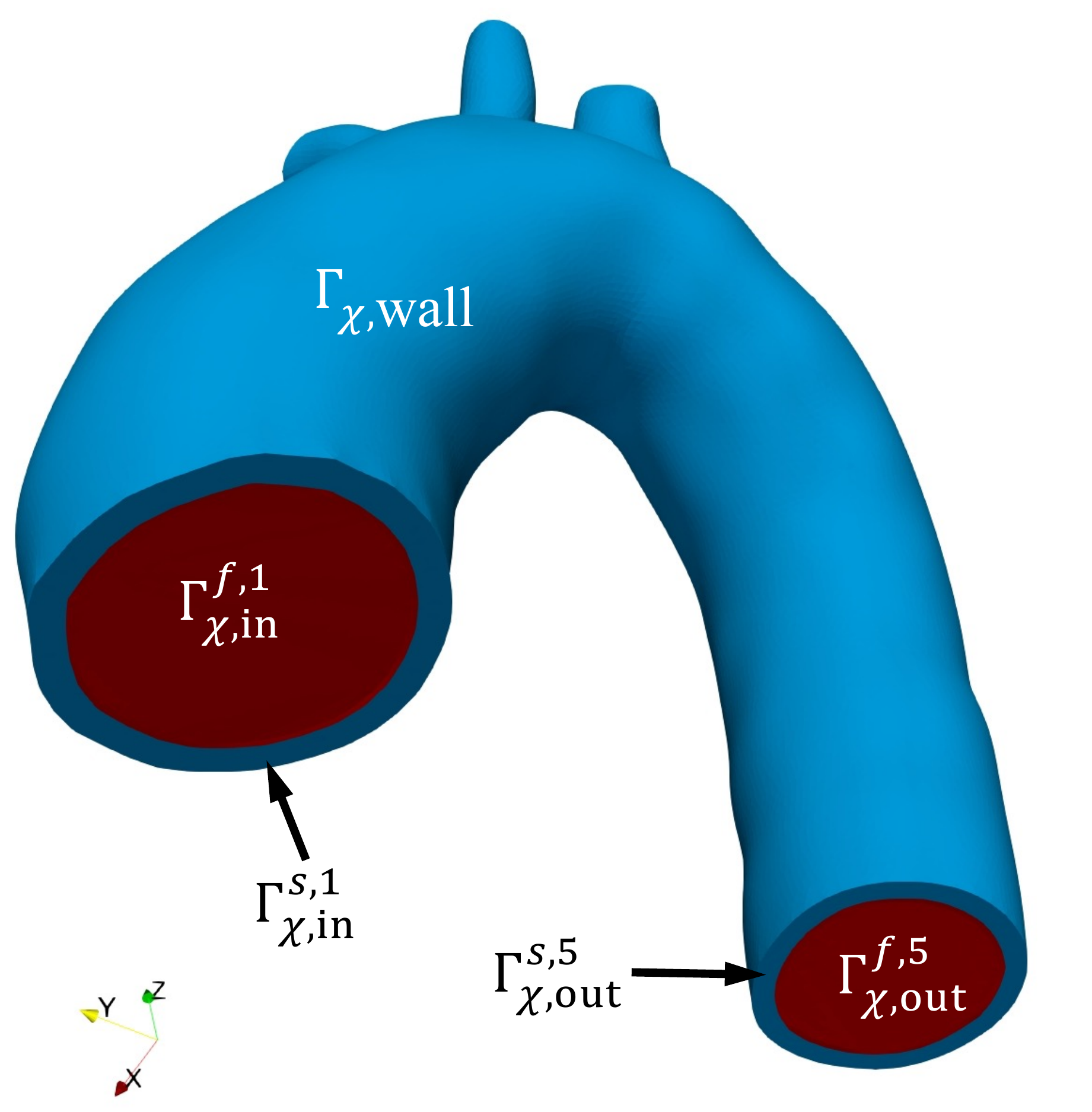}
\includegraphics[angle=0, trim=0 0 430 0, clip=true, scale=0.36]{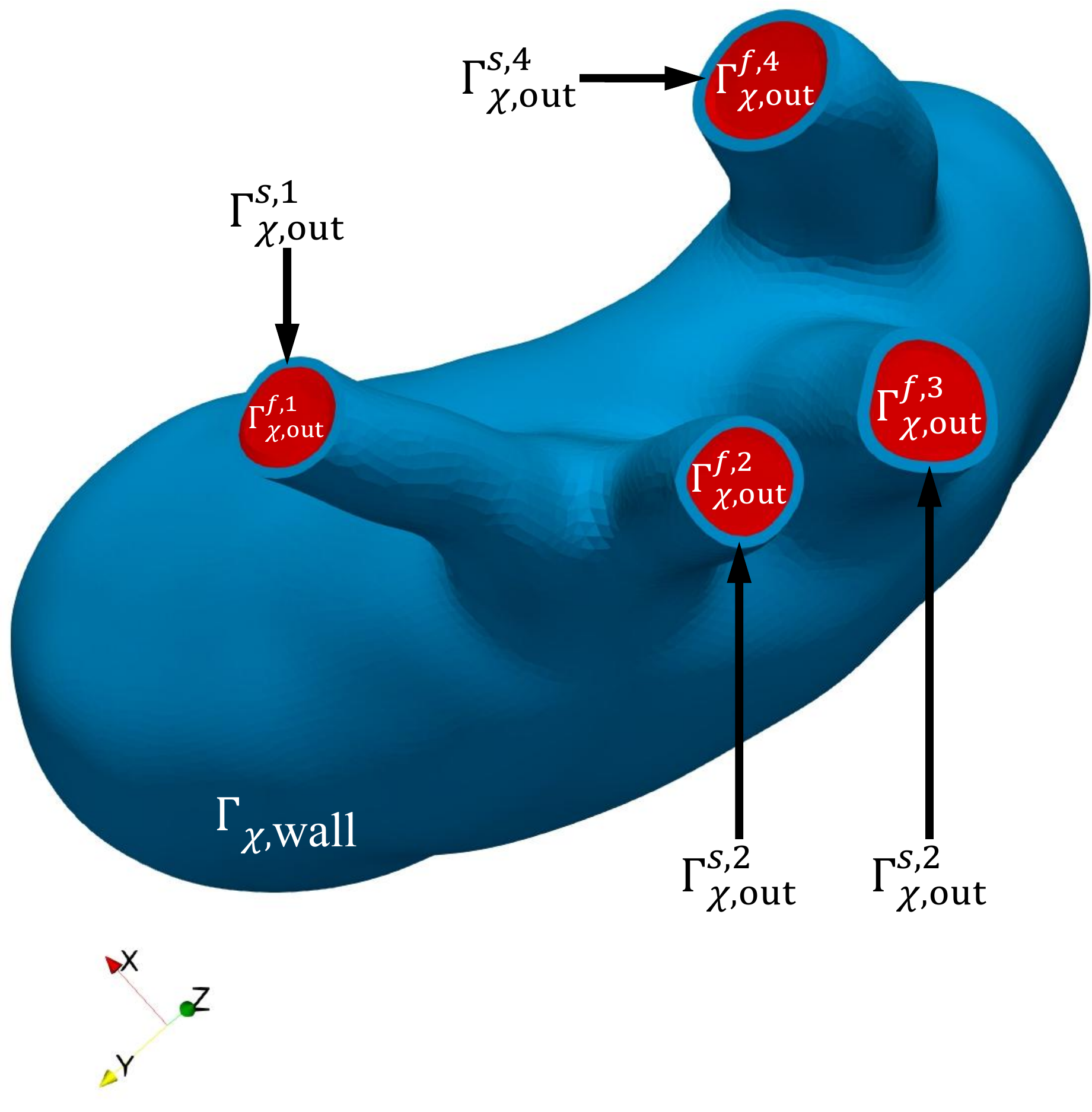}
\end{tabular}
\end{center}
\caption{Illustration of the FSI subdomains.}
\label{fig:illustration-fsi-subdomains}
\end{figure}

In this section, we present the initial-boundary value problem for FSI problems. The time interval of interest is denoted as $(0,T)$, with $T > 0$. We assume that there exists a referential frame $\Omega_{\bm \chi} \in \mathbb R ^3$, representing the domain occupied by the continuum body of interests. The referential configuration is fixed in time and is conceptually associated with a computational mesh. It also admits a non-overlapping subdivision by two subdomains $\Omega^f_{\bm \chi}$ and $\Omega^s_{\bm \chi}$,
\begin{align*}
\overline{\Omega}_{\bm \chi} = \overline{\Omega^f_{\bm \chi} \cup  \Omega^s_{\bm \chi}}, \quad \emptyset = \Omega^f_{\bm \chi} \cap \Omega^s_{\bm \chi}.
\end{align*}
The subdomains $\Omega^f_{\bm \chi}$ and $\Omega^s_{\bm \chi}$ are occupied by the fluid and solid materials, respectively. There is a two-dimensional manifold $\Gamma^I_{\bm \chi}$ representing the interface between $\Omega^f_{\bm \chi}$ and $\Omega^s_{\bm \chi}$,
\begin{align*}
\overline{\Gamma}^{I}_{\bm \chi} = \overline{\Omega}^f_{\bm \chi} \cap \overline{\Omega}^s_{\bm \chi}.
\end{align*}
On $\Gamma^{I}_{\bm \chi}$, the unit outward normal vectors to $\Omega^f_{\bm \chi}$ and $\Omega^s_{\bm \chi}$ are denoted as $\bm N^s$ and $\bm N^f$, respectively. The two unit outward normal vectors differ by a sign. The boundary of $\Omega_{\bm \chi}$ is denoted as $\Gamma_{\bm \chi} := \partial \Omega_{\bm \chi}$, and it enjoys a non-overlapping subdivision as
\begin{gather}
\label{eq:decomposition_gamma_chi_1}
\Gamma_{\bm \chi} = \Gamma_{\bm \chi,\mathrm{wall}} \cup \Gamma_{\bm \chi,\mathrm{in}} \cup \Gamma_{\bm \chi,\mathrm{out}}, \\
\Gamma_{\bm \chi,\mathrm{in}} = \Gamma^f_{\bm \chi,\mathrm{in}} \cup \Gamma^s_{\bm \chi,\mathrm{in}}, \quad \Gamma^f_{\bm \chi,\mathrm{in}} = \bigcup_{k=1}^{\mathrm{n}_{\mathrm{in}}} \Gamma^{f,k}_{\bm \chi,\mathrm{in}}, \quad \Gamma^s_{\bm \chi,\mathrm{in}} = \bigcup_{k=1}^{\mathrm{n}_{\mathrm{in}}} \Gamma^{s,k}_{\bm \chi,\mathrm{in}}, \\
\label{eq:decomposition_gamma_chi_2}
\Gamma_{\bm \chi,\mathrm{out}} = \Gamma^f_{\bm \chi,\mathrm{out}} \cup \Gamma^s_{\bm \chi,\mathrm{out}}, \quad \Gamma^f_{\bm \chi,\mathrm{out}} = \bigcup_{k=1}^{\mathrm{n}_{\mathrm{out}}} \Gamma^{f,k}_{\bm \chi,\mathrm{out}}, \quad \Gamma^{s}_{\bm \chi,\mathrm{out}} = \bigcup_{k=1}^{\mathrm{n}_{\mathrm{out}}} \Gamma^{s,k}_{\bm \chi,\mathrm{out}}.
\end{gather}
In \eqref{eq:decomposition_gamma_chi_1}, the boundary $\Gamma_{\bm \chi,\mathrm{wall}}$ represents the exterior wall surface that encloses the solid; $\Gamma_{\bm \chi,\mathrm{in}}$ represents the union of inlet surfaces; $\Gamma_{\bm \chi,\mathrm{out}}$ represents the union of outlet surfaces. On the boundary $\Gamma_{\bm \chi}$, there is a unit outward normal vector to $\Omega_{\bm \chi}$ and is denoted as $\bm N$. We demand that the fluid and solid surfaces on the inlets and outlets are co-planar, which means that the unit outward normal vectors on $\Gamma^{f,k}_{\bm \chi, \mathrm{out}}$ and $\Gamma^{s,k}_{\bm \chi, \mathrm{out}}$  ($\Gamma^{f,k}_{\bm \chi, \mathrm{in}}$ and $\Gamma^{s,k}_{\bm \chi, \mathrm{in}}$) are identical constant vectors for the same $k$ index. A schematic illustration of the subdomains, as well as the boundary subdivision, is given in Figure \ref{fig:illustration-fsi-subdomains}.

On the referential configuration $\Omega_{\bm \chi}$, there is a referential-to-Eulerian map $\bm x = \hat{\bm \varphi}(\bm \chi, t)$ defined, which maps $\Omega_{\bm \chi}$ to the current configuration $\Omega_{\bm x}(t)$. We assume that the map $\hat{\bm \varphi}$ is smooth for all arguments and is one-to-one for each $t\in (0,T)$. Correspondingly, the current domain admits a non-overlapping subdivision by two subdomains: $\Omega^{f}_{\bm x}(t)$ and $\Omega^{s}_{\bm x}(t)$. The interface between $\Omega^{f}_{\bm x}(t)$ and $\Omega^{s}_{\bm x}(t)$ is denoted as $\Gamma^{I}_{\bm x}(t) := \hat{\bm \varphi}(\Gamma^{I}_{\bm \chi}, t)$. On $\Gamma^{I}_{\bm x}(t)$, the unit outward normal vectors to $\Omega^{f}_{\bm x}(t)$ and $\Omega^{s}_{\bm x}(t)$ are denoted as $\bm n^f$ and $\bm n^s$, respectively. The notations for the boundary of $\Omega_{\bm x}(t)$ and its decomposition follow a similar fashion as \eqref{eq:decomposition_gamma_chi_1}-\eqref{eq:decomposition_gamma_chi_2} with the subscripts $\bm \chi$ replaced by $\bm x$. The unit outward normal vector on $\Gamma_{\bm x}$ is denoted by $\bm n$. We have to keep in mind that the boundary $\Gamma_{\bm x}(t)$ and its decomposition move in time with $\Omega_{\bm x}(t)$. 

In this work, we demand that the current configuration at time $t=0$ coincides with the referential configuration, which implies $\hat{\bm \varphi}(\cdot, 0) = \bm{id}(\cdot)$. Recalling that a fixed computational mesh is associated with $\Omega_{\bm \chi}$, the mapping $\hat{\bm \varphi}(\bm \chi, t)$ can be regarded as a description of the mesh deformation, and it induces
\begin{align*}
\hat{\bm U}(\bm \chi, t) := \hat{\bm \varphi}(\bm \chi, t) - \hat{\bm \varphi}(\bm \chi, 0) = \hat{\bm \varphi}(\bm \chi, t) - \bm \chi, \quad \hat{\bm V}(\bm \chi, t) := \left. \frac{\partial \hat{\bm \varphi}(\bm \chi, t)}{\partial t} \right|_{\bm \chi} = \left. \frac{\partial \hat{\bm U}(\bm \chi, t)}{\partial t} \right|_{\bm \chi},
\end{align*}
which are known as the mesh displacement and mesh velocity, respectively. The shifter \cite{Marsden1994} induced by $\hat{\bm \varphi}$ transports the vectors $\hat{\bm U}$ and $\hat{\bm V}$ to the current configuration,
\begin{align*}
\hat{\bm u}(\bm x, t) := \hat{\bm U}(\hat{\bm \varphi}^{-1}(\bm x , t), t), \quad \hat{\bm v}(\bm x, t) := \hat{\bm V}(\hat{\bm \varphi}^{-1}(\bm x , t), t).
\end{align*}

\begin{figure}[htb]
\begin{center}
\begin{tabular}{c}
\includegraphics[angle=0, trim=0 320 260 0, clip=true, scale = 0.55]{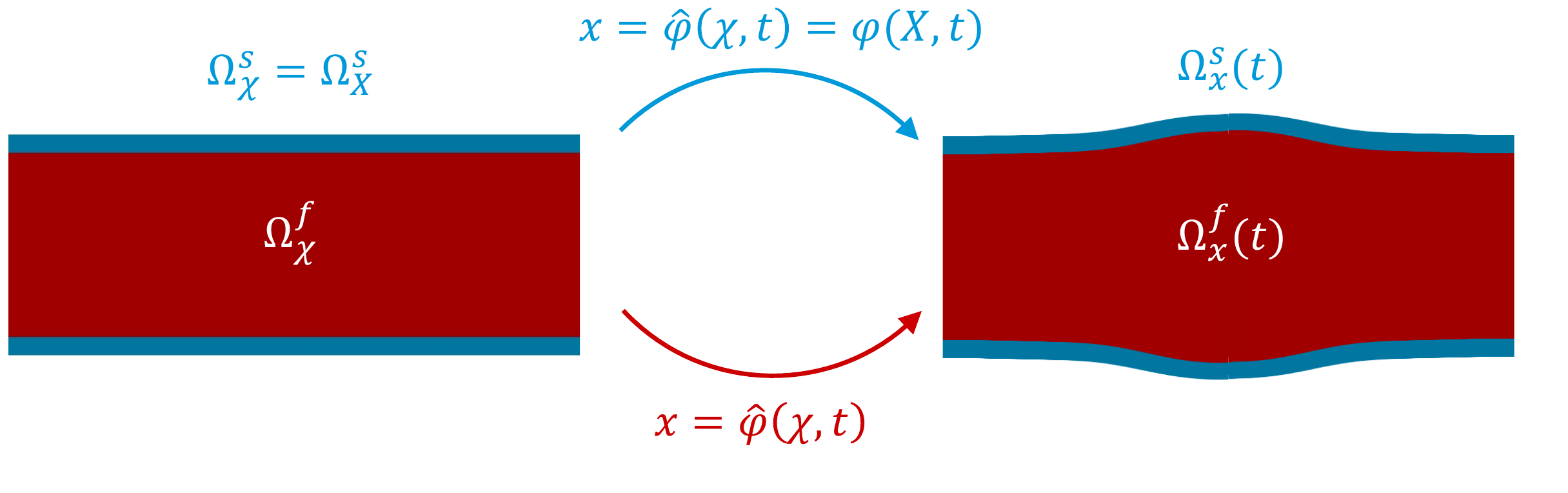} 
\end{tabular}
\end{center}
\caption{Illustration of the mappings between the configurations.}
\label{fig:fsi_mappings}
\end{figure}

Let $\Omega_{\bm X}$ denote the Lagrangian or material configuration, with a Lagrangian-to-Eulerian map $\bm x = \bm \varphi(\bm X,t)$ given. Similar to the notations used above, there is a non-overlapping subdivision of domain $\Omega_{\bm X}$, which is represented by the fluid subdomain $\Omega^{f}_{\bm X}$ and solid subdomain $\Omega^{s}_{\bm X}$. If we demand that the Lagrangian-to-Eulerian map is an identity map at time $t=0$ (i.e., $\bm \varphi(\bm X, 0) = \bm X$), the material particle's displacement and velocity can be defined as
\begin{align*}
\bm U(\bm X, t) := \bm \varphi(\bm X, t) - \bm \varphi(\bm X,0) = \bm \varphi(\bm X, t) - \bm X, \quad \bm V(\bm X, t) := \left. \frac{\partial \bm \varphi}{\partial t}\right|_{\bm X}= \left. \frac{\partial \bm U}{\partial t}\right|_{\bm X} = \frac{d\bm U}{dt},
\end{align*}
in which $d(\cdot)/dt$ designates a total time derivative. Similarly, the shifter induced by $\bm \varphi$ transports $\bm U$ and $\bm V$ to the current configuration by
\begin{align*}
\bm u(\bm x, t) := \bm U(\bm \varphi^{-1}(\bm x , t), t), \quad \bm v(\bm x, t) := \bm V(\bm \varphi^{-1}(\bm x , t), t).
\end{align*}
The deformation gradient, Jacobian determinant, and right Cauchy-Green deformation tensor of the material particle initially located at $\bm X$ are defined as
\begin{align*}
\bm F:= \frac{\partial \bm \varphi}{\partial \bm X}, \quad J := \textup{det}\left(\bm F \right), \quad \bm C := \bm F^T \bm F.
\end{align*}
The distortional parts of $\bm F$ and $\bm C$ are defined as
\begin{align*}
\tilde{\bm F} := J^{-\frac13} \bm F, \quad \tilde{\bm C} := J^{-\frac23} \bm C.
\end{align*}
Because solid materials exhibit the finite deformation property, it is often convenient to choose the referential configuration $\Omega^s_{\bm \chi}$ to be $\Omega^s_{\bm X}$. In doing so, the mesh motion coincides with the solid material particle's motion. An illustration of the mappings is provided in Figure \ref{fig:fsi_mappings}.

\subsection{Balance equations}
Under the isothermal hypothesis, the energy equation is decoupled from the mechanical system, and the FSI system can be viewed as a two-component continuum body governed by the following momentum and mass balance equations,
\begin{align}
\label{eq:continuum_momentum_eqn}
\bm 0 &= \rho(p) \left.  \frac{\partial \bm v}{\partial t}\right|_{\bm \chi}  + \rho(p) \left( \nabla_{\bm x} \bm v \right) \left( \bm v - \hat{\bm v} \right)  - \nabla_{\bm x} \cdot \bm \sigma_{\mathrm{dev}} + \nabla_{\bm x} p - \rho(p) \bm b, && \mbox{ in } \Omega_{\bm x}(t), \\
\label{eq:continuum_mass_eqn}
0 &= \left. \beta_{\theta}(p)\frac{\partial p}{\partial t}\right|_{\bm \chi} + \beta_{\theta}(p) \left( \bm v - \hat{\bm v} \right) \cdot \nabla_{\bm x} p + \nabla_{\bm x} \cdot \bm v, && \mbox{ in } \Omega_{\bm x}(t).
\end{align}
In the above, $\rho$ is the material density, $\bm \sigma_{\mathrm{dev}}$ is the deviatoric part of the Cauchy stress, $\bm b$ is the body force per unit mass, $\beta_{\theta}$ is the isothermal compressibility factor. The Cauchy stress $\bm \sigma$ is given by $\bm \sigma_{\mathrm{dev}} - p \bm I$. The constitutive relations for a specific material are dictated by a Gibbs free energy $G(\tilde{\bm C}, p)$, which had been shown to adopt a decoupled form \cite[p.~559]{Liu2018},
\begin{align*}
G(\tilde{\bm C}, p) = G_{\mathrm{ich}}(\tilde{\bm C}) + G_{\mathrm{vol}}(p),
\end{align*}
where $G_{\mathrm{ich}}$ and $G_{\mathrm{vol}}$ represent the isochoric and volumetric parts of the free energy, respectively. Notably, the pressure acts as an independent variable, and the volumetric part of the free energy $G_{\mathrm{vol}}$ is written as a function of pressure $p$ here. Given the free energy, the constitutive relations can be represented as
\begin{gather*}
\rho(p) = \left( \frac{d G_{\mathrm{vol}}}{dp} \right)^{-1}, \quad \beta_{\theta}(p) := \frac{1}{\rho} \frac{d\rho}{d p} = -\frac{d^2 G_{\mathrm{vol}}}{d p^2} / \frac{d G_{\mathrm{vol}}}{d p}, \displaybreak[2] \\
\bm \sigma_{\mathrm{dev}} = J^{-1} \tilde{\bm F} \left( \mathbb P : \tilde{\bm S} \right) \tilde{\bm F}^T + 2\bar{\mu} \textup{dev}[\bm d], \quad
\mathbb P := \mathbb I - \frac13 \bm C^{-1} \otimes \bm C, \quad
\tilde{\bm S} = 2 \frac{\partial \left(\rho_0 G \right) }{\partial \tilde{\bm C} }, \quad
\bm d := \frac12 \left(\nabla_{\bm x} \bm v + \nabla_{\bm x} \bm v^T \right),
\end{gather*}
where $\mathbb I$ is the fourth-order identity tensor, $\rho_0$ represents the density in the Lagrangian domain, $\bar{\mu}$ is the dynamic viscosity, and $\bm d$ is the rate-of-deformation tensor. Given an analytical form of the free energy $G$, one may complete the description of the governing system. It is worth pointing out that the above balance equations are written in the \textit{advective} form, meaning that the time and spatial derivatives are separately associated with the referential and current configurations. It is known that this form is amenable to the semi-discrete discretization approach \cite{Bazilevs2008} and automatically satisfies the geometric conservation laws \cite{Scovazzi2007}.

\subsubsection{Solid subproblem}
In the solid subdomain, we ignore the viscous effect (i.e., $\bar{\mu} = 0$) in this study. The Cauchy stress of the solid is denoted by $\bm \sigma^s := \bm \sigma^s_{\mathrm{dev}} - p \bm I$, and the constitutive relation for $\bm \sigma^s_{\mathrm{dev}}$ is given by
\begin{align*}
\bm \sigma^{s}_{\mathrm{dev}} = J^{-1} \tilde{\bm F} \left( \mathbb P : \tilde{\bm S} \right) \tilde{\bm F}^T.
\end{align*} 
Making use of the fact that $\hat{\bm v} = \bm v$ in $\Omega^s_{\bm x}(t)$, the balance equations \eqref{eq:continuum_momentum_eqn}-\eqref{eq:continuum_mass_eqn} are specialized to the following form,
\begin{align}
\label{eq:continuum_momentum_eqn_solid}
\bm 0 &= \rho^s(p) \left.  \frac{\partial \bm v}{\partial t}\right|_{\bm \chi}  - \nabla_{\bm x} \cdot \bm \sigma^s_{\mathrm{dev}} + \nabla_{\bm x} p - \rho^s(p) \bm b, && \mbox{ in } \Omega^s_{\bm x}(t), \displaybreak[2] \\
\label{eq:continuum_mass_eqn_solid}
0 &= \left. \beta^s_{\theta}(p)\frac{\partial p}{\partial t}\right|_{\bm \chi} + \nabla_{\bm x} \cdot \bm v, && \mbox{ in } \Omega^s_{\bm x}(t).
\end{align}
The motion of $\Omega^s_{\bm x}(t)$ is determined from the following,
\begin{align}
\label{eq:continuum_kinematics_solid}
\bm v = \frac{d \bm u}{dt}, \quad \mbox{ in } \Omega^s_{\bm x}(t).
\end{align}
We mention that the above formulation is less frequent to see in the solid mechanics community, partly because of the missing link between the compressibility factor $\beta^s_{\theta}$ and the classical constitutive theory. The relation between $\beta^s_{\theta}$ and the volumetric energy is elucidated through a Legendre transformation performed on the Helmholtz free energy with respect to $J$. \cite[pp.~562-564]{Liu2018}. In the incompressible limit, the mass equation \eqref{eq:continuum_mass_eqn_solid} reduces to a divergence-free condition for the velocity field. In this regard, there were some existing formulations \cite{Hoffman2011,Idelsohn2008} that bear some similarity to \eqref{eq:continuum_momentum_eqn_solid}-\eqref{eq:continuum_kinematics_solid}, and in those works, stress rates were used in expressing the constitutive relation. Although the rate-form or the hypoelastic-based constitutive theory suffers from the lack of a thermomechanical basis, nonphysical dissipative behavior in the absence of inelastic models, and challenges in numerical treatment \cite{Hughes1980,Simo1984}, it is still adopted nowadays likely due to the simplicity of its form and the inertia in using mature computer codes. In the last, it is known that the divergence-free velocity condition is equivalent to the constraint $J=1$ at the continuum level. The latter one is more commonly used with the Lagrange-multiplier method to derive a mixed formulation for finite elasticity \cite{Holzapfel2000}. The stability analysis of that type of mixed formulation is based on linearization \cite{Auricchio2005}, and it remains unclear if it is \textit{nonlinearly} stable. On the other side, an a priori stability estimate has been obtained for the system \eqref{eq:continuum_momentum_eqn_solid}-\eqref{eq:continuum_kinematics_solid} in the incompressible limit with the pressure shown to be bounded by invoking the inf-sup condition \cite{Liu2019,Liu2021}. These properties distinguish the presented formulation from the previously existing ones and are crucial for guaranteeing reliable numerical predictions \cite{Liu2018}.

\subsubsection{Fluid subproblem}
Assuming the free energy adopts the form $G(\tilde{\bm C}, p) = p/\rho^f$ in the fluid subdomain, the constitutive relations subsequently become $\rho(p) = \rho^f$, $\beta_{\theta}(p) = 0$,  $\bm \sigma^f_{\mathrm{dev}} = 2 \bar{\mu} \mathrm{dev}[\bm d]$,
and the balance equations \eqref{eq:continuum_momentum_eqn}-\eqref{eq:continuum_mass_eqn} can be instantiated as
\begin{align}
\label{eq:continuum_momentum_eqn_fluid}
\bm 0 &= \rho^f \left.  \frac{\partial \bm v}{\partial t}\right|_{\bm \chi}  + \rho^f \left( \nabla_{\bm x} \bm v \right) \left( \bm v - \hat{\bm v} \right)  - \nabla_{\bm x} \cdot \bm \sigma^f_{\mathrm{dev}} + \nabla_{\bm x} p - \rho^f \bm b, && \mbox{ in } \Omega^f_{\bm x}(t), \displaybreak[2] \\
\label{eq:continuum_mass_eqn_fluid}
0 &= \nabla_{\bm x} \cdot \bm v, && \mbox{ in } \Omega^f_{\bm x}(t).
\end{align}
The Cauchy stress here is given by $\bm \sigma^f := \bm \sigma^f_{\mathrm{dev}} - p \bm I$. The free energy here represents a fully incompressible model without elastic contribution. The system \eqref{eq:continuum_momentum_eqn_fluid}-\eqref{eq:continuum_mass_eqn_fluid} constitutes the incompressible Navier-Stokes equations written in the ALE coordinate system. To close the system, we need to provide a set of equations describing the motion of $\Omega^f_{\bm x}(t)$, or $\hat{\bm v}$ in \eqref{eq:continuum_momentum_eqn_fluid}. Typically, the motion is described via solving a system of elliptic differential equations equipped with proper boundary conditions to ensure smooth motions of the mesh with adherence to the interface $\Gamma^{I}_{\bm \chi}$. In this work, we determine $\hat{\bm U}$ as a harmonic extension of the trace of $\bm U^s$ on $\Gamma^I_{\bm \chi}$, that is,
\begin{align}
\label{eq:mesh-strong-eqn}
-\nabla_{\bm \chi} \cdot \left( \nabla_{\bm \chi} \hat{\bm U} \right) &= \bm 0, && \mbox{ in } \Omega^f_{\bm \chi}, \displaybreak[2] \\
\label{eq:mesh-strong-bc-1}
\hat{\bm U} &= \bm U^s, && \mbox{ on } \Gamma^{I}_{\bm \chi}, \displaybreak[2] \\
\label{eq:mesh-strong-bc-2}
\hat{\bm U} &= \bm 0, && \mbox{ on } \Gamma^{f}_{\bm \chi, \mathrm{in}} \cup \Gamma^{f}_{\bm \chi, \mathrm{out}}.
\end{align}
With the mesh displacement $\hat{\bm U}$ determined from solving the above equations, the mesh velocity can then be specified by
\begin{align}
\hat{\bm V} = \left. \frac{\partial \hat{\bm U}}{\partial t} \right|_{\bm \chi} \quad \mbox{ in } \Omega^f_{\bm \chi}.
\end{align}
This approach is essentially solving a Laplace equation for each component of the nodal displacement. The cost of solving this system resides in the assembly of the matrix, setup of its preconditioner, and matrix-vector multiplications in an iterative solver. In fact, the matrix only needs to be assembled once since it remains identical over time. In this regard, the harmonic extension algorithm is rather cost-effective. Yet, for problems with very large mesh deformation (e.g. the vortex-induced cantilever vibration benchmark \cite{Liu2018}), this method delivers ALE meshes with poor regularity, jeopardizing the robustness of the overall FSI formulation. It is therefore necessary to consider alternative methods such as the pseudo-linear-elasticity algorithm with fictitious Lam\'e parameters associated with the sizes of elements in the current mesh \cite{Johnson1994,Zeng2016}. In our experience, the harmonic extension algorithm is sufficiently robust in maintaining mesh regularity in tubular geometries, such as \textit{vascular} problems. It is thus the choice adopted in this work.

\subsection{Initial, boundary, and coupling conditions}
To complete the description of the FSI problem, the initial and boundary conditions need to be specified. Given the initial data $\bm v^f_0(\bm x)$, $p^f_0(\bm x)$, $\bm v^s_0(\bm x)$, and $p^s_0(\bm x)$, the initial conditions for the coupled system are given as follows,
\begin{align*}
& \bm v^f(\bm x, 0) = \bm v^f_0(\bm x), && p^f(\bm x, 0) = p^f_0(\bm x), && \mbox{for} \quad \bm x \in \Omega^f_{\bm x}(0), \displaybreak[2] \\
& \bm v^s(\bm x, 0) = \bm v^s_0(\bm x), && p^s(\bm x, 0) = p^s_0(\bm x), && \mbox{for} \quad \bm x \in \Omega^s_{\bm x}(0).
\end{align*}
Notice that in the fluid subdomain, the mesh motion equations \eqref{eq:mesh-strong-eqn} do not involve a time derivative, and there is thus no need to specify initial conditions for that static problem. In the solid subdomain, we have already chosen the Lagrangian-to-Eulerian map to be an identity map at time $t=0$, which directly implies that the initial displacement of the solid has to be zero, that is, $\bm u^s(\bm x, 0) = \bm 0$.

In terms of the boundary conditions, we impose homogeneous displacement boundary conditions on $\Gamma^{s}_{\bm X,\mathrm{in}} = \Gamma^{s}_{\bm \chi,\mathrm{in}}$ and $\Gamma^{s}_{\bm X,\mathrm{out}} = \Gamma^{s}_{\bm \chi,\mathrm{out}}$. On the exterior tissue surface $\Gamma^{s}_{\bm X,\mathrm{wall}}=\Gamma^{s}_{\bm \chi,\mathrm{wall}}$, a stress-free boundary condition $\bm \sigma^s \bm n = \bm 0$ is imposed. On each inlet surface $\Gamma^{f,k}_{\bm x,\mathrm{in}}$, the velocity field is specified by a prescribed profile $\bm v^{k}_{\mathrm{in}}(\bm x,t)$, i.e.,
\begin{align*}
\bm v = \bm v^{k}_{\mathrm{in}}(\bm x,t), \quad \mbox{ on } \quad \Gamma^{f,k}_{\bm x,\mathrm{in}}, \quad \mbox{for} \quad k = 1, \cdots, \mathrm{n}_{\mathrm{in}}.
\end{align*}
The inflow velocity $\bm v^{k}_{\mathrm{in}}$ is generated based on a parabolic or Womersley profile \cite{Takizawa2010} scaled by the volumetric flow rate. When the experimentally measured velocity field becomes available, it is also recommended using these data to impose the inflow. A recent validation study suggests the idealized parabolic profile may fail to capture local flow features \cite{Lan2022}. On each outlet surface $\Gamma^{f,k}_{\bm x,\mathrm{out}}$, a traction boundary condition is imposed,
\begin{align}
\label{eq:bc-fluid-outlet-general}
\bm \sigma^f \bm n = - P^{k}(t) \bm n, \quad \mbox{ on } \quad \Gamma^{f,k}_{\bm x,\mathrm{out}},
\end{align}
where $P^k(\cdot) : \mathbb R_{+} \rightarrow \mathbb R$ is a scalar function of time, representing the pressure on the $k$-th outlet surface. In \eqref{eq:bc-fluid-outlet-general}, the viscous contribution to the boundary traction is neglected as it is typically several orders of magnitude smaller than the pressure in hemodynamics. The function $P^k$ is typically given implicitly through the flow rate $Q^k(t)$ defined as
\begin{align*}
Q^k(t) := \int_{\Gamma^{f,k}_{\bm x,\mathrm{out}}} \bm v \cdot \bm n d\Gamma, \quad \mbox{for} \quad k = 1, \cdots \mathrm n_{\mathrm{out}}.
\end{align*}
In general, $P^k$ is determined by $Q^k$ through solving a reduced model, such as a one- or zero-dimensional model \cite{Vignon-Clementel2006a}. The reduced model provides the response of the downstream vasculature and is coupled with the three-dimensional problem via the above traction boundary condition. This is known as the Dirichlet-to-Neumann approach, belonging to the more general family of geometric multiscale coupling strategies \cite{Quarteroni2016}. The relation between $P^k$ and $Q^k$ is given by differential-algebraic equations in general, and in this work we use $P^k(t) = \mathcal F^k\left( Q^k(t) \right)$ to represent the \textit{functional} relationship between the two on the $k$-th outlet surface $\Gamma^{f,k}_{\bm x,\mathrm{out}}$.

\begin{remark}
The boundary conditions presented above can be properly generalized to account for more physiologically realistic conditions. The homogeneous displacement boundary condition imposed on $\Gamma^{s}_{\bm X,\mathrm{in}} \cup \Gamma^{s}_{\bm X,\mathrm{out}}$ can be replaced by applying inclined `roller' boundary conditions to allow motion of the vascular wall within the cut plane of the inlet and outlet surfaces \cite{Bazilevs2010,Griffiths1990,Hsu2014,Lan2022}. The stress-free boundary condition applied on $\Gamma^{s}_{\bm X,\mathrm{wall}}$ can be replaced by a Robin-type boundary condition to mimic the support from surrounding tissues \cite{Lan2022,Lan2022a,Moireau2012}. Moreover, the time-varying velocity fields imposed on $\Gamma^{f,k}_{\bm x,\mathrm{in}}$ can be modeled by the geometric multiscale approach through the Neumann-to-Dirichlet method \cite{Moghadam2013,Vignon-Clementel2006a}, which effectively models the upstream circulatory system.
\end{remark}

For the two subproblems, the following conditions are imposed over the fluid-solid interface to make them properly coupled and constitute an FSI problem. First, we demand that the velocity $\bm v$ is continuous across $\Gamma^{I}_{\bm x}$. This is known as the \textit{kinematic} coupling condition that equates the fluid velocity with solid velocity. This condition implicitly ensures the displacement is continuous across the fluid-solid interface as well. Second, we demand that the traction exerted by the fluid to the solid and the traction exerted by the solid to the fluid have the same magnitude but are in opposite directions, i.e.,
\begin{align}
\label{eq:dynamic-coupling-condition-strong}
\bm \sigma^f \bm n^f + \bm \sigma^s \bm n ^s = \bm 0, \quad \mbox{ on } \quad \Gamma^{I}_{\bm x}.
\end{align}
This is known as the \textit{dynamic} coupling condition. We mention specifically that the pressure typically experiences a jump across the fluid-solid interface. This can be explained by analyzing the stress distribution in a pressurized thick-walled cylinder \cite[Chapter~8]{Saad2009}. Although we solve for pressure as an independent variable in the unified formulation, we do not impose a continuity condition for it on $\Gamma^{I}_{\bm x}$.

\section{Numerical formulation}
\label{sec:numerical-formulation}
In this section, we present the numerical formulation designed for discretizing the FSI problem given in Section \ref{sec:unified-continuum-formulation-FSI}. Our numerical strategies include the VMS formulation for spatial discretization, the generalized-$\alpha$ scheme for temporal discretization, and the segregated algorithm based on the Newton-Raphson method for solving the nonlinear problem in a monolithically coupled approach. The above techniques combined provide an effective way to deal with the incompressibility constraint, turbulence modeling, and inertial-type structural dynamics integration. It also paves the way for designing a scalable iterative solution method, to be discussed in the next section.

\subsection{Solid subproblem}
Let $\mathcal S^s_{\bm u}$, $\mathcal S^{s}_{\bm v}$, and $\mathcal S^{s}_{p}$ denote the finite dimensional trial solution spaces for the solid displacement, velocity, and pressure in the current solid subdomain $\Omega^s_{\bm x}(t)$; let $\mathcal V^s_{\bm v}$, and $\mathcal V^s_{p}$ represent the test function spaces. The Dirichlet boundary conditions for the solid displacement and velocity are built into the definition of the discrete function spaces. The spatial discretization of \eqref{eq:continuum_momentum_eqn_solid}-\eqref{eq:continuum_kinematics_solid} is based on the VMS formulation \cite{Liu2018,Zeng2017} with the purpose of providing pressure stabilization for equal-order interpolations. The formulation can be stated as follows. Find
\begin{align*}
\bm y_h^s(t) := \left\lbrace  \bm u_h^s(t), p_h^s(t), \bm v_h^s(t)\right\rbrace^T \in \mathcal S_{\bm u}^s \times \mathcal S_{p}^s \times \mathcal S_{\bm v}^s
\end{align*}
such that
\begin{align}
& \mathbf B^s_{k}\left( \dot{\bm y}_h^s, \bm y_h^s \right) = \bm 0, \displaybreak[2] \\
& \mathbf B^s_{m}\left( \bm w^s; \dot{\bm y}_h^s, \bm y_h^s \right) = 0, && \forall \bm w^s \in \mathcal V_{\bm v}^s, \displaybreak[2] \\
& \mathbf B^s_{p}\left( w^s; \dot{\bm y}_h^s, \bm y_h^s \right) = 0, && \forall w^s \in \mathcal V_{p}^s,
\end{align}
wherein
\begin{align}
& \mathbf B^s_{k}\left( \dot{\bm y}_h^s, \bm y_h^s \right) := \frac{d\bm u_h^s}{dt} - \bm v_h^s, \displaybreak[2] \\
\label{eq:fsi_solid_momentum}
& \mathbf B^s_{m}\left( \bm w^s; \dot{\bm y}_h^s, \bm y_h^s \right) := \int_{\Omega_{\bm x}^s(t)} \bm w^s \cdot \rho^s(p^s_h) \frac{d\bm v_h^s}{dt} + \nabla_{\bm x} \bm w^s : \bm \sigma^s_{\mathrm{dev}}(\bm u^s_h) - \nabla_{\bm x} \cdot \bm w^s p_h^s - \bm w^s \cdot \rho^s(p^s_h) \bm b d\Omega_{\bm x} \nonumber \displaybreak[2]  \\
& \hspace{15mm} + \int_{\Omega_{\bm x}^{\prime s}(t)} \nabla_{\bm x} \cdot \bm w^s \tau_{\mathrm{C}}^s \left(\beta_{\theta}^s(p^s_h) \frac{dp_h^s}{dt} + \nabla_{\bm x} \cdot \bm v_h^s \right) d\Omega_{\bm x}, \\
& \mathbf B^s_{p}\left( w^s; \dot{\bm y}_h^s, \bm y_h^s \right) := \int_{\Omega_{\bm x}^s(t)} w^s \left( \beta_{\theta}^s(p^s_h) \frac{dp_h^s}{dt} + \nabla_{\bm x} \cdot \bm v_h^s \right) d\Omega_{\bm x} \nonumber \displaybreak[2]  \\
& \hspace{15mm} + \int_{\Omega_{\bm x}^{\prime s}(t)} \nabla_{\bm x} w^s \cdot \bm \tau_{\mathrm{M}}^s \left( \rho^s(p_h^s) \frac{d\bm v_h^s}{dt} - \nabla_{\bm x} \cdot \bm \sigma^s_{\mathrm{dev}}(\bm u^s_h) + \nabla_{\bm x} p^s_h - \rho^s(p^s_h) \bm b \right) d\Omega_{\bm x}.
\end{align}
In the above VMS formulation, the parameters $\bm \tau_{\mathrm{M}}^s$ and $\tau_{\mathrm{C}}^s$ characterize the subgrid-scale velocity and pressure, respectively. In this work, they are defined as
\begin{align}
\label{eq:fsi_solid_tau_designs}
\bm \tau_{\mathrm{M}}^s = \tau_{\mathrm{M}}^s \bm I, \quad \tau_{\mathrm{M}}^s = c_{\mathrm{m}} \frac{\Delta x}{c\rho^s}, \quad \tau_{\mathrm{C}}^s = c_{\mathrm{c}} c \Delta x \rho^s,
\end{align}
where $\Delta x$ is the diameter of the smallest sphere circumscribing the tetrahedral element, $c_{\mathrm{m}}$ and $c_{\mathrm{c}}$ are non-dimensional constants, and $c$ is the maximum wave speed in the solid material. The value of $c$ can be conveniently estimated from a small-strain theory \cite[p.~572]{Liu2018}, which is the choice made in this work. Alternatively, one may obtain the maximum wave speed from an eigenvalue analysis for the nonlinear elastodynamic problem \cite{Bonet2015}. The design of the stabilization parameters \eqref{eq:fsi_solid_tau_designs} was first proposed in the setting of linear elastodynamics \cite{Hughes1988}, and recent investigations show that it is also effective in a range of nonlinear solid problems \cite{Abboud2018,Abboud2021,Liu2018,Liu2020b,Zeng2017}.

\subsection{Fluid subproblem}
Let $\mathcal S^f_{\bm u}$ denote the trial solution space of the mesh displacement $\hat{\bm u}^f_h$ defined on the fluid current domain $\Omega^f_{\bm x}(t)$, with the boundary condition \eqref{eq:mesh-strong-bc-1}-\eqref{eq:mesh-strong-bc-2} built into the definition of $\mathcal S^f_{\bm u}$; let $\mathcal S^f_{\bm v}$ and $\mathcal S^f_{p}$ denote the trial solution spaces of the fluid velocity and pressure; let $\mathcal V^f_{p}$ and $\mathcal V^f_{\bm v}$ be the test function spaces. The Dirichlet boundary condition is built into the definition of $\mathcal S^f_{\bm v}$. The VMS formulation for the fluid subproblem can be stated as follows. Find
\begin{align*}
\bm y_h^f(t):= \left\lbrace \hat{\bm u}_h(t), p_h^f(t), \bm v_h^f(t) \right\rbrace^T \in \mathcal S_{\bm u}^f \times \mathcal S^f_{p} \times \mathcal S^f_{\bm v}
\end{align*}
such that
\begin{align}
& \mathbf B^f_{k}\left(  \hat{\bm w}^f ; \dot{\bm y}_h^f, \bm y_h^f \right) = 0, && \forall \hat{\bm w}^f \in \mathcal V^f_{\bm u}, \displaybreak[2] \\
& \mathbf B^f_m \left( \bm w^f ;  \dot{\bm y}_h^f, \bm y_h^f \right) = 0, && \forall \bm w^f \in \mathcal V^f_{\bm v}, \displaybreak[2] \\
& \mathbf B^f_p\left( w^f; \dot{\bm y}^f_h, \bm y_h^f \right) = 0, && \forall w^f \in \mathcal V^f_{p},
\end{align}
where
\begin{align}
\label{eq:fsi_fluid_mesh}
& \mathbf B^f_{k}\left(  \hat{\bm w}^f ; \dot{\bm y}_h^f, \bm y_h^f \right) := \int_{\Omega_{\bm \chi}^f} \nabla_{\bm \chi} \hat{\bm w}^f \cdot \nabla_{\bm \chi} \hat{\bm u}_h d\Omega_{\bm \chi}, \displaybreak[2] \\
\label{eq:fsi_fluid_residual_based_vms_momentum}
& \mathbf B^f_m \left( \bm w^f ;  \dot{\bm y}_h^f, \bm y_h^f \right) := \int_{\Omega_{\bm x}^f(t)} \bm w^f \cdot \left( \left. \rho^f \frac{\partial \bm v_h^f}{\partial t} \right|_{\bm \chi} + \rho^f \left( \nabla_{\bm x} \bm v_h^f \right) \left(\bm v_h^f - \hat{\bm v}_h \right) - \rho^f \bm b \right) d\Omega_{\bm x} \nonumber \displaybreak[2] \\
& \hspace{3mm} - \int_{\Omega_{\bm x}^f(t)} \nabla_{\bm x} \cdot \bm w^f p_h^f d\Omega_{\bm x} + \int_{\Omega_{\bm x}^f(t)} \nabla_{\bm x} \bm w^f : \bm \sigma_{\mathrm{dev}}^f(\bm v^f_h) d\Omega_{\bm x} + \sum_{k=1}^{\mathrm{n}_{\mathrm{out}}}\int_{\Gamma^{f,k}_{\bm x,\mathrm{out}}(t)} \bm w^f \cdot \bm n \mathcal{F}^k\left( Q^k \right) d\Gamma_{\bm x} \nonumber \displaybreak[2] \\
& \hspace{3mm} - \int_{\Omega_{\bm x}^{\prime f}(t)} \nabla_{\bm x} \bm w^f : \left( \rho^f \bm v^{f\prime} \otimes \left( \bm v_h^f - \hat{\bm v}_h \right) \right) d\Omega_{\bm x} + \int_{\Omega_{\bm x}^{\prime f}(t)} \nabla_{\bm x} \bm v^f_h : \left( \rho^f \bm w^f \otimes \bm v^{f\prime} \right) d\Omega_{\bm x} \nonumber \displaybreak[2] \\
& \hspace{3mm} - \int_{\Omega_{\bm x}^{\prime f}(t)} \nabla_{\bm x} \bm w^f : \left( \rho^f \bm v^{f\prime} \otimes \bm v^{f\prime} \right) d\Omega_{\bm x} - \int_{\Omega_{\bm x}^{\prime f}(t)} \nabla_{\bm x} \cdot \bm w^f p^{f\prime} d\Omega_{\bm x}, \displaybreak[2] \\
\label{eq:fsi_fluid_residual_based_vms_mass}
& \mathbf B^f_p\left( w^f ; \dot{\bm y}_h^f, \bm y_h^f \right) := \int_{\Omega_{\bm x}^f(t)} w^f \nabla_{\bm x} \cdot \bm v_h^f d\Omega_{\bm x}  - \int_{\Omega_{\bm x}^{\prime f}(t)} \nabla_{\bm x} w^f \cdot  \bm v^{f\prime} d\Omega_{\bm x}, \displaybreak[2] \\
& \bm v^{f\prime} := -\bm \tau_{M}^f \left( \left. \rho^f \frac{\partial \bm v_h^f}{\partial t} \right|_{\bm \chi} + \rho^f \left( \nabla_{\bm x} \bm v_h^f \right) \left(\bm v_h^f - \hat{\bm v}_h \right) + \nabla_{\bm x} p_h^f - \nabla_{\bm x} \cdot \bm \sigma^f_{\mathrm{dev}}(\bm v^f_h) - \rho^f \bm b \right), \displaybreak[2] \\
& p^{f\prime} := -\tau^f_C \nabla_{\bm x} \cdot \bm v_h^f.
\end{align}
In the above, $\bm v^{f\prime}$ and $p^{f\prime}$ are the fine-scale velocity and pressure, which are proportional to the corresponding coarse-scale residuals. The proportionality factors, $\bm \tau^f_{\mathrm{M}}$ and $\tau^f_{\mathrm{C}}$, were obtained by scaling arguments in the numerical analysis of stabilized finite element methods and adopt the following form,
\begin{align}
\label{eq:fsi_fluid_def_tau_m}
& \bm \tau^f_M := \tau^f_M \bm I, \quad \tau^f_M := \frac{1}{\rho^f}\left( \frac{C_T}{\Delta t^2} + \left(\bm v_h^f - \hat{\bm v}^m_h \right) \cdot \bm G \left(\bm v_h^f - \hat{\bm v}^m_h \right) + C_I \left( \frac{\bar{\mu}}{\rho^f} \right)^2 \bm G : \bm G \right)^{-\frac12}, \quad \tau^f_C := \frac{1}{\tau_M \textup{tr}\bm G}, \displaybreak[2] \\
\label{eq:def_K_for_scale_G}
& G_{ij} := \sum_{k=1}^{3} \frac{\partial \xi_k}{\partial x_i} M_{kl} \frac{\partial \xi_l}{\partial x_j}, \quad \bm M = [ M_{kl} ] = \frac{\sqrt[3]{2}}{2}\begin{bmatrix}
2 & 1 & 1 \\
1 & 2 & 1 \\
1 & 1 & 2
\end{bmatrix}, \quad \bm G : \bm G := \sum_{i,j=1}^{3} G_{ij} G_{ij}, \quad \mathrm{tr}\bm G := \sum_{i=1}^{3} G_{ii}.
\end{align}
In the above, $\bm \xi = \left\lbrace \xi_i \right\rbrace_{i=1}^{3}$ are the coordinates of an element in the parent domain. The parameter $C_I$ relies on the polynomial order of the interpolation basis functions and takes the value of $36$ for linear interpolations \cite{Figueroa2006a,Franca1992}. The value of $C_T$ is taken to be $4$ \cite{Bazilevs2007,Shakib1991}. In \eqref{eq:def_K_for_scale_G}, $\bm M$ is introduced for simplex elements because the standard reference element is non-symmetric, and nodal permutations lead to variance in the definition of $\bm G$ without $\bm M$. The entries of $\bm M$ are determined by mapping the reference element to a regular simplex without changing the element volume \cite{Pauli2017,Danwitz2019}.

\begin{figure}[htb]
\begin{center}
\begin{tabular}{ccc}
\includegraphics[angle=0, trim=0 130 310 0, clip=true, scale = 0.28]{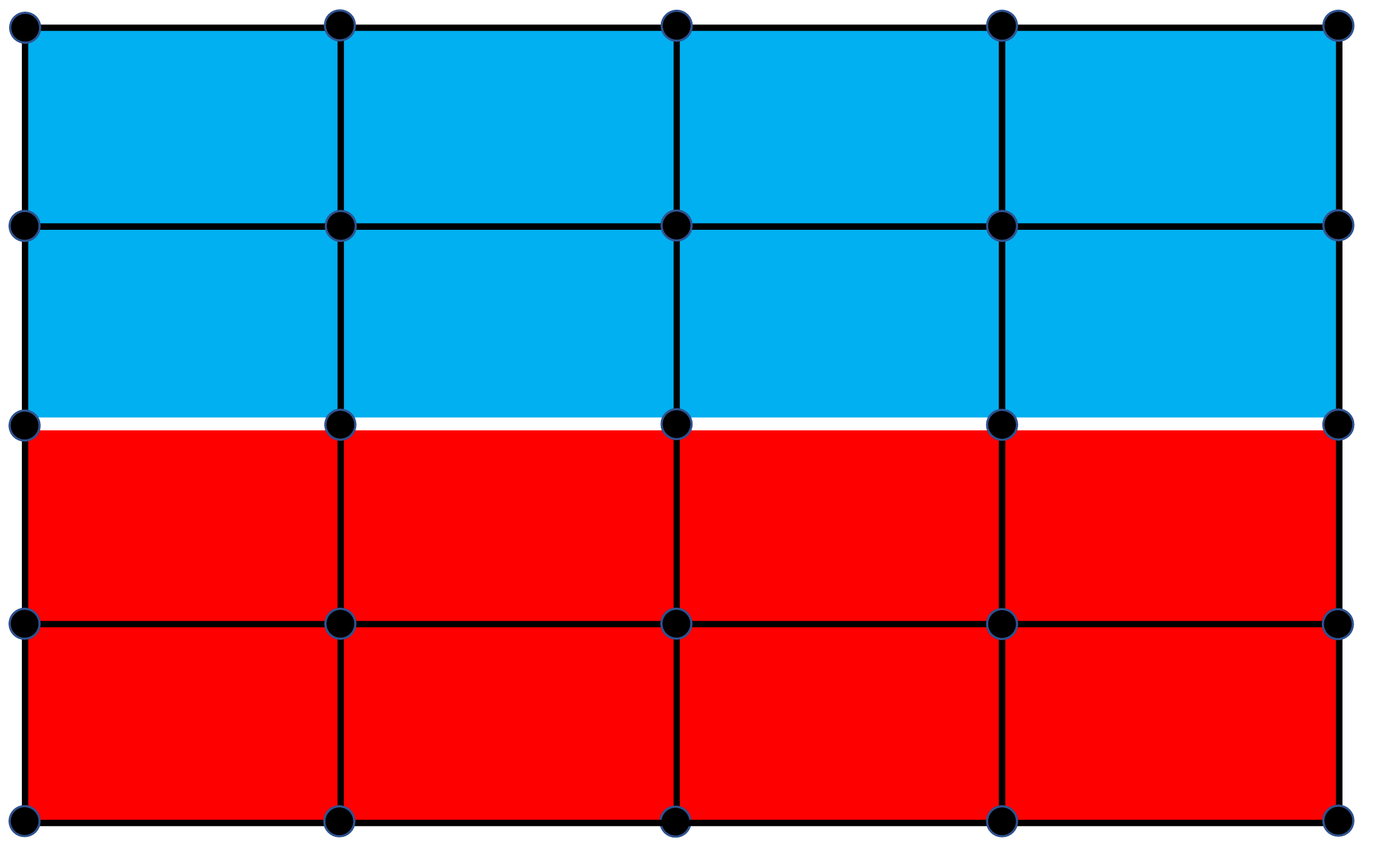} &
\includegraphics[angle=0, trim=0 130 310 0, clip=true, scale = 0.28]{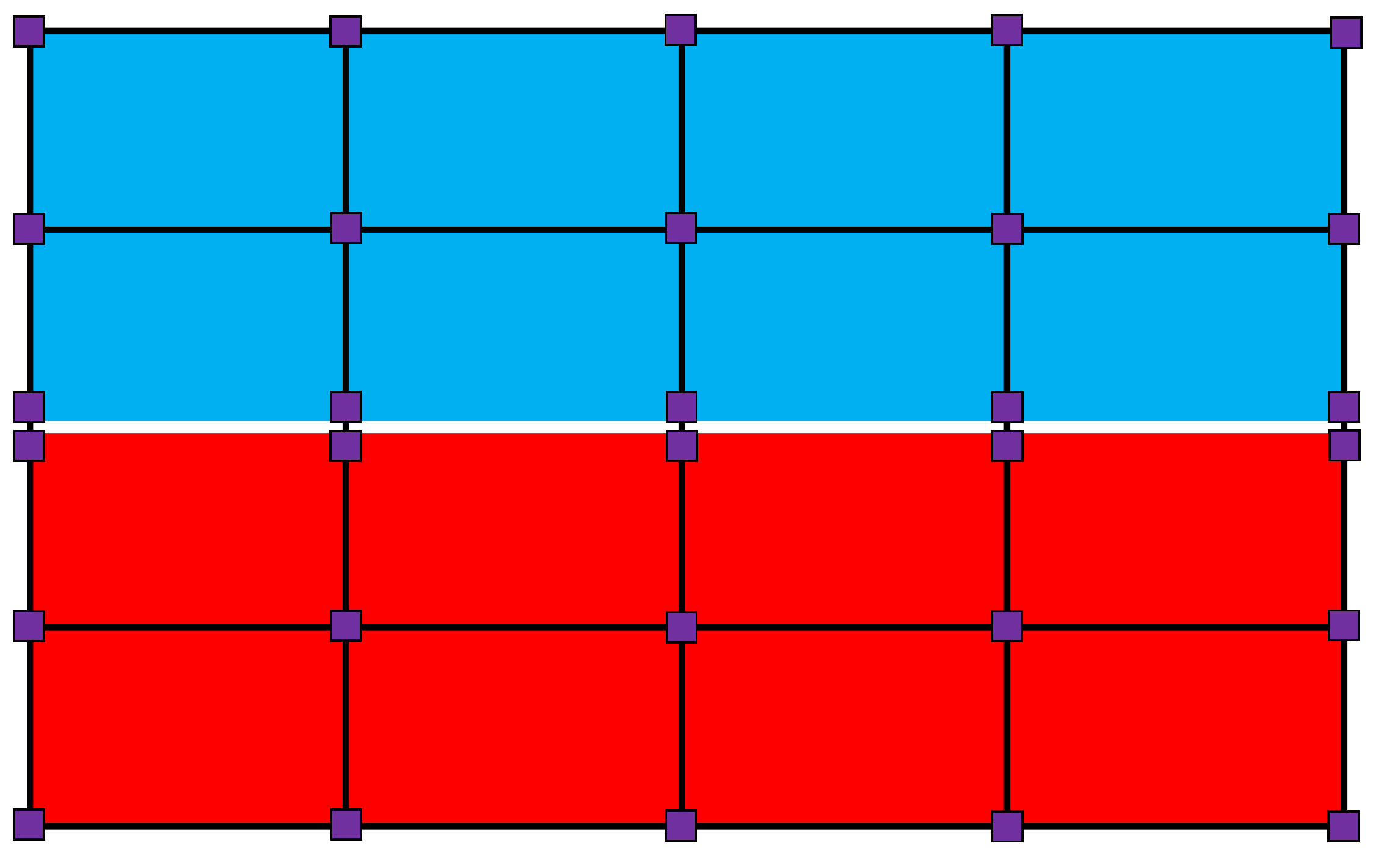} \\
\multicolumn{2}{c}{
\includegraphics[angle=0, trim=0 450 120 0, clip=true, scale = 0.28]{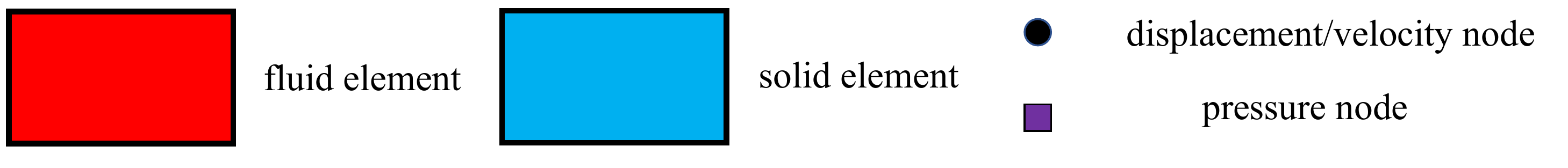}
}
\end{tabular}
\end{center}
\caption{Illustrartion of an FSI mesh. There is a set of additional pressure nodes on the fluid-solid interface.}
\label{fig:fsi_mesh}
\end{figure}

\subsection{Semi-discrete formulation for the FSI coupled problem}
\label{sec:semi-discrete-FSI}
The above semi-discrete formulations are defined on two subdomains separately. In each subdomain, there are seven DOFs to be solved. The solid subdomain involves the displacement, velocity, and pressure, while the fluid subdomain involves the mesh displacement, velocity, and pressure. To ensure proper coupling of the two systems, we consider a single set of mesh generated for the whole domain, with the fluid and solid elements sharing the same set of nodes across the interface. The basis functions associated with the kinematic fields are assumed to be at least $C^0$-continuous. Let
\begin{align*}
\bm u_h(t) :=
\begin{cases}
 \bm u_h^s(t) \in \mathcal S_{\bm u}^s \\
 \hat{\bm u}_h(t) \in \mathcal S^f_{\bm u}
\end{cases} \quad \mbox{ and } \quad
\bm v_h(t) :=
\begin{cases}
 \bm v_h^s(t) \in \mathcal S^s_{\bm v} \\
 \bm v_h^f(t) \in \mathcal S^f_{\bm v}
\end{cases}
\end{align*}
be the displacement and velocity for the continuum body which are represented by the above basis functions and satisfy the essential boundary conditions. The corresponding test function
\begin{align*}
\bm w_h :=
\begin{cases}
 \bm w^s_h \in \mathcal V^s_{\bm u} \\
 \bm w^f_h \in \mathcal V^f_{\bm u}
\end{cases}
\end{align*}
is written by the same set of basis functions and satisfies homogeneous essential boundary conditions. The $C^0$-continuity of the chosen basis functions automatically guarantees that
\begin{align}
\label{eq:coupling-condition-kinematics}
\bm u^s_h(t) = \hat{\bm u}_h(t), \quad \bm v^s_h(t) = \bm v^f_h(t), \quad \bm w^s_h = \bm w^f_h, \quad \mbox{ on } \Gamma^I_{\bm x}.
\end{align}
The first two of \eqref{eq:coupling-condition-kinematics} indicate that the kinematic coupling condition is guaranteed to be satisfied by construction. For the discretization of pressure, we consider
\begin{align*}
p_h(t) :=
\begin{cases}
 p^s_h(t) \in \mathcal S^s_{p} \\
 p^f_h(t) \in \mathcal S^f_{p}
\end{cases},
\end{align*}
whose continuity is released on the fluid-solid interface $\Gamma^I_{\bm x}$, allowing a pressure jump across $\Gamma^I_{\bm x}$. This can be achieved by directly working on the aforesaid mesh. We modify the element-node connectivity array (i.e., the IEN array \cite[Chapter~2]{Hughes1987}) for the solid element such that the nodes on the interface are associated with the newly generated indices. In other words, there is a set of additional variables and equations differentiating the solid pressure from the fluid pressure on the fluid-solid interface, even when using equal-order interpolations. A schematic illustration of the FSI mesh is given in Figure \ref{fig:fsi_mesh}. Let us now denote the solution vector in $\Omega_{\bm x}(t)$ as
\begin{align}
\bm y_h(t) :=
\begin{cases}
\bm y_h^s(t) = \lbrace  \bm u_h^s(t), p_h^s(t), \bm v_h^s(t) \rbrace^T \in \mathcal S_{\bm u}^s \times \mathcal S_{p}^s \times \mathcal S_{\bm v}^s \displaybreak[2] \\
\bm y_h^f(t) = \lbrace  \hat{\bm u}_h(t), p_h^f(t), \bm v_h^f(t) \rbrace^T \in \mathcal S^f_{\bm u} \times \mathcal S_{p}^f \times \mathcal S_{\bm v}^f
\end{cases},
\end{align}
which has seven DOFs in both subdomains. The coupled FSI problem can be stated as follows. Find $\bm y_h(t)$ such that
\begin{align}
\label{eq:spatial_form_FSI_k}
& \mathbf B^s_k(\dot{\bm y}^s_h, \bm y^s_h) = \bm 0, \quad \mbox{ in } \Omega^s_{\bm x}(t), \quad \mathbf B^f_k\left(\hat{\bm w}; \hat{\bm u}_h \right) = 0, \quad \forall \hat{\bm w} \in \mathcal V^f_{\bm u}, \quad \mbox{ in } \Omega^f_{\bm x}(t), \\
\label{eq:spatial_form_FSI_m}
& \mathbf B_m( \bm w; \dot{\bm y}_h, \bm y_h ) := \mathbf B^s_m( \bm w^s; \dot{\bm y}^s_h, \bm y^s_h ) + \mathbf B^f_m( \bm w^f; \dot{\bm y}^f_h, \bm y^f_h ) = 0, && \forall \bm w^s \in \mathcal V^s_{\bm v}, \bm w^f \in \mathcal V^f_{\bm v}, \\
\label{eq:spatial_form_FSI_p}
& \mathbf B_p( w; \dot{\bm y}_h, \bm y_h) := \mathbf B^s_p( w^s; \dot{\bm y}^s_h, \bm y^s_h ) + \mathbf B^f_p( w^f; \dot{\bm y}^f_h, \bm y^f_h) = 0, && \forall w^s \in \mathcal V^s_{p}, w^f \in \mathcal V^f_{p}.
\end{align}
Deriving the Euler-Lagrange equations for \eqref{eq:spatial_form_FSI_m} reveals that, in addition to the satisfaction of the balance equations as well as the traction boundary conditions in \eqref{eq:spatial_form_FSI_m}, one also has the following satisifed on the interface $\Gamma^I_{\bm x}$,
\begin{align*}
\int_{\Gamma^I_{\bm x}} \bm w^s \cdot \left( \bm \sigma^s_{\mathrm{dev}} - p^s_h\bm I \right) \bm n^s + \bm w^f \cdot \left( \bm \sigma^f_{\mathrm{dev}} - p^f_h\bm I \right) \bm n^f d\Gamma_{\bm x} = 0.
\end{align*}
With \eqref{eq:coupling-condition-kinematics}$_{3}$ and the fact that $\bm n^s = - \bm n^f$, the above can be rewritten as
\begin{align}
\label{eq:dynamic-coupling-condition}
\int_{\Gamma^I_{\bm x}} \bm w^s \cdot \left( \left( \bm \sigma^s_{\mathrm{dev}} - p^s_h\bm I \right) \bm n^s -  \left( \bm \sigma^f_{\mathrm{dev}} - p^f_h\bm I \right) \bm n^s \right) d\Gamma_{\bm x} = 0,
\end{align}
which indicates a weak satisfaction of the dynamic coupling condition \eqref{eq:dynamic-coupling-condition-strong}. This and the kinematic coupling condition \eqref{eq:coupling-condition-kinematics} constitute a complete characterization of the FSI coupling condition in the semi-discrete formulation \eqref{eq:spatial_form_FSI_k}-\eqref{eq:spatial_form_FSI_p}, which is inherited at the fully discrete level.

\begin{remark}
In this work, we apply equal-order interpolations for all fields using piecewise linear basis functions defined on an unstructured tetrahedral mesh. The VMS formulation applied introduces a pressure stabilization mechanism that helps circumvent the Ladyzhenskaya-Babu\v{s}ka-Brezzi condition in both subproblems. In fluid subproblem, the VMS formulation also provides a turbulence model for large eddy simulation of incompressible flows.
\end{remark}

\begin{remark}
The release of pressure continuity across the interface $\Gamma^{I}_{\bm x}$ is critical for maintaining a proper coupling condition. Otherwise, a continuous pressure interpolation across $\Gamma^{I}_{\bm x}$ will lead to $p^s_h = p^f_h$. The pressure continuity condition combined with \eqref{eq:dynamic-coupling-condition} further results in
\begin{align*}
\int_{\Gamma^{I}_{\bm x}} \bm w^s \cdot \left( \sigma^s_{\mathrm{dev}} \bm n^s - \sigma^f_{\mathrm{dev}} \bm n^s \right)d\Gamma_{\bm x} = 0.
\end{align*}
The above conditions are apparently stronger than \eqref{eq:dynamic-coupling-condition} and non-physical. Releasing the pressure continuity is also necessary for generating prestress in the vascular tissue (see Section~\ref{subsec:prestressing}).
\end{remark}

\begin{remark}
We notice that the kinematic equations \eqref{eq:spatial_form_FSI_k} in the solid subdomain are not written in a weighted residual form, simply because they do not involve a spatial differential operator. Indeed, one may weight those equations by a set of properly chosen test functions in an integral form and will eventually get a mass matrix on both sides of the equations. After applying the inverse of the mass matrices, the equations \eqref{eq:spatial_form_FSI_k} will be recovered. Readers may refer to Section 3.3 of \cite{Liu2019a} for more details.
\end{remark}

\subsection{Temporal discretization}
In the semi-discrete formulation, the balance laws are stated in the advective form in the sense that the time derivatives are taken with the referential coordinate $\bm \chi$ held fixed, while the spatial derivatives are taken with respect to the current spatial coordinate $\bm x$. Let $\mathcal N_A(\bm \chi)$ denote a basis function defined on the referential configuration $\Omega_{\bm \chi}$. For an arbitrary function $F(\bm \chi, t)$ defined on $\Omega_{\bm \chi}$ in terms of the basis functions, we may express it as
\begin{align*}
F(\bm \chi, t) = \sum_{A} F_A(t) \mathcal N_A(\bm \chi).
\end{align*}
The basis function in the current configuration can be obtained by pushing $\mathcal N_A(\bm \chi)$ forward as $N_A(\bm x, t) = \mathcal N_A( \hat{\bm \varphi}^{-1}(\bm x, t) )$. Correspondingly, the function $f(\bm x,t) := F(\hat{\bm \varphi}^{-1}_t(\bm x, t), t)$ can be represented in terms of $N_A(\bm x, t)$ as
\begin{align*}
f(\bm x, t) = \sum_{A} F_A(t) \mathcal N_A( \hat{\bm \varphi}^{-1}(\bm x, t) ) = \sum_{A} F_A(t) N_A(\bm x, t).
\end{align*}
The referential time derivative of $f$ and the spatial gradient can be calculated, respectively, by
\begin{align}
\label{eq:temporal_ale_derivatives}
\left. \frac{\partial f}{\partial t} \right|_{\bm \chi} = \sum_{A} \frac{dF_A}{dt}(t) N_A(\bm x, t), \quad
\nabla_{\bm x} f = \sum_{A} F_A(t) \nabla_{\bm x}N_A(\bm x, t).
\end{align}
We perform temporal discretization of the semi-discrete problem \eqref{eq:spatial_form_FSI_k}-\eqref{eq:spatial_form_FSI_p} by collocating the solution variables $\bm y_h(t)$ and their time derivatives $\dot{\bm y}_h(t)$ in time by the generalized-$\alpha$ method \cite{Jansen2000}. The above expressions suggest that the referential time derivative $\dot{\bm y}_h(t)$ can be calculated conveniently using \eqref{eq:temporal_ale_derivatives}$_{1}$ on the current configuration. The time interval $(0,T)$ is subdivided into $n_{\mathrm{ts}}$ sub-intervals $(t_n, t_{n+1})$ of size $\Delta t_n := t_{n+1} - t_n$. Let $\bm y_n$ and $\dot{\bm y}_n$ denote the approximations of the semi-discrete solutions $\bm y_h(t)$ and its time derivative $\dot{\bm y}_h(t)$ at time $t_n$. Let $N_A(\bm x,t)$ and $M_A(\bm x,t)$ denote the basis functions of the velocity components and pressure on the current configuration, respectively. The solution variables in the two subdomains can be represented as
\begin{align*}
\bm y^s_n := \lbrace \bm u^s_n, p^s_n, \bm v^s_n \rbrace^T, \quad \bm y^f_n := \lbrace \hat{\bm u}_n, p^f_n, \bm v^f_n \rbrace^T, \quad \dot{\bm y}^s_n := \lbrace \dot{\bm u}^s_n, \dot{p}^s_n, \dot{\bm v}^s_n \rbrace^T, \quad \mbox{and} \quad \dot{\bm y}^f_n := \lbrace \hat{\bm v}_n, \dot{p}^f_n, \dot{\bm v}^f_n \rbrace^T.
\end{align*}
Notice that we have the discrete mesh displacement and mesh velocity stored in $\bm y^f_n$ and $\dot{\bm y}^f_n$, respectively. Let $\bm e_i$ denote the $i$-th Cartesian basis vector. We define the residual vectors as
\begin{align*}
\mathbf R_k(\dot{\bm y}_n, \bm y_n) :=&
\begin{cases}
\dot{\bm u}^s_n - \bm v^s_n \\
\left\lbrace \mathbf B^f_k( N_A \bm e_i; \dot{\bm y}_n, \bm y_n ) \right \rbrace
\end{cases}, \displaybreak[2] \\
\mathbf R_m(\dot{\bm y}_n, \bm y_n) :=& \left\lbrace \mathbf B_m( N_A \bm e_i; \dot{\bm y}_n, \bm y_n )  \right\rbrace, \displaybreak[2] \\
\mathbf R_p(\dot{\bm y}_n, \bm y_n) :=& \left\lbrace \mathbf B_p( M_A; \dot{\bm y}_n, \bm y_n )  \right\rbrace,
\end{align*}
in which the vector entries are collected such that the index $A$ spans the index set of test functions and the index $i$ spans the dimension of space (i.e. $3$ in this work). The fully discrete formulation can be stated as follows. At time step $t_{n+1}$, given $\bm y_n$ and $\dot{\bm y}_n$, find $\bm y_{n+1}$ and $\dot{\bm y}_{n+1}$ such that
\begin{gather}
\label{eq:temporal-scheme-1}
\mathbf R_k(\dot{\bm y}_{n+\alpha_m}, \bm y_{n+\alpha_f}) = \bm 0, \quad \mathbf R_m(\dot{\bm y}_{n+\alpha_m}, \bm y_{n+\alpha_f}) = \bm 0, \quad \mathbf R_p(\dot{\bm y}_{n+\alpha_m}, \bm y_{n+\alpha_f}) = \bm 0, \displaybreak[2] \\
\dot{\bm y}_{n+\alpha_m} = \dot{\bm y}_n + \alpha_m \left( \dot{\bm y}_{n+1} - \dot{\bm y}_n \right), \quad \bm y_{n+\alpha_f} = \bm y_n + \alpha_f \left( \bm y_{n+1} - \bm y_n \right), \displaybreak[2] \\
\label{eq:temporal-scheme-3}
\bm y_{n+1} = \bm y_n + \Delta t_n \dot{\bm y}_n + \gamma \Delta t_n \left( \dot{\bm y}_{n+1} - \dot{\bm y}_n \right).
\end{gather}
The parameters $\alpha_m$, $\alpha_f$, and $\gamma$ define the above temporal integration scheme, and they are parameterized by $\varrho_{\infty}$, the spectral radius of the amplification matrix at the highest mode, as follows,
\begin{align*}
\alpha_m = \frac12 \left( \frac{3-\varrho_{\infty}}{1+\varrho_{\infty}} \right), \quad \alpha_f = \frac{1}{1+\varrho_{\infty}}, \quad \gamma = \frac{1}{1+\varrho_{\infty}}.
\end{align*}
The above scheme is known as the \textit{first-order} generalized-$\alpha$ scheme because it was originally designed for first-order ordinary differential equations in CFD analysis \cite{Jansen2000}. Interestingly, its parameterization is different from that of its \textit{second-order} counterpart, which was designed for hyperbolic problems \cite{Chung1993}. This discrepancy has led to a dilemma in the conventional FSI formulation because one has to deal with a first-order time derivative in the fluid subproblem and a second-order time derivative in the solid subproblem. Using two different parameterizations inevitably results in a mismatch of the discrete acceleration on the fluid-solid interface. On the other hand, using a uniform parameterization will sacrifice the dynamic behavior in one subdomain. Interested readers may refer to \cite[pp.~119-120]{Bazilevs2012} for a discussion, and the structural dynamics was chosen to be sacrificed there in the conventional FSI formulation. We highlight that the aforesaid issue does not arise here in \eqref{eq:temporal-scheme-1}-\eqref{eq:temporal-scheme-3}, because both subproblems are written as first-order systems and can thus be integrated in time using the optimal parameterization. One may argue that the cost of writing the elastodynamics as a first-order system is the introduction of an additional set of velocity DOFs. It will be shown in Section \ref{sec:segregated-predictor-multi-corrector} that a segregation algorithm is designed for the velocity variables, and they will be segregatedly updated through an explicit approach. In other words, the additional cost is merely saving the DOFs in memory and an explicit update in each multi-corrector iteration. Furthermore, it was recently discovered that the first-order generalized-$\alpha$ scheme does not exhibit the overshoot phenomenon in a numerical test of structural dynamics \cite{Kadapa2017}. We confirm that observation by theoretically analyzing the amplification matrix, which is detailed in \ref{sec:overshoot-analysis}. In the last, we point out that the pressure variable is collocated at the intermediate time step, rather than at $t_{n+1}$ \cite{Bazilevs2008,Bazilevs2012}, in the above scheme to maintain the second-order temporal accuracy \cite{Liu2021a}. 

We mention that the geometric multiscale coupling term 
\begin{align*}
\sum_{k=1}^{\mathrm{n}_{\mathrm{out}}}\int_{\Gamma^{f,k}_{\bm x,\mathrm{out}}(t)} \bm w^f \cdot \bm n \mathcal{F}^k\left( Q^k \right) d\Gamma_{\bm x}
\end{align*}
in \eqref{eq:fsi_fluid_residual_based_vms_momentum} is discretized in time as
\begin{align}
\label{eq:reduced-model-temporal-discretization}
\sum_{k=1}^{\mathrm{n}_{\mathrm{out}}}\int_{\Gamma^{f,k}_{\bm x,\mathrm{out}}(t)} \bm w^f \cdot \bm n \left( (1-\alpha_f) \mathcal{F}^k\left( Q^k_{n} \right) + \alpha_f \mathcal{F}^k\left( Q^k_{n+1} \right) \right) d\Gamma_{\bm x},
\end{align}
where
\begin{align*}
Q^k_n := \int_{\Gamma^{f,k}_{\bm x,\mathrm{out}}(t)} \bm v_n \cdot \bm n d\Gamma_{\bm x}, \quad \mbox{and} \quad Q^k_{n+1} := \int_{\Gamma^{f,k}_{\bm x,\mathrm{out}}(t)} \bm v_{n+1} \cdot \bm n d\Gamma_{\bm x}.
\end{align*}
The discretization choice \eqref{eq:reduced-model-temporal-discretization} for the geometric multiscale coupling has been also verified to guarantee the second-order temporal accuracy \cite{Liu2021a}. The reduced model is typically represented through a set of algebraic-differential equations, making an analytical solution unattainable in general. Consequently, the evaluation of the functional relations $\mathcal{F}^k( Q^k_n )$ and $\mathcal{F}^k( Q^k_{n+1} )$ requires a separated time integration within each time step. In this work, this is performed by an explicit fourth-order Runge-Kutta method and subdividing each time interval $(t_n, t_{n+1})$ into equal-sized subintervals. Readers may refer to Algorithm 1 in \cite{Liu2020} for details about integrating the reduced models.

\subsection{Segregated predictor multi-corrector algorithm}
\label{sec:segregated-predictor-multi-corrector}
In the solution of the above fully discrete system, we need to solve a set of nonlinear algebraic equations in each time step. The solution vector can be obtained by an iterative algorithm, and we denote $\bm y_{n+1,(l)}$ as the solution vector at the time step $t_{n+1}$ and the iteration step $0 \leq l \leq l_{\mathrm{max}}$. The nonlinear algebraic equations evaluated at the $l$-th iteration are denoted as
\begin{align}
\label{eq:consistent_residual}
& \mathbf R_{(l)} := \left\lbrace \mathbf R_{k,(l)}; \mathbf R_{m,(l)}; \mathbf R_{p,(l)} \right\rbrace, \displaybreak[2] \\
& \mathbf R_{k,(l)} := \mathbf R_k( \dot{\bm y}_{n+\alpha_m, (l)}, \bm y_{n+\alpha_f, (l)} ), \displaybreak[2] \nonumber \\
& \mathbf R_{m,(l)} := \mathbf R_m( \dot{\bm y}_{n+\alpha_m, (l)}, \bm y_{n+\alpha_f, (l)} ), \displaybreak[2] \nonumber \\
& \mathbf R_{p,(l)} := \mathbf R_p( \dot{\bm y}_{n+\alpha_m, (l)}, \bm y_{n+\alpha_f, (l)} ). \nonumber
\end{align}
The consistent tangent matrix of the nonlinear system can be written as
\begin{align}
\label{eq:consistent_tangent_3x3}
\boldsymbol{\mathrm K}_{(l)} =
\begin{bmatrix}
\boldsymbol{\mathrm K}_{k, (l), \dot{\bm u}} & \boldsymbol{\mathrm O} & \boldsymbol{\mathrm K}_{k, (l), \dot{\bm v}} \\[0.3mm]
\boldsymbol{\mathrm K}_{m, (l), \dot{\bm u}} & \boldsymbol{\mathrm K}_{m, (l), \dot{p}} & \boldsymbol{\mathrm K}_{m, (l), \dot{\bm v}} \\[0.3mm]
\boldsymbol{\mathrm K}_{p, (l), \dot{\bm u}} & \boldsymbol{\mathrm O} & \boldsymbol{\mathrm K}_{p, (l), \dot{\bm v}}
\end{bmatrix},
\end{align}
wherein
\begin{align*}
\boldsymbol{\mathrm K}_{(\cdot), (l), \dot{\bm u}} := \alpha_m \frac{\partial \mathbf R_{(\cdot)}( \dot{\bm y}_{n+\alpha_m, (l)}, \bm y_{n+\alpha_f, (l)} )}{\partial \dot{\bm u}_{n+\alpha_m}} + \alpha_f \gamma \Delta t_n \frac{\partial \mathbf R_{(\cdot)}( \dot{\bm y}_{n+\alpha_m, (l)}, \bm y_{n+\alpha_f, (l)} )}{\partial \bm u_{n+\alpha_f}}, \displaybreak[2] \\
\boldsymbol{\mathrm K}_{m, (l), \dot{\bm p}} := \alpha_m \frac{\partial \mathbf R_{m}( \dot{\bm y}_{n+\alpha_m, (l)}, \bm y_{n+\alpha_f, (l)} )}{\partial \dot{\bm p}_{n+\alpha_m}} + \alpha_f \gamma \Delta t_n \frac{\partial \mathbf R_{m}( \dot{\bm y}_{n+\alpha_m, (l)}, \bm y_{n+\alpha_f, (l)} )}{\partial \bm p_{n+\alpha_f}}, \displaybreak[2] \\
\boldsymbol{\mathrm K}_{(\cdot), (l), \dot{\bm v}} := \alpha_m \frac{\partial \mathbf R_{(\cdot)}( \dot{\bm y}_{n+\alpha_m, (l)}, \bm y_{n+\alpha_f, (l)} )}{\partial \dot{\bm v}_{n+\alpha_m}} + \alpha_f \gamma \Delta t_n \frac{\partial \mathbf R_{(\cdot)}( \dot{\bm y}_{n+\alpha_m, (l)}, \bm y_{n+\alpha_f, (l)} )}{\partial \bm v_{n+\alpha_f}},
\end{align*}
and the symbol $(\cdot)$ in the subscripts can be replaced by $k$, $m$, or $p$. The explicit definitions of the residual vector \eqref{eq:consistent_residual} and consistent tangent matrix \eqref{eq:consistent_tangent_3x3}  take different forms in the fluid and solid subdomains. We use a superscript $f$ or $s$ to refer to the entries belonging to specific subdomains.

Restricting to the solid subproblem, the first row of \eqref{eq:consistent_tangent_3x3} is specialized to
\begin{align*}
\boldsymbol{\mathrm K}^{s}_{k, (l), \dot{\bm u}} = \alpha_m \bm I, \qquad \boldsymbol{\mathrm K}^{s}_{k, (l), \dot{\bm v}} = -\alpha_f \gamma \Delta t_n \bm I,
\end{align*}
and the tangent matrix \eqref{eq:consistent_tangent_3x3} enjoys a block factorization (see \cite[Sec.~4.4]{Liu2018} for details). This factorization results in the following segregated algorithm without losing the consistency of the Newton-Raphson procedure. One may first determine $\Delta \dot{\bm v}^{s}_{n+1,(l)}$ and $\Delta \dot{p}^{s}_{n+1,(l)}$ from solving the following system,
\begin{align}
\label{eq:segregated_matrix_solid}
\begin{bmatrix}
\boldsymbol{\mathrm K}^{s}_{m, (l), \dot{\bm v}} + \frac{\alpha_f \gamma \Delta t_n}{\alpha_m} \boldsymbol{\mathrm K}^{s}_{m, (l), \dot{\bm u}} & \boldsymbol{\mathrm K}^{s}_{m, (l), \dot{\bm p}}  \\[1.2mm]
\boldsymbol{\mathrm K}^{s}_{p, (l), \dot{\bm v}} + \frac{\alpha_f \gamma \Delta t_n}{\alpha_m} \boldsymbol{\mathrm K}^{s}_{p, (l), \dot{\bm u}} & \boldsymbol{\mathrm K}^{s}_{p, (l), \dot{\bm p}} 
\end{bmatrix}
\begin{bmatrix}
\Delta \dot{\bm v}^{s}_{n+1,(l)} \\[1.2mm]
\Delta \dot{p}^{s}_{n+1,(l)}
\end{bmatrix}
= -
\begin{bmatrix}
\mathbf R^{s}_m - \frac{1}{\alpha_m} \boldsymbol{\mathrm K}^{s}_{m, (l), \dot{\bm u}} \mathbf R^{s}_{k,(l)}  \\[1.2mm]
\mathbf R^{s}_p - \frac{1}{\alpha_m} \boldsymbol{\mathrm K}^{s}_{p, (l), \dot{\bm u}} \mathbf R^{s}_{k,(l)} 
\end{bmatrix},
\end{align}
and the incremental $\Delta \dot{\bm u}^{s}_{n+1,(l)}$ can be obtained as
\begin{align}
\label{eq:segregard_disp_vec_solid}
\Delta \dot{\bm u}^{s}_{n+1,(l)} = \frac{\alpha_f \gamma \Delta t_n}{\alpha_m} \Delta \dot{\bm v}^{s}_{n+1,(l)} - \frac{1}{\alpha_m} \mathbf R^{s}_{k,(l)}.
\end{align}
Furthermore, it has been proven that $\mathbf R^{s}_{k,(l)} = \bm 0$ for $l \geq 2$ in Proposition 5 of \cite{Liu2018}. In practice, we choose $\mathbf R^{s}_{k,(l)} = \bm 0$  for all $l \geq 1$ in \eqref{eq:segregated_matrix_solid} to simplify the formation of the right-hand side of the linear system. Both theoretical analysis \cite[Appendix~B]{Liu2018} and numerical experiences \cite{Liu2019,Liu2019a,Lan2022b} suggest that this choice does not harm the overall algorithmic robustness.  It is worthy of mentioning that the update of the displacement incremental in \eqref{eq:segregard_disp_vec_solid} still necessitates a consistent definition of $\mathbf R^{s}_{k,(l)}$.

Considering the fluid subproblem, the consistent tangent matrix enjoys the following form,
\begin{align}
\label{eq:consistent_tangent_3x3_fluid}
\boldsymbol{\mathrm K}^f_{(l)} =
\begin{bmatrix}
\boldsymbol{\mathrm K}^{f}_{k, (l), \dot{\bm u}} & \boldsymbol{\mathrm O} & \boldsymbol{\mathrm O}  \\[0.3mm]
\boldsymbol{\mathrm K}^{f}_{m, (l), \dot{\bm u}} & \boldsymbol{\mathrm K}^{f}_{m, (l), \dot{p}} & \boldsymbol{\mathrm K}^{f}_{m, (l), \dot{\bm v}} \\[0.3mm]
\boldsymbol{\mathrm K}^{f}_{p, (l), \dot{\bm u}} & \boldsymbol{\mathrm K}^{f}_{p, (l), \dot{p}} & \boldsymbol{\mathrm K}^{f}_{p, (l), \dot{\bm v}}
\end{bmatrix}.
\end{align}
A segregation strategy can also be performed for linear problems associated with matrices of the form \eqref{eq:consistent_tangent_3x3_fluid}. One may determine the mesh velocity incremental from the equation
\begin{align}
\label{eq:segregated_fluid_mesh_velo_incremental}
\boldsymbol{\mathrm K}^{f}_{k, (l), \dot{\bm u}} \Delta \hat{\bm v}_{n+1,(l)} = - \mathbf R^{f}_{k,(l)},
\end{align}
and update the mesh velocity by the solved incremental. From $\Delta \hat{\bm v}_{n+1,(l)}$, we may conveniently obtain the incremental for the mesh displacement as $\Delta \hat{\bm u}_{n+1,(l)} = \gamma \Delta t_n \Delta \hat{\bm v}_{n+1,(l)}$, which can be used to update the mesh displacement. After that, the values of $\Delta \dot{\bm v}^{f}_{n+1,(l)}$ and $\Delta \dot{p}^{f}_{n+1,(l)}$ are obtained from solving the following system,
\begin{align}
\label{eq:segregated_matrix_fluid}
\begin{bmatrix}
\boldsymbol{\mathrm K}^{f}_{m, (l), \dot{\bm v}} & \boldsymbol{\mathrm K}^{f}_{m, (l), \dot{\bm p}}  \\[1.2mm]
\boldsymbol{\mathrm K}^{f}_{p, (l), \dot{\bm v}}  & \boldsymbol{\mathrm K}^{f}_{p, (l), \dot{p}}
\end{bmatrix}
\begin{bmatrix}
\Delta \dot{\bm v}^{s}_{n+1,(l)} \\[1.2mm]
\Delta \dot{p}^{s}_{n+1,(l)}
\end{bmatrix}
= -
\begin{bmatrix}
\mathbf R^{f}_m - \boldsymbol{\mathrm K}^{f}_{m, (l), \dot{\bm u}} \Delta \hat{\bm v}_{n+1,(l)}  \\[1.2mm]
\mathbf R^{f}_p - \boldsymbol{\mathrm K}^{f}_{p, (l), \dot{\bm u}} \Delta \hat{\bm v}_{n+1,(l)}
\end{bmatrix}.
\end{align}
Inspired from the quasi-direct coupling strategy \cite[Chapter~6]{Bazilevs2012}, we replace the right-hand side of \eqref{eq:segregated_matrix_fluid} by
\begin{align*}
-\begin{bmatrix} \mathbf R^{f}_{m,(l)} ; \mathbf R^{f}_{p,(l)}\end{bmatrix}^{T}.
\end{align*}
This simplification ignores the contribution of $\Delta \hat{\bm v}_{n+1,(l)}$ to the right-hand side of \eqref{eq:segregated_matrix_fluid}. The assembly of the left-hand side matrix in \eqref{eq:segregated_matrix_fluid} still necessitates the newest mesh displacement. 

It is worthy of noting that the consistent linearization of \eqref{eq:reduced-model-temporal-discretization} leads to a non-standard matrix in the definition of $\boldsymbol{\mathrm K}^{f}_{m, (l), \dot{\bm v}}$, adopting the following form,
\begin{align}
\label{eq:geometric_multiscale_coupling_matrix}
\alpha_f \gamma \Delta t_n \sum_{k=1}^{\mathrm{n}_{\mathrm{out}}} \left( \mathfrak m^k_{(l)} \bm a^k \bm a^{kT} \right), \quad \mbox{ with } \quad \mathfrak{m}^k_{n+1,(l)} := \frac{\partial P^{k}_{n+1,(l)}}{\partial Q^{k}_{n+1,(l)}},
\end{align}
and the component of the vector $\bm a^k$ is defined as
\begin{align*}
\bm a^k_{Ai} := \int_{\Gamma^{f,k}_{\bm x,\mathrm{out}}(t)} \bm N_A \bm n_i  d\Gamma_{\bm x}.
\end{align*}
Recall that the indices $A$ and $i$ span the set of test functions and the spatial dimension, respectively. Since the evaluation of $P^k_{n+1,(l)}$ is achieved through an algorithm, the value of $\mathfrak{m}^k_{(l)}$ has to be obtained algorithmically as well. The approximation of $\mathfrak{m}^k_{n+1,(l)}$ is obtained via a difference formula. One first obtain
\begin{align}
\label{eq:tilde_P_and_hat_P}
\tilde{P}^k_{n+1,(l)} := \mathcal{F}^k(\bm Q^k_{n+1,(l)} +\epsilon/2) \quad \mbox{ and } \quad \hat{P}^k_{n+1,(l)} := \mathcal{F}^k(\bm Q^k_{n+1,(l)} - \epsilon/2)
\end{align}
with $\epsilon := \max\lbrace \epsilon_{\mathrm{abs}}, \epsilon_{\mathrm{rel}} |Q^k_{n+1,(l)}|  \rbrace$ by calling the aforementioned Runge-Kutta integration scheme and then calculates
\begin{align}
\label{eq:mathfrak_m_difference}
\mathfrak{m}^k_{n+1,(l)} \approx \frac{\tilde{P}^k_{n+1,(l)} - \hat{P}^k_{n+1,(l)}}{\epsilon}.
\end{align}
Following \cite{Liu2020}, we set $\epsilon_{\mathrm{abs}} = 10^{-8}$ and $\epsilon_{\mathrm{rel}} = 10^{-5}$ in this study.
 
\begin{remark}
In the above analysis, the displacement field is segregated from the balance equations within the nonlinear solution procedure. The displacement undergone by a solid material particle is governed by \eqref{eq:segregard_disp_vec_solid} while the mesh displacement in the fluid subdomain is updated by solving \eqref{eq:segregated_fluid_mesh_velo_incremental}. Conceptually, the two can be unified by viewing the mesh displacement as a field variable governed by the Laplacian operator where the mesh displacement is set to satisfy a non-homogeneous Dirichlet ``boundary condition" in the solid subdomain specified by \eqref{eq:segregard_disp_vec_solid}. This unified viewpoint may help simplify the implementation of the solution method.
\end{remark}

\begin{figure}
\begin{center}
\begin{tabular}{c}
\includegraphics[angle=0, trim=0 0 350 0, clip=true, scale = 0.72]{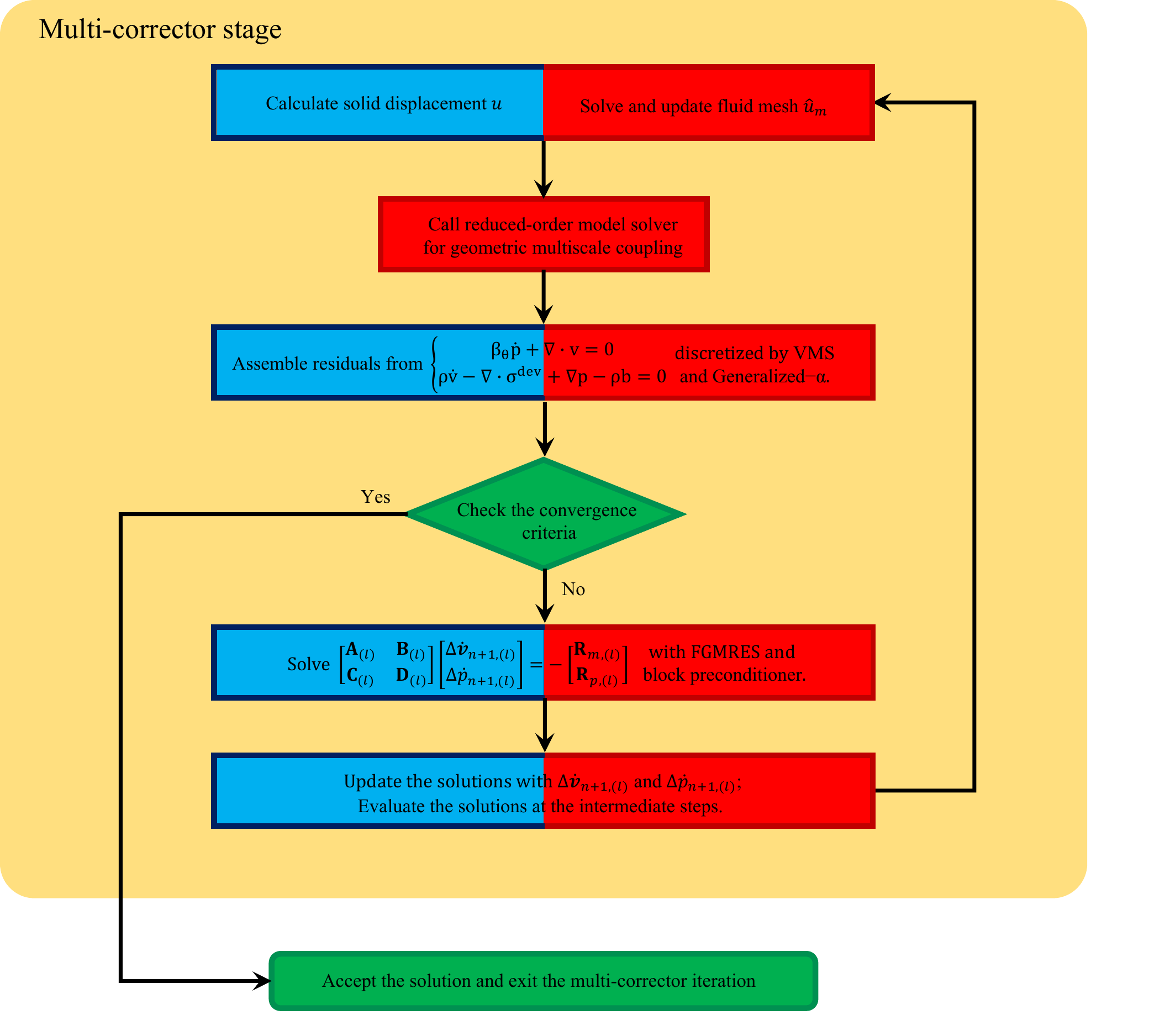} \\[1.0em]
\includegraphics[angle=0, trim=0 500 450 0, clip=true, scale = 0.72]{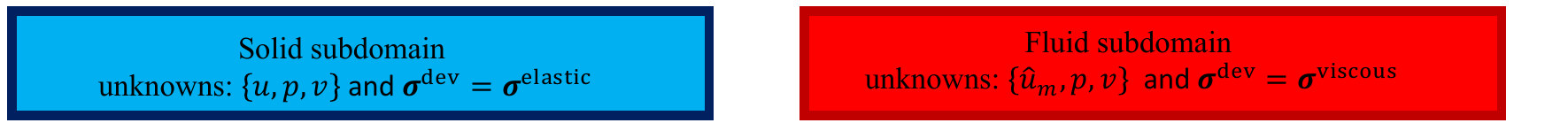}
\end{tabular}
\end{center}
\caption{Flowchart of the multi-corrector stage.}
\label{fig:FSI-flowchart}
\end{figure}

Combining our above analysis, the proposed algorithm is endowed with the following attributes. In both fluid and solid subdomains, the velocity and pressure incrementals are determined by a matrix problem with the following two-by-two block structure,
\begin{align*}
\boldsymbol{\mathcal A}_{(l)} :=
\begin{bmatrix}
\boldsymbol{\mathrm A}_{(l)} & \boldsymbol{\mathrm B}_{(l)} \\[0.3mm]
\boldsymbol{\mathrm C}_{(l)} & \boldsymbol{\mathrm D}_{(l)}
\end{bmatrix}.
\end{align*}
The four blocks adopt different definitions in the two subdomains. In the fluid subdomain, we have
\begin{align}
\label{eq:ABCD_block_matrices_fluid}
\boldsymbol{\mathrm A}_{(l)} := \boldsymbol{\mathrm K}^{f}_{m, (l), \dot{\bm v}}, \quad \boldsymbol{\mathrm B}_{(l)} := \boldsymbol{\mathrm K}^{f}_{m, (l), \dot{\bm p}}, \quad \boldsymbol{\mathrm C}_{(l)} := \boldsymbol{\mathrm K}^{f}_{p, (l), \dot{\bm v}}, \quad \boldsymbol{\mathrm D}_{(l)} := \boldsymbol{\mathrm K}^{f}_{p, (l), \dot{p}},
\end{align}
and, in the solid subdomain, we have
\begin{align}
& \boldsymbol{\mathrm A}_{(l)} := \boldsymbol{\mathrm K}^{s}_{m, (l), \dot{\bm v}} + \frac{\alpha_f \gamma \Delta t_n}{\alpha_m} \boldsymbol{\mathrm K}^{s}_{m, (l), \dot{\bm u}}, && \boldsymbol{\mathrm B}_{(l)} := \boldsymbol{\mathrm K}^{s}_{m, (l), \dot{\bm p}}, \\
\label{eq:ABCD_block_matrices_solid}
& \boldsymbol{\mathrm C}_{(l)} :=  \boldsymbol{\mathrm K}^{s}_{p, (l), \dot{\bm v}} + \frac{\alpha_f \gamma \Delta t_n}{\alpha_m} \boldsymbol{\mathrm K}^{s}_{p, (l), \dot{\bm u}}, && \boldsymbol{\mathrm D}_{(l)} := \boldsymbol{\mathrm K}^{s}_{p, (l), \dot{\bm v}}.
\end{align}
The right-hand side of the linear system is a block vector $\begin{bmatrix} \mathbf R_{m,(l)}; \mathbf R_{p,(l)} \end{bmatrix}^{T}$, which is part of the original residual vector \eqref{eq:consistent_residual}. The solid displacement and fluid mesh motion are determined segregatedly by \eqref{eq:segregard_disp_vec_solid} and \eqref{eq:segregated_fluid_mesh_velo_incremental}, respectively. Summarizing the above discussion, we state a predictor multi-corrector algorithm for solving the FSI nonlinear system as follows. The multi-corrector stage is also illustrated in Figure \ref{fig:FSI-flowchart}.

\begin{myenv}{Segregated predictor multi-corrector algorithm}
\noindent \textbf{Predictor stage:}
\begin{enumerate}
\item Set
\begin{align*}
\bm y_{n+1,(0)} = \bm y_n, \quad \dot{\bm y}_{n+1,(0)} = \frac{\gamma -1}{\gamma }\dot{\bm y}_n.
\end{align*}

\item Evaluate the solutions at the intermediate steps as
\begin{align*}
\bm y_{n+\alpha_f,(1)} = \bm y_{n} + \alpha_f \left( \bm y_{n+1,(0)} - \bm y_n \right), \quad \dot{\bm y}_{n+\alpha_m,(1)} = \dot{\bm y}_{n} + \alpha_m \left( \dot{\bm y}_{n+1,(0)} - \dot{\bm y}_n \right).
\end{align*}
\end{enumerate}

\noindent \textbf{Multi-corrector stage} Repeat the following steps for $l=1,\cdots, l_{\mathrm{max}}$.
\begin{enumerate}
\item Obtain the incremental for the solid displacement as
\begin{align*}
\Delta \dot{\bm u}^s_{n+1,(l)} = \frac{\alpha_f \gamma \Delta t_n}{\alpha_m} \Delta \dot{\bm v}^s_{n+1,(l)} - \frac{1}{\alpha_m} \mathbf R_{k,(l)}.
\end{align*}

\item Update the solid displacement as
\begin{align*}
\dot{\bm u}^s_{n+1,(l)} = \dot{\bm u}^s_{n+1,(l-1)} + \Delta \dot{\bm u}^s_{n+1,(l)}, \quad \bm u^s_{n+1,(l)}=\bm u^s_{n+1,(l-1)} + \gamma \Delta t_n \Delta \dot{\bm u}^s_{n+1,(l)}.
\end{align*}

\item Solve the mesh motion equation to obtain $\hat{\bm u}_{n+1,(l)}$.

\item Update the mesh velocity by
\begin{align*}
\hat{\bm v}_{n+1,(l)} = \frac{\hat{\bm u}_{n+1,(l)} - \hat{\bm u}_n}{\gamma \Delta t_n} + \frac{\gamma -1}{\gamma}\hat{\bm v}_{n}.
\end{align*}

\item Calculate the flow rate
\begin{align*}
Q^k_{n+1,(l)} := \int_{\Gamma^{f,k}_{\bm x,\mathrm{out}}(t)} \bm v_{n+1,(l)} \cdot \bm n d\Gamma_{\bm x}.
\end{align*}

\item Calculate the pressure $P^{k}_{n+1,(l)} = \mathcal F^{k}(Q^{k}_{n+1,(l)})$ from the reduced model; calculate $\tilde{P}^{k}_{n+1,(l)}$ and $\hat{P}^{k}_{n+1,(l)}$ from \eqref{eq:tilde_P_and_hat_P}; calculate $\mathfrak m^k_{n+1,(l)}$ from \eqref{eq:mathfrak_m_difference}. 

\item Assemble the residual vectors of the nonlinear system using the solution vectors evaluated at the intermediate steps:
\begin{align*}
\mathbf R_{m,(l)} := \mathbf R_m( \dot{\bm y}_{n+\alpha_m, (l)}, \bm y_{n+\alpha_f, (l)} ), \quad \mathbf R_{p,(l)} := \mathbf R_p( \dot{\bm y}_{n+\alpha_m, (l)}, \bm y_{n+\alpha_f, (l)} ).
\end{align*}
\item Let $\| \left( \mathbf R_{m,(l)}; \mathbf R_{p,(l)} \right)\|_{\mathfrak l^2}$ denote the $\mathfrak l^2$-norm of the residual vector. Let $\mathrm{tol}_A$ and $\mathrm{tol}_R$ be the prescribed tolerances. If either one of the stopping criteria
\begin{align}
\label{eq:predictor-multi-corrector-stopping-criteria}
\frac{\| \left( \mathbf R_{m,(l)}; \mathbf R_{p,(l)} \right)\|_{\mathfrak l^2}}{\| \left( \mathbf R_{m,(0)}; \mathbf R_{p,(0)} \right)\|_{\mathfrak l^2}} \leq \mathrm{tol}_R, \qquad \| \left( \mathbf R_{m,(l)}; \mathbf R_{p,(l)} \right)\|_{\mathfrak l^2} \leq \mathrm{tol}_A,
\end{align}
is satisfied, set the solution vector at time step $t_{n+1}$ as $\bm y_{n+1} = \bm y_{n+1,(l)}$ and $\dot{\bm y}_{n+1} = \dot{\bm y}_{n+1,(l)}$ and exit the multi-corrector stage; otherwise proceed to step 9.
\item Assemble the tangent matrix $\boldsymbol{\mathcal A}_{(l)}$ according to \eqref{eq:ABCD_block_matrices_fluid}-\eqref{eq:ABCD_block_matrices_solid} and solve the linear system
\begin{align}
\label{eq:predictor-multi-corrector-linear-system}
\begin{bmatrix}
\boldsymbol{\mathrm A}_{(l)} & \boldsymbol{\mathrm B}_{(l)} \\[0.3mm]
\boldsymbol{\mathrm C}_{(l)} & \boldsymbol{\mathrm D}_{(l)}
\end{bmatrix}
\begin{bmatrix}
\Delta \dot{\bm v}_{n+1,(l)} \\[1.2mm]
\Delta \dot{p}_{n+1,(l)}
\end{bmatrix}
= -
\begin{bmatrix}
\mathbf R_{m,(l)}  \\[1.2mm]
\mathbf R_{p,(l)}  
\end{bmatrix}.
\end{align}
\item Update the iterates as
\begin{align*}
\dot{\bm v}_{n+1,(l)} &= \dot{\bm v}_{n+1,(l-1)} + \Delta \dot{\bm v}_{n+1,(l)}, \quad \bm v_{n+1,(l)} = \bm v_{n+1,(l-1)} + \gamma \Delta t_n \Delta \dot{\bm v}_{n+1,(l)}, \displaybreak[2] \\
\dot{p}_{n+1,(l)} &= \dot{p}_{n+1,(l-1)} + \Delta \dot{p}_{n+1,(l)}, \quad p_{n+1,(l)} = p_{n+1,(l-1)} + \gamma \Delta t_n \Delta \dot{p}_{n+1,(l)}.
\end{align*}
\item Evaluate the solutions at the intermediate steps as
\begin{align*}
\bm y_{n+\alpha_f,(l+1)} = \bm y_{n} + \alpha_f \left( \bm y_{n+1,(l)} - \bm y_n \right), \quad \dot{\bm y}_{n+\alpha_m,(l+1)} = \dot{\bm y}_{n} + \alpha_m \left( \dot{\bm y}_{n+1,(l)} - \dot{\bm y}_n \right).
\end{align*}
\end{enumerate}
\end{myenv}

\begin{remark}
In the above predictor multi-corrector algorithm, the segregation is performed within the multi-corrector as a means of solving the nonlinear algebraic equations. The stopping criteria \eqref{eq:predictor-multi-corrector-stopping-criteria} are enforced such that the residual of the balance equations are driven below the prescribed tolerances in each time step. Therefore, the proposed implicit solution algorithm belongs to the category of strongly-coupled FSI formulation.
\end{remark}

\begin{remark}
The predictor used in the above algorithm is known as the same-Y predictor. It is a rather robust option. There are other predictor candidates, which can be advantageous in certain scenarios. Readers may refer to \cite[pp.311-312]{Jansen2000} for a discussion on this point.
\end{remark}

\section{Vascular modeling}
\label{sec:vascular-modeling}
In this section, we introduce a combined suite of modeling technologies, including the geometric modeling of the arterial walls, the generation of the prestress in the vascular wall, and an initialization procedure for vascular FSI simulations.

\subsection{Geometric modeling and mesh generation}
\label{sec:geometric_modeling_mesh_generation}
We first propose a pipeline for generating high-quality FSI meshes from medical images. In this approach, three-dimensional angiographic data obtained from computed tomography (CT) or magnetic resonance (MR) imaging is used to construct the geometric model of the lumen and vascular tissue. The designing goal is to maintain mesh continuity across the fluid-solid interface, a critical property for ensuring FSI coupling.

\begin{figure}
\begin{center}
\begin{tabular}{c}
\includegraphics[angle=0, trim=0 0 0 0, clip=true, scale = 0.45]{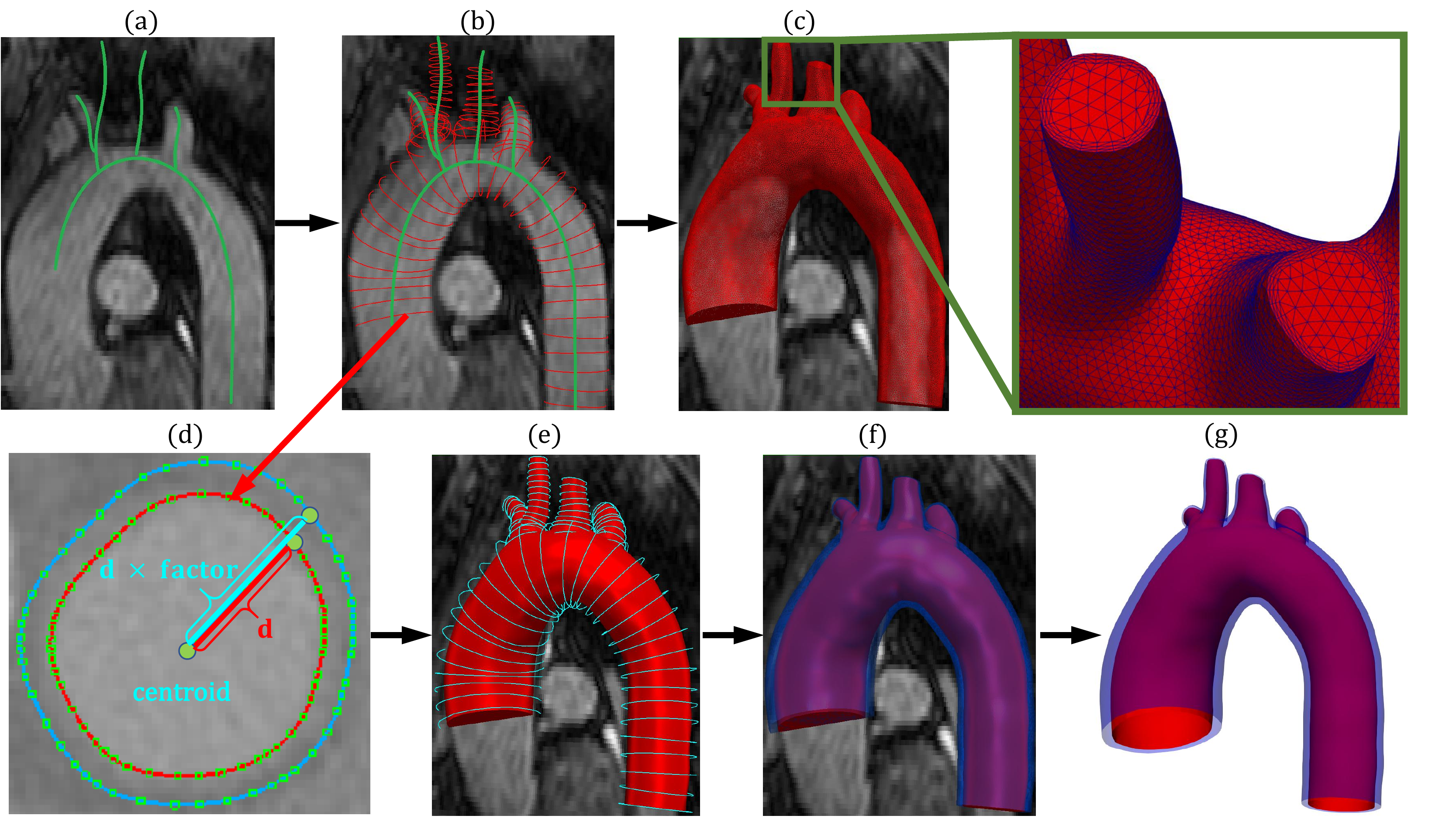} \\
\includegraphics[angle=0, trim=0 0 0 0, clip=true, scale = 0.45]{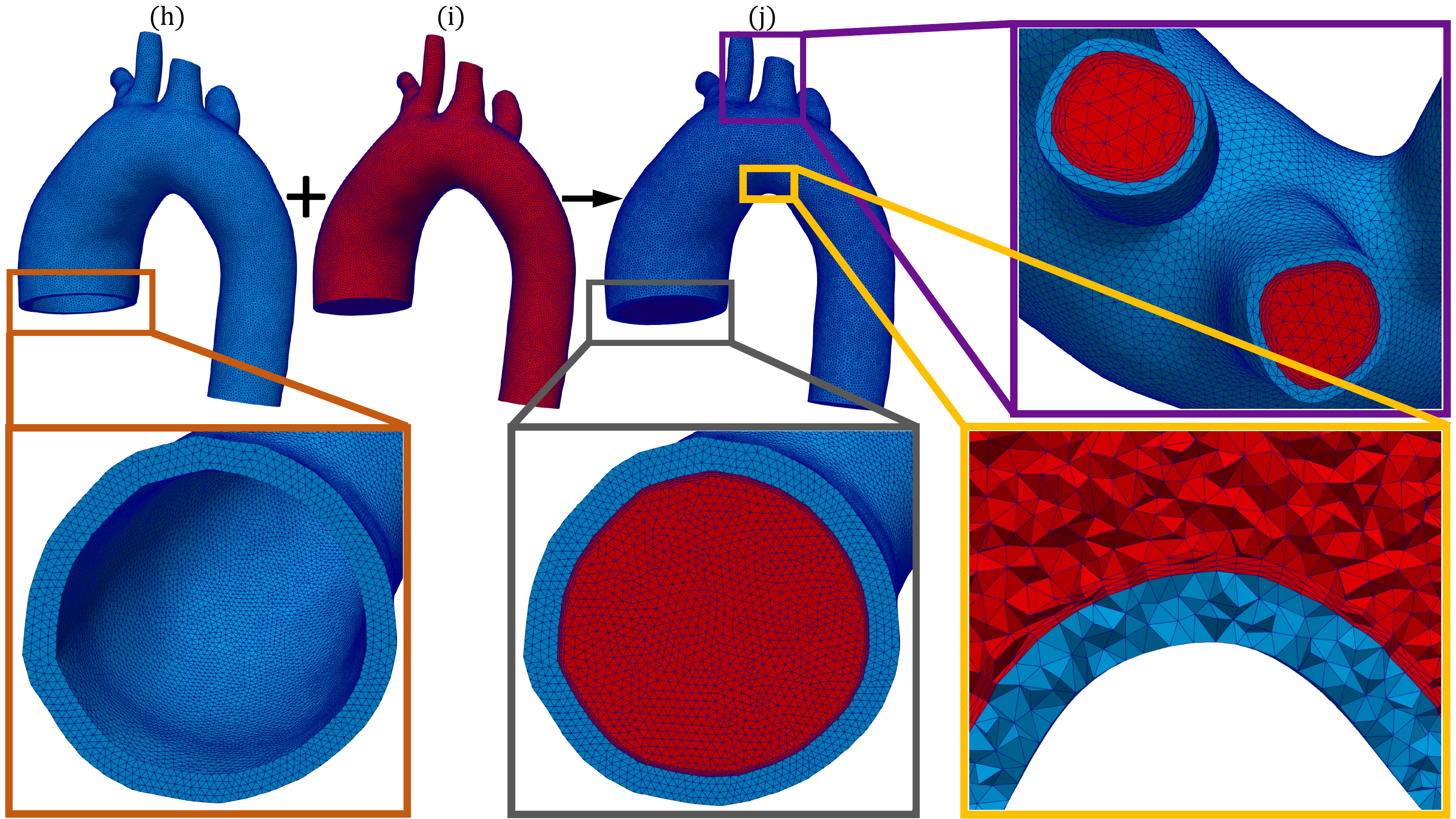}
\end{tabular}
\end{center}
\caption{The fluid model and mesh are generated by (a) creating pathlines along the vessels of interest, (b) creating two-dimensional segmentations along each of these pathlines, and (c) lofting the segmentations to create a geometric model and generating the mesh with boundary layers. The solid model and mesh are constructed by (d \& e) scaling the contour points with respect to the centroid point to define the contours for the exterior wall surface, (f) lofting the segmentations to create the exterior wall surface with the necessary surface fairing performed, (g) combining the newly generated exterior wall surface and the lumen wall surface to define the geometric model of the vascular tissue, (h) closing the annular-shaped plane surfaces and generating the volumetric mesh for the vessel wall, and (i \& j) combining meshes generated in the two subdomains to finalize the vascular FSI mesh generation.}
\label{fig:mesh-generation-pipeline}
\end{figure}

\paragraph{Lumen modeling}
Our pipeline starts with medical images obtained from different modalities, such as CT or MR imaging. The images are typically stored in the DICOM (DIgital COmmunication in Medicine) format. This data format stores image data and acquisition information that allow reconstruction of the acquired volume. To demonstrate our proposed approach, we adopt a thoracic aortic model of a healthy 23-years old male from a set of MR images in the vascular model repository \cite{VMR}. There has been a variety of software that can visualize and process the DICOM images \cite{Arthurs2021,Updegrove2017,Yushkevich2006}, and the open source software SimVascular is utilized to fulfill this job in this work \cite{Updegrove2017}. With the images acquired and visualized, we first manually create pathlines for the vessels of interest (Figure \ref{fig:mesh-generation-pipeline} (a)). A smoothing procedure is performed to prevent abrupt changes in the pathline direction. Two-dimensional plane images perpendicular to the pathlines can be extracted then. On the plane images, segmentation can be performed by a variety of methods (e.g., analytic curves, the thresholding method, the level set method, etc.). The output of the segmentation is represented through a set of control points interpolated by splines (Figure \ref{fig:mesh-generation-pipeline} (b)). After that, the boundary surface of each vessel can be created in terms of a non-uniform rational B-splines surface through interpolation along the vessel length. The boundary surfaces of vessels can then be connected through boolean operations, and the resulting combined surface represents $\Omega^{I}_{\bm \chi}$ for the vascular domain \cite{Wang2001}. Importantly, the resulting wall surface may contain undesired features coming from ignored secondary vessels and sharp transitions of the vessel radius at the branching location. The surface, therefore, needs to be modified using locally smoothing algorithms \cite{Antiga2002,Updegrove2016} to guarantee physiologically realistic geometries, which is critical for delivering high-quality computational analysis.  The ends of the combined vessels are $\mathrm{n}_{\mathrm{in}} + \mathrm{n}_{\mathrm{out}}$ planar surfaces delineated by the segmented curves. The luminal volume can be created by generating planar surfaces at the inlets and outlets.

\paragraph{Tissue modeling}
The next step is to determine the geometric model of the vessel wall, which is bounded by the lumen surface, the wall exterior surface, and $\mathrm{n}_{\mathrm{in}} + \mathrm{n}_{\mathrm{out}}$ annular surfaces. Considering that most existing imaging modalities cannot resolve the wall surface except the black blood MR imaging technique, we are often unable to delineate the exterior wall surface directly from CT or MRI images. Thus, we choose to generate an artificial segmentation of the vascular wall by exploiting existing physiological knowledge of the vessel wall thickness. The aorta wall thickness is approximately 2 millimeters, which is around 8\% of its lumen diameter; the thickness in large arteries reduces to around 1 millimeter, which is about 10\% of its lumen diameter \cite{Adame2006,Caro2012}. As a demonstrative case, we assume the vessel wall thickness is $10\%$ of its local diameter uniformly along the vessel. Of course, this can be conveniently adjusted manually on each plane image to accommodate local pathological situations \cite{Humphrey2002,Martufi2009}. We also hypothesize that the wall thickness is uniform in all radial directions with respect to the luminal centroid. With the two assumptions, a hypothetical contour of the exterior wall surface on each slice can be generated based on the existing lumen segmentation. The centroid is determined by calculating the mean position of all spatial coordinates of the contour points of the lumen segmentation. The contour points of the lumen segmentation are scaled by a factor of $110\%$ to define those of the outer wall surface (Figure \ref{fig:mesh-generation-pipeline} (d) and (e)). Subsequently, the geometric model based on the exterior surface segmentation can be generated by lofting the newly generated contours (Figure \ref{fig:mesh-generation-pipeline} (f) and (g)). Whenever the image can resolve the vessel wall, the aforesaid step should be replaced by performing segmentation of the exterior wall directly. Again, there could be undesirable surface features, especially near the branching regions. Surface fairing and smoothing need to be performed for the exterior wall surface. The resulting geometric model defines the exterior wall, from which the inlet and outlet boundary curves can be specified. The $\mathrm{n}_{\mathrm{in}} + \mathrm{n}_{\mathrm{out}}$ annular-shaped plane surfaces need to be closed to complete the definition of the tissue volume. 

\paragraph{Mesh generation}
With the geometric model for the two subdomains created, mesh generation will be performed to discretize the subject-specific subdomains into well-shaped elements. In this work, we use the linear tetrahedral element by leveraging its flexibility in discretizing complex geometries and robustness in numerical simulations. There are two special requirements in meshing the vascular FSI subdomains. First, the mesh across the two subdomains needs to maintain $C^0$-continuity at least, according to our discussion in Section \ref{sec:semi-discrete-FSI}. Second, the hemodynamic calculations require thin layers of elements near the lumen wall in the fluid subdomain to capture important biomarkers, such as wall shear stress, oscillatory shear index, etc. The advancing layer method \cite{Garimella2000,Marchandise2013} is often used for creating the boundary layer mesh. Although local surface editing has been performed for the luminal surface in the geometric modeling stage, the surface mesh is usually inadequate for generating the boundary layer mesh. Thus, the advancing layer algorithm involves a re-triangulation procedure to perform local mesh smoothing. In this work, we use the algorithm implemented in Gmsh \cite{Geuzaine2009,Marchandise2013} to generate the mesh in the fluid subdomain first (Figure \ref{fig:mesh-generation-pipeline} (c)). With the updated luminal surface mesh, the boundary representation of the tissue subdomain can be fixed, and the volumetric mesh of the tissue can be generated through the three-dimensional mesh generation algorithm provided by Gmsh (Figure \ref{fig:mesh-generation-pipeline} (h)). Combining the two meshes completes the FSI mesh generation (Figure \ref{fig:mesh-generation-pipeline} (j)).

\subsection{Tissue prestressing and an initialization procedure}
\label{subsec:prestressing}
In patient-specific modeling, the vascular wall is imaged at a state under physiological loading. Thereby, the imaged configuration is not the \textit{natural} configuration for the wall, and there is an internal stress, known as the prestress, distributed in the wall that balances the in vivo blood pressure and wall shear stress. With the prestress $\bm \sigma_0$, the momentum balance equations \eqref{eq:fsi_solid_momentum} are modified into the following form,
\begin{align}
& \mathbf B^s_{m}\left( \bm w^s; \dot{\bm y}_h^s, \bm y_h^s \right) := \int_{\Omega_{\bm x}^s(t)} \bm w^s \cdot \rho^s(p^s_h) \frac{d\bm v_h^s}{dt} + \nabla_{\bm x} \bm w^s : \left( \bm \sigma^s_{\mathrm{dev}}(\bm u^s_h) + \bm \sigma_0 \right) - \nabla_{\bm x} \cdot \bm w^s p_h^s - \bm w^s \cdot \rho^s(p^s_h) \bm b d\Omega_{\bm x} \nonumber \displaybreak[2]  \\
& \hspace{15mm} -\int_{\Gamma_{\bm x,h}^{s}(t)}\bm w^s \cdot \bm h^s d\Gamma_{\bm x} + \int_{\Omega_{\bm x}^{\prime s}(t)} \nabla_{\bm x} \cdot \bm w^s \tau_{\mathrm{C}}^s \left(\beta_{\theta}^s(p^s_h) \frac{dp_h^s}{dt} + \nabla_{\bm x} \cdot \bm v_h^s \right) d\Omega_{\bm x},
\end{align}
where the prestress $\bm \sigma_0$ is added to $\bm \sigma^s_{\mathrm{dev}}$. In order to determine the value of $\bm \sigma_0$, a rigid-walled simulation needs to be performed to obtain the physiological intramural pressure and wall shear stress. This can be conveniently achieved by performing an FSI simulation with the displacement and velocity DOFs in $\Omega^s_{\bm x}$ fixed. Once a steady state flow profile $\bm y^f_{\mathrm{steady}}$ is obtained, one may load the obtained solution as the input data for the prestress generation. The prestress generation algorithm stated below is inspired from the prior work developed based on the pure displacement formulation \cite{Hsu2011}. There are alternative approaches to determine the prestress. For example, it was proposed to determine the zero-pressure configuration first, which deforms to the imaged configuration under physiological loading \cite{Tezduyar2007,Takizawa2010a}; it was also suggested to calculate a hypothetical deformation gradient to determine the prestress \cite{Gee2009,Gee2010}. We favor the approach that directly determines the prestress field because it is now widely accepted that the prestress is generated from tissue growth and remodeling rather than elastic deformation \cite{Humphrey2002}. Calculating the prestress from elastic deformation may contradict the facts from mechanobiology.

\begin{myenv}{Prestress generation algorithm}
\noindent \textbf{Initialization:} Set $\bm \sigma_{0, (0)} = \bm 0$, $\bm y^s_{h} = \bm 0$, and $\bm y^f_h = \bm y^f_{\mathrm{steady}}$ \footnote{Notice that the steady-state fluid solution is obtained via a rigid-wall assumption which gives no-slip velocity on the fluid-solid interface. In the meantime, the pressure is discontinuous across the interface. Thereby, this initialization of $\bm y^s_{h} = \bm 0$ and $\bm y^f_h = \bm y^f_{\mathrm{steady}}$  causes no ambiguity.}

\noindent \textbf{Fixed-point iteration:} Repeat the following steps for $m=0, 1, ..., m_{\mathrm{max}}$.
\begin{enumerate}
	\item Set $\bm \sigma_0 = \bm \sigma_{0, (m)}$, $\bm y^s_h = \bm 0$, and $\bm y^f_h = \bm y^f_{\mathrm{steady}}$.
    \item From $t_m$ to $t_{m+1}$, solve the FSI coupled problem using the backward Euler method for temporal integration.
    \item Update the prestress tensor as $\bm \sigma_{0, (m+1)} = \bm \sigma^s_{\mathrm{dev}}(\bm u^s_{m+1}) - p^s_{m+1}\bm I + \bm \sigma_{0, (m)}$.
    \item Let $\mathrm{tol}_{\mathrm{P}}$ denote a prescribed tolerance. If the stopping criteria $\| \bm u^s_{m+1} \|_{\mathfrak l_{\infty}} \leq \mathrm{tol}_{\mathrm{P}}$ is satisfied, set $\bm \sigma_0 = \bm \sigma_{0,(m+1)}$ and exit the fixed-point iteration; otherwise continue the iteration.
\end{enumerate}
\end{myenv}

\paragraph{The initialization procedure for vascular FSI simulations}
From the above prestress generation algorithm, it can be seen that an initialization procedure is needed for vascular FSI simulations. We perform a rigid-walled simulation to determine the fully developed flow field using the FSI mesh. It is reasonable to hypothesize that the images are acquired during diastole and the obtained flow profile at diastole can be utilized to determine the prestress $\bm \sigma_{0}$ in the vascular wall by invoking the algorithm given above. Of course, the above-mentioned hypothesis is unnecessary if the image acquisition is gated and synchronized to the cardiac cycle. The flow profile and the prestress $\bm \sigma_0$ constitute the initial data to be loaded for pulsatile FSI simulations.

\section{Parallelization and preconditioner design}
\label{sec:preconditioning-technique}
The advantage of the unified framework can be appreciated by designing a parallel solver strategy. In this section, we first discuss the parallel implementation based on the domain decomposition approach and then devise a preconditioning technique for the linear system \eqref{eq:predictor-multi-corrector-linear-system}.

\subsection{Parallelization}
The parallelization of the solution method described in Section \ref{sec:numerical-formulation} is achieved by the domain decomposition concept for distributed-memory parallelization. The data structure of the whole FSI domain is partitioned into non-overlapping subdomains\footnote{Here we use ``subdomain" to refer to the region associated with a separate computing processor, which is different from the terminology ``subdomain" referring regions occupied by the fluid or solid materials in earlier parts of this article. Readers should be able to tell the difference based on the context.}, each of which is assigned to a processor. The generation of the subdomains for parallel finite element analysis includes the distribution of the element- and node-related data structures across computing processors. In this work, we convert the finite element mesh into a nodal or dual graph and invoke a graph-theory-based algorithm for the mesh partitioning \cite{Karypis1998}. The generated distribution ensures that the load is well-balanced among processors and the inter-processor communication is minimized.

\begin{figure}
\begin{center}
\begin{tabular}{cc}
\includegraphics[angle=0, trim=0 260 250 0, clip=true, scale = 0.28]{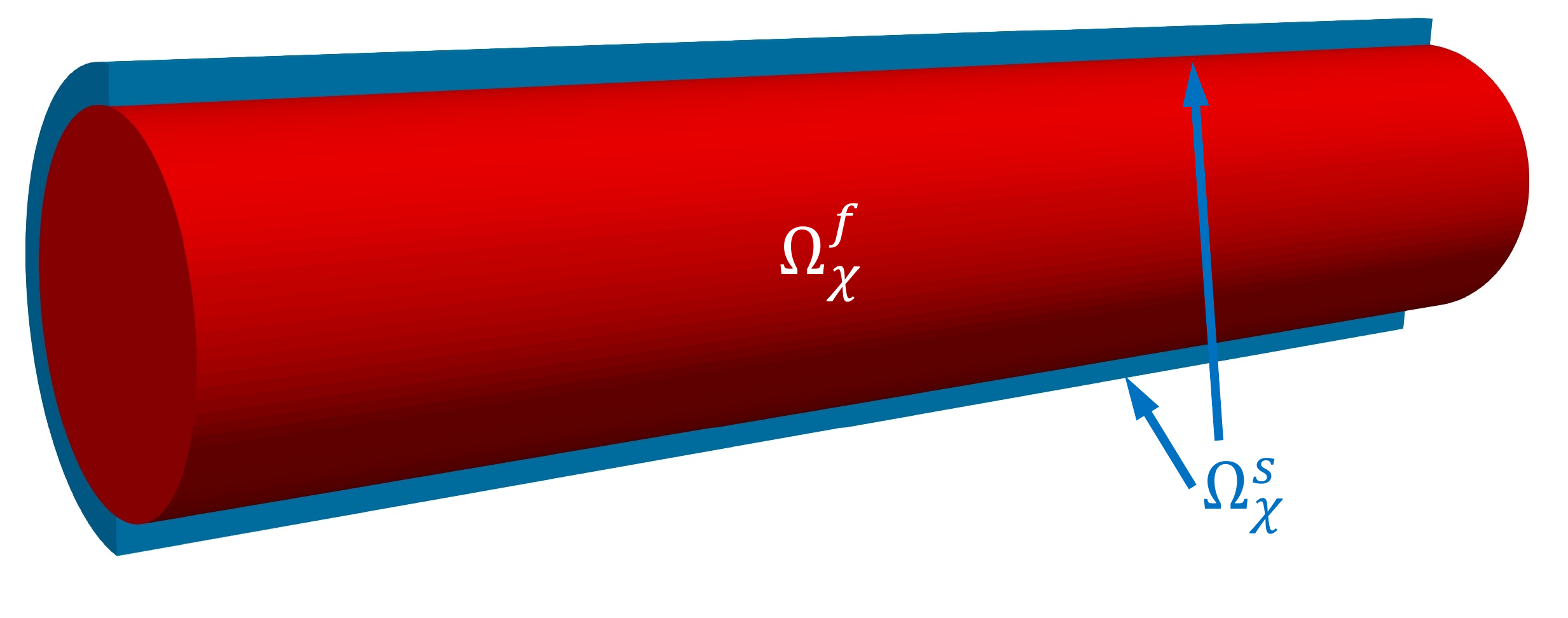} &
\includegraphics[angle=0, trim=0 260 250 0, clip=true, scale = 0.28]{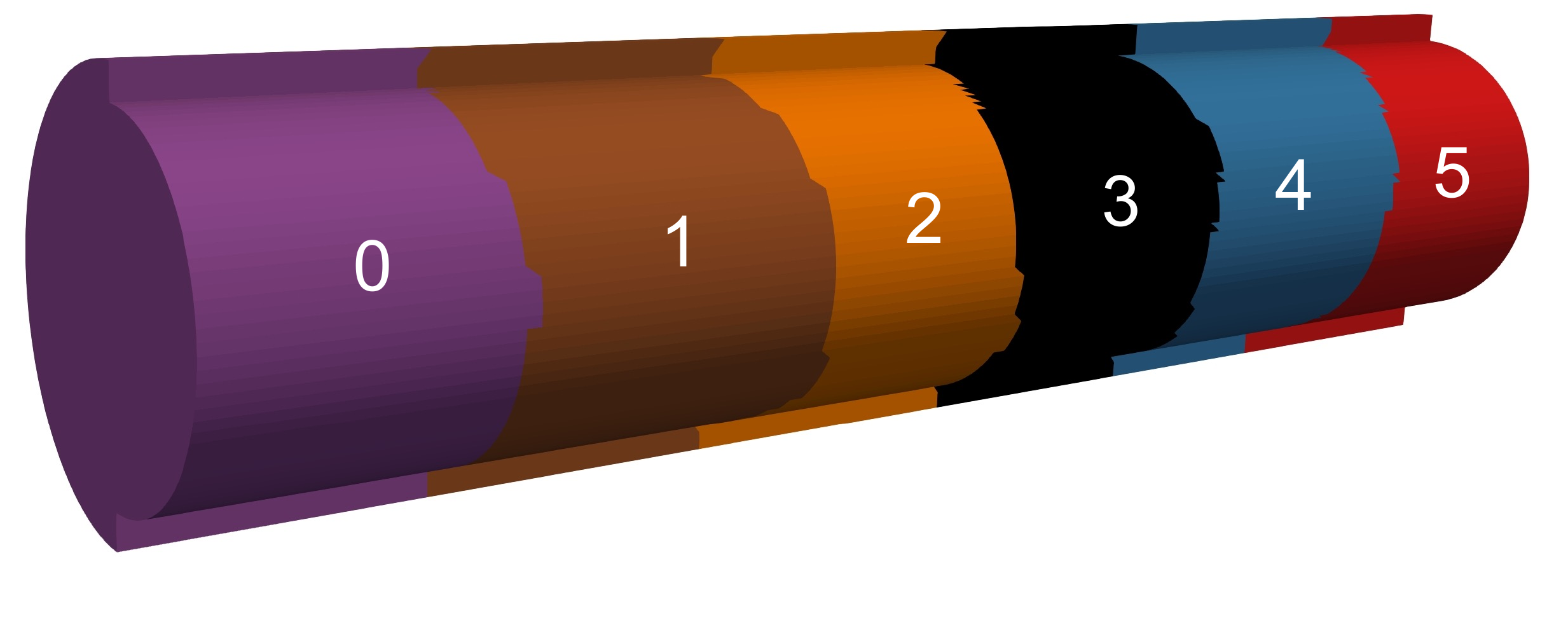}
\end{tabular}
\end{center}
\caption{Domain decomposition of the Greenshields-Weller benchmark. The left figure depicts the fluid and solid subdomains, where the solid subdomain is cut by a plane passing the central axis for visualization purposes. The right figure shows the domain decomposition generated by the mesh partition algorithm. }
\label{fig:fsi_beam_domain_decomposition}
\end{figure}

In the conventional FSI formulation where the solid subproblem is governed by the pure displacement formulation \cite{Bazilevs2008}, one needs to exert extra care in mesh partitioning. In that formulation, the fluid subdomain involves seven DOFs (i.e., the fluid velocity, pressure, and mesh displacement), while in the solid subdomain only the displacement field needs to be solved. Therefore, the elements and nodes need to be correspondingly weighted during mesh partitioning in order to achieve desired parallel efficiency. Sometimes, when the solid subproblem is modeled by thin-walled structures, practitioners assign the solid subproblem to a single processor and the fluid subdomain is partitioned as a single-physics problem \cite{Bazilevs2017}. In the unified continuum formulation, the issue is significantly alleviated because both subdomains involve seven DOFs in the solution vectors. The mesh solver deals with the displacement DOFs, and the two-by-two matrix solver involves velocity and pressure DOFs in both subdomains. Therefore, the load assigned in the two subdomains is close to each other,\footnote{Strictly speaking, the amount of floating point operations in the assembly of the matrices and vectors in the fluid and solid subdomain is not the same. We can only guarantee that the number of elements and the generated distributed matrix problem size are close among processors.} and one may partition the FSI mesh as a whole in a straightforward way without worrying about the load balancing issue arising in the conventional FSI formulations. In Figure \ref{fig:fsi_beam_domain_decomposition}, the domain decomposition of an elastic pipe for six processors is demonstrated. In the partitioned mesh, each subdomain may contain only fluid elements, only solid elements, or elements of both types.

\subsection{Preconditioning techniques}
\label{subsec:preconditioning}
In this section, we present a suite of preconditioning techniques for the linear systems arising in the predictor multi-corrector algorithm given in Section \ref{sec:segregated-predictor-multi-corrector}. The segregation algorithm breaks the three-by-three block Jacobian matrix into two smaller linear systems. One is the two-by-two block matrix problem \eqref{sec:segregated-predictor-multi-corrector} associated with the mass and linear momentum balance equations, and the other one is associated with the mesh update. Since the AMG method is an ideal candidate for discrete elliptic operators in robustness, scalability, and efficiency, we strive to leverage its superior properties by performing block factorization on the two-by-two block matrix. The AMG preconditioner in this study is adopted from the BoomerAMG implementation in the Hypre package \cite{Falgout2002}, whose detailed settings are listed in Table \ref{table:amg_settings}.
\begin{table}[htbp]
\small
\begin{center}
\renewcommand{\arraystretch}{1.1}
\begin{tabular}{p{8.5cm} p{5.5cm} }
\hline
Cycle type &  V-cycle \\
Coarsening method & HMIS \cite{Sterck2006} \\
Interpolation method & Extended method (ext+i) \cite{Sterck2008} \\
Smoother & Hybrid Gauss-Seidel \cite{Baker2011} \\
Truncation factor for the interpolation & $0.3$ \\
Threshold for being strongly connected & $0.5$ \\
Maximum number of elements per row for interp. & $5$ \\
The number of levels for aggressive coarsening & $2$ \\
\hline
\end{tabular}
\end{center}
\caption{The settings of the BoomerAMG \cite{Falgout2002} preconditioner used in this work. These settings are selected based on trial-and-error with the goal of achieving a balance between robustness and efficiency.}
\label{table:amg_settings}
\end{table}

\subsubsection{Preconditioning techniques for the two-by-two block matrix problem}
In the predictor multi-corrector algorithm, the most time-consuming part is associated with solving the linear system \eqref{eq:predictor-multi-corrector-linear-system}. To facilitate our subsequent discussion, we neglect the subscript $(l)$ and focus on the linear system $\boldsymbol{\mathcal A} \bm x = \bm r$, where the matrix $\boldsymbol{\mathcal A}$ and vectors enjoy the following block structure,
\begin{align*}
\boldsymbol{\mathcal A} :=
\begin{bmatrix}
\boldsymbol{\mathrm A} & \boldsymbol{\mathrm B} \\[0.3mm]
\boldsymbol{\mathrm C} & \boldsymbol{\mathrm D}
\end{bmatrix}, \quad
\bm x :=
\begin{bmatrix}
\bm x_{\bm v} \\[0.3em]
\bm x_{p}
\end{bmatrix},
\quad
\bm r :=
\begin{bmatrix}
\bm r_{\bm v} \\[0.3em]
\bm r_{p}
\end{bmatrix}.
\end{align*}
The matrix $\boldsymbol{\mathcal A}$ can be factorized into a lower triangular, a diagonal, and an upper triangular matrices,
\begin{align}
\label{eq:A_LDU_block_factorization}
\boldsymbol{\mathcal A} =
\begin{bmatrix}
\boldsymbol{\mathrm I} & \boldsymbol{\mathrm O} \\[0.3em]
\boldsymbol{\mathrm C} \boldsymbol{\mathrm A}^{-1} & \boldsymbol{\mathrm I}
\end{bmatrix}
\begin{bmatrix}
\boldsymbol{\mathrm A} & \boldsymbol{\mathrm O} \\[0.3em]
\boldsymbol{\mathrm O} & \boldsymbol{\mathrm S}
\end{bmatrix}
\begin{bmatrix}
\boldsymbol{\mathrm I} & \boldsymbol{\mathrm A}^{-1} \boldsymbol{\mathrm B} \\[0.3em]
\boldsymbol{\mathrm O} & \boldsymbol{\mathrm I}
\end{bmatrix},
\end{align}
in which $\boldsymbol{\mathrm I}$ is the identity matrix, $\boldsymbol{\mathrm O}$ is the zero matrix, and $\boldsymbol{\mathrm S} := \boldsymbol{\mathrm D} - \boldsymbol{\mathrm C} \boldsymbol{\mathrm A}^{-1} \boldsymbol{\mathrm B}$ is the Schur complement of the block $\boldsymbol{\mathrm A}$. Leveraging the above factorization, the inverse of $\boldsymbol{\mathcal A}$ can be conveniently represented as
\begin{align}
\label{eq:A_inverse_LDU_block_factorization}
\boldsymbol{\mathcal A}^{-1} =
\begin{bmatrix}
\boldsymbol{\mathrm I} & -\boldsymbol{\mathrm A}^{-1} \boldsymbol{\mathrm B} \\[0.3em]
\boldsymbol{\mathrm O} & \boldsymbol{\mathrm I}
\end{bmatrix}
\begin{bmatrix}
\boldsymbol{\mathrm A}^{-1} & \boldsymbol{\mathrm O} \\[0.3em]
\boldsymbol{\mathrm O} & \boldsymbol{\mathrm S}^{-1}
\end{bmatrix}
\begin{bmatrix}
\boldsymbol{\mathrm I} & \boldsymbol{\mathrm O} \\[0.3em]
-\boldsymbol{\mathrm C} \boldsymbol{\mathrm A}^{-1} & \boldsymbol{\mathrm I}
\end{bmatrix}.
\end{align}
The above formula suggests the so-called Schur complement reduction (SCR) procedure \cite{Benzi2005,May2008} that solves the linear problem $\boldsymbol{\mathcal A} \bm x = \bm r$ by solving smaller block matrices $\boldsymbol{\mathrm A}$ and $\boldsymbol{\mathrm S}$. If the block matrices are not solved to very high precision, the SCR procedure can be utilized as a preconditioner denoted by $\boldsymbol{\mathcal P_\mathrm {\textup{SCR}}}$. When using the SCR as a preconditioner, one needs to invoke the Flexible Generalized Minimal REsidual method (FGMRES) \cite{Saad1993} as the Krylov iteration method because the preconditioner varies over iterations. We refer to the FGMRES iteration as the \textit{outer level} in this study, and its stopping criteria involve the relative tolerance $\delta$ and the maximum number of iterations $\mathrm{n}^{\mathrm{max}}$. The action of the SCR preconditioner on a vector is referred to as the \textit{intermediate level} in this work. It involves solving the systems associated with $\boldsymbol{\mathrm A}$ and $\boldsymbol{\mathrm S}$. We invoke GMRES to solve the two block matrices preconditioned by $\boldsymbol{\mathrm P}_{\mathrm A}$ and $\boldsymbol{\mathrm P}_{\mathrm S}$, respectively. The stopping criteria of the two solvers at the intermediate level include the relative tolerances $\delta_{\mathrm A}$ for $\boldsymbol{\mathrm A}$ and $\delta_{\mathrm S}$ for $\boldsymbol{\mathrm S}$ and the corresponding maximum iteration numbers $\mathrm{n}^{\mathrm{max}}_{\mathrm A}$ and $\mathrm{n}^{\mathrm{max}}_{\mathrm S}$. The action of the SCR preconditioner is detailed as Algorithm \ref{algorithm:exact_block_factorization} in \ref{sec:schur_complement_reduction_PC}.

The block matrix $\boldsymbol{\mathrm A}$ is a discrete convection-diffusion-reaction operator with the geometric multiscale coupling term \eqref{eq:geometric_multiscale_coupling_matrix} in the fluid subdomain \cite{Liu2020} or a discrete elasticity operator in the solid subdomain \cite{Liu2019a}. One advisable choice is to use the AMG method \cite{Ieary2000,Falgout2002} based on $\boldsymbol{\mathrm A}$ to construct $\boldsymbol{\mathrm P}_{\mathrm A}$. We mention that prior works also investigated the possibility of using the additive Schwarz method \cite{Deparis2016} and the Jacobi preconditioner \cite{Liu2020}.

The block matrices $\boldsymbol{\mathrm B}$ and $\boldsymbol{\mathrm C}$ are the discrete gradient and divergence operators, respectively, and the block matrix $\boldsymbol{\mathrm D}$ involves contributions from the VMS subgrid-scale modeling terms as well as the pressure time derivative term in \eqref{eq:continuum_mass_eqn}. With the above observation, the significance of $\boldsymbol{\mathrm S}$ can be better revealed by simplifying $\boldsymbol{\mathrm A}$ as a lumped mass matrix, and the Schur complement can be subsequently interpreted as an algebraic analog of the Laplace operator acting on the pressure variable. The SCR preconditioning technique can thus be viewed as wrapping the projection method \cite{Chorin1968,Kim1985,Teman1969} within the Krylov iterations \cite{Turek1999}. However, it is burdensome to work directly with $\boldsymbol{\mathrm S}$, which is a dense matrix due to the appearance of the inverse of $\boldsymbol{\mathrm A}$ in its definition. While the application of iterative methods like GMRES for $\boldsymbol{\mathrm S}$ can still be achieved via a matrix-free strategy, the construction of $\boldsymbol{\mathrm P}_{\mathrm S}$ is desirable to accelerate the convergence and is typically based on a sparse approximation of $\boldsymbol{\mathrm S}$ \cite{Cyr2012,Elman2008}. A simple and effective choice is $\hat{\boldsymbol{\mathrm S}}:=\boldsymbol{\mathrm D} - \boldsymbol{\mathrm C} \left( \mathrm{diag}\left(\boldsymbol{\mathrm A}\right)\right)^{-1} \boldsymbol{\mathrm B}$. The matrix-free definition of the action of  $\boldsymbol{\mathrm S}$ used in GMRES iteration is stated as Algorithm \ref{algorithm:matrix_free_mat_vec_for_S} in \ref{sec:schur_complement_reduction_PC}. In this algorithm, the action of  $\boldsymbol{\mathrm A}^{-1}$ is represented by a solution procedure associated with $\boldsymbol{\mathrm A}$ by GMRES preconditioned by $\boldsymbol{\mathrm P}_{\mathrm I}$. This solver is referred to be residing at the \textit{inner level} with stopping criteria involving the relative tolerance $\delta_{\mathrm I}$ and the maximum number of iterations $\mathrm{n}^{\mathrm{max}}_{\mathrm I}$. In this work, we construct $\boldsymbol{\mathrm P}_{\mathrm S}$ using the AMG method based on $\hat{\boldsymbol{\mathrm S}}$, and both $\boldsymbol{\mathrm P}_{\mathrm A}$ and $\boldsymbol{\mathrm P}_{\mathrm I}$ are constructed based on $\boldsymbol{\mathrm A}$ using the AMG method.

We mention that the above-described preconditioning technique can be simplified to the well-known SIMPLE preconditioner $\boldsymbol{\mathcal P}_{\textup{SIMPLE}}$ by using $\hat{\boldsymbol{\mathrm S}}$ to replace the Schur complement and using a diagonal approximation of $\boldsymbol{\mathrm A}$ in the upper diagonal matrix, that is,
\begin{align*}
\boldsymbol{\mathcal P}_{\textup{SIMPLE}} := 
\begin{bmatrix}
\boldsymbol{\mathrm I} & \boldsymbol{\mathrm O} \\[0.3em]
\boldsymbol{\mathrm C} \boldsymbol{\mathrm A}^{-1} & \boldsymbol{\mathrm I}
\end{bmatrix}
\begin{bmatrix}
\boldsymbol{\mathrm A} & \boldsymbol{\mathrm O} \\[0.3em]
\boldsymbol{\mathrm O} & \hat{ \boldsymbol{\mathrm S} }
\end{bmatrix}
\begin{bmatrix}
\boldsymbol{\mathrm I} & \left(\textup{diag}\left(\boldsymbol{\mathrm A}\right)\right)^{-1} \boldsymbol{\mathrm B} \\[0.3em]
\boldsymbol{\mathrm O} & \boldsymbol{\mathrm I}
\end{bmatrix}
=
\begin{bmatrix}
\boldsymbol{\mathrm A} & \boldsymbol{\mathrm A}\textup{diag}\left(\boldsymbol{\mathrm A}\right)^{-1}\boldsymbol{\mathrm B} \\[0.3em]
\boldsymbol{\mathrm C} & \boldsymbol{\mathrm D}
\end{bmatrix}.
\end{align*}
If the diagonal matrix is instead lumped by the absolute values of the entries of $\boldsymbol{\mathrm A}$, one recovers the SIMPLEC preconditioner \cite{Elman2008}. The implementation of $ \boldsymbol{\mathcal P}_{\textup{SIMPLE}}^{-1}$ and its variants can be conveniently realized within the framework of Algorithm 2.

\begin{remark}
In multiphysics simulations, block factorization has been demonstrated to an effective tool in designing solver technologies. Here we utilized the block structure to reduce the three-by-three block system into two smaller linear systems by invoking a simplifying assumption on the right-hand side of \eqref{eq:segregated_matrix_fluid}. Past experience suggested this assumption does not deteriorate the performance of the nonlinear solver \cite{Bazilevs2013a}. The linear system with a two-by-two block structure is factorized further to inspire the preconditioner design within the Krylov iterations. The repeated applications of the block factorization at both nonlinear and linear solver stages constitute the primary designing ingredient in our solver technology. There are other possibilities that are worthy of investigation. For example, the three-by-three block structure can be maintained in the nonlinear solver and thus renders the formulation to be of the ``fully-direct" type, according to the taxonomy introduced in \cite{Tezduyar2006}. The factorization of the three-by-three block matrix can be performed at the linear solver stage to facilitate a new preconditioner design.
\end{remark}

\subsubsection{Preconditioning technique for the mesh motion equations}
In the predictor multi-corrector algorithm,  in addition to solving the Jacobian matrix \eqref{eq:predictor-multi-corrector-linear-system}, one needs to solve the mesh motion equation. In a general situation, the mesh motion is governed by a system of elliptic differential equations, and an ideal choice is using the conjugate gradient method preconditioned by the AMG method. In this work, the mesh motion is governed by the harmonic extension algorithm and the discrete system is a system of discrete Laplacian equations. The mesh update procedure can be further simplified because one may assemble the matrix and setup its preconditioner before entering into the time stepping algorithms. In each time step, the solution of the mesh motion equation (i.e., Step 3 in the multi-corrector stage) only invokes the iterative method without updating the matrix or the preconditioner. Therefore, the mesh motion in this work is achieved in a rather cost-effective manner.

\section{Numerical examples}
\label{sec:numerical_examples}
In this section, we examine the proposed vascular FSI framework using the benchmark designed in \cite{Greenshields2005} as well as a patient-specific example. Unless otherwise specified, this work adopts centimeter-gram-second units. The computations reported in this work were made on the BSCC-T6 and Tai-Yi supercomputers. The former one is a supercomputer equipped with 2680 compute nodes, each of which consists of 2 Xeon Platinum 9242 48-core CPUs with 2.3 GHz nominal clock rate and 384 GB memory \cite{BSCC-T6-details}; the latter one is equipped with 815 compute nodes, each of which has 2 Xeon Gold 6148 20-core CPUs with 2.4 GHz nominal clock rate and 192 GB memory \cite{Taiyi-machine-details}. In the following numerical examples, the settings for the two-by-two block matrix problem are documented in Table \ref{table:gw_strong_scaling_solver_setting}; the stopping criteria of the mesh solver involve the relative tolerance and the maximum number of iterations, whose values are set to be $10^{-12}$ and $500$, respectively. 

\begin{table}[H]
\small
\begin{center}
\tabcolsep=0.19cm
\renewcommand{\arraystretch}{1.3}
\begin{tabular}{@{\extracolsep{3pt}}P{1.5cm} P{1.5cm} P{1.5cm} P{0.7cm} P{1.5cm} P{0.7cm} P{1.5cm} P{0.7cm} P{1.5cm} P{0.7cm} @{}}
\hline
\multicolumn{2}{c}{Nonlinear solver} & \multicolumn{2}{c}{Outer-level solver} & \multicolumn{4}{c}{Intermediate-level solver} & \multicolumn{2}{c}{Inner-level solver} \\
\cline{1-2} \cline{3-4} \cline{5-8} \cline{9-10}
$\mathrm{tol}_R$ & $\mathrm{tol}_A$ & $\delta$ & $\mathrm{n}^{\mathrm{max}}$ & $\delta_{\mathrm A}$ & $\mathrm{n}_{\mathrm{A}}^{\mathrm{max}}$ & $\delta_{\mathrm S}$ & $\mathrm{n}_{\mathrm{S}}^{\mathrm{max}}$ & $\delta_{\mathrm I}$ & $\mathrm{n}_{\mathrm{I}}^{\mathrm{max}}$ \\
\hline
$1.0\times 10^{-6}$ & $1.0\times 10^{-6}$ & $1.0\times 10^{-6}$ & $200$ & $1.0\times 10^{-3}$ & $200$ & $1.0 \times 10^{-3}$ & $200$ & $1.0\times 10^{-3}$ & $200$ \\
\hline
\end{tabular}
\end{center}
\caption{Settings of the nonlinear and linear solvers used in the fixed-size scalability study. The definitions of the solver parameters listed above can be found in Section \ref{sec:segregated-predictor-multi-corrector}, Section \ref{subsec:preconditioning}, and \ref{sec:schur_complement_reduction_PC}. In the SIMPLE preconditioner, the inner-level solver is not invoked, and the rest settings were the same.}
\label{table:gw_strong_scaling_solver_setting}
\end{table}

\subsection{The Greenshields-Weller benchmark}
The Greenshields-Weller benchmark was designed as a model problem in verifying FSI models \cite{Greenshields2005,Passerini2013,Liu2018,Bazilevs2006}. The problem is posed in an axisymmetric, deformable tube filled with fluid. The cylindrical tube length is 10 cm, its radius is 1 cm, and the wall thickness is 0.2 cm. A propagating wave is initiated in an initially quiescent system by imposing a pressure jump of 5 kPa at the inlet, while the exterior wall and the outlet of the tube are kept to be stress free. On the inlet and outlet surfaces, the solid wall is constrained in the axial direction but is free to move in the rest two directions. The material properties are chosen to mimic blood flows in large arteries. Although a linear elastic model was adopted in the original benchmark \cite{Greenshields2005}, we used the neo-Hookean model to characterize the wall. Its Gibbs free energy can be expressed as follows,
\begin{align*}
G(\widetilde{\bm{C}},p)=\frac{1}{2}\frac{\mu}{\rho_{0}^{s}}({\rm tr}\widetilde{\bm{C}}-3)-\frac{\kappa}{\rho_{0}^{s}}{\rm ln}(\frac{\kappa}{p+\kappa}).
\end{align*}
The detailed material properties are summarized in Table \ref{table:gw_material_properties}. Since the solid material is compressible in this case, we set $c_m$ and $c_c$ in \eqref{eq:fsi_solid_tau_designs} to $0$ to turn off the numerical dissipation mechanism in the solid subproblem.

\begin{table}[htbp]
\small
\begin{center}
\renewcommand{\arraystretch}{1.0}
\begin{tabular}{l l}
\hline
Solid properties \\
Shear modulus & $\mu = 3.85\times{10^{6}}$ dyn/cm$^{2}$ \\
Bulk modulus  & $\kappa = 8.33 \times 10^6$ dyn/cm$^2$ \\
Reference density & $\rho_{0}^{s}=1.00$ g/cm$^{3}$ \\[0.5em]
Fluid properties \\
Dynamic viscosity & $\bar{\mu}= 4.00 \times 10^{-2}$ poise  \\
Density & $\rho_{0}^{f}=1.00$ g/cm$^{3}$ \\
\hline
\end{tabular}
\end{center}
\caption{Material parameters used in the Greenshields-Weller benchmark. The shear and bulk moduli correspond to Young's modulus $E = 1.0\times 10^7$ dyn/cm$^2$ and Poisson's ratio $\nu = 0.3$ in the small-strain limit. }
\label{table:gw_material_properties}
\end{table}

\begin{figure}
\begin{center}
\begin{tabular}{cc}
\includegraphics[angle=0, trim=80 80 130 110, clip=true, scale = 0.32]{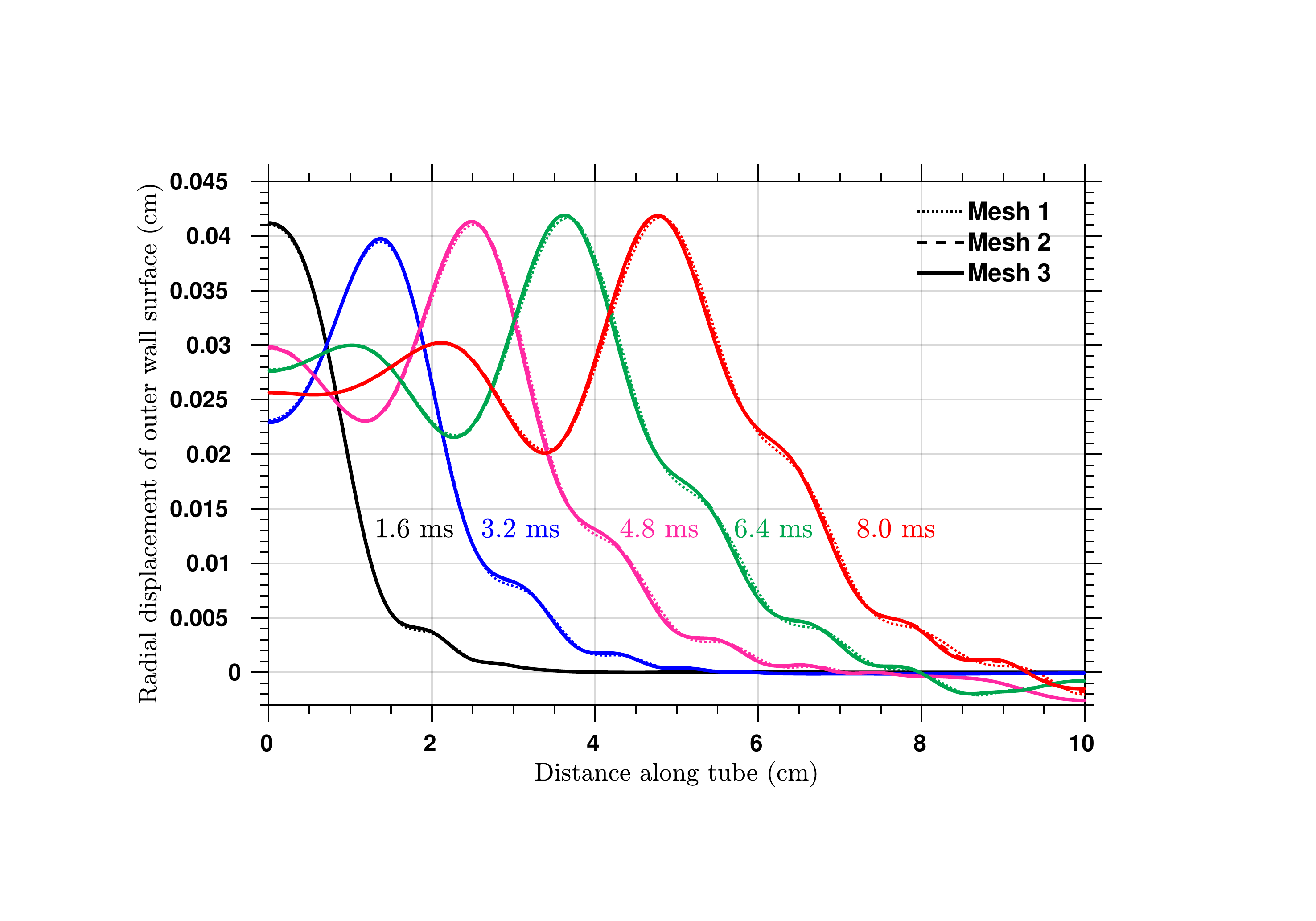} &
\includegraphics[angle=0, trim=80 80 130 110, clip=true, scale = 0.32]{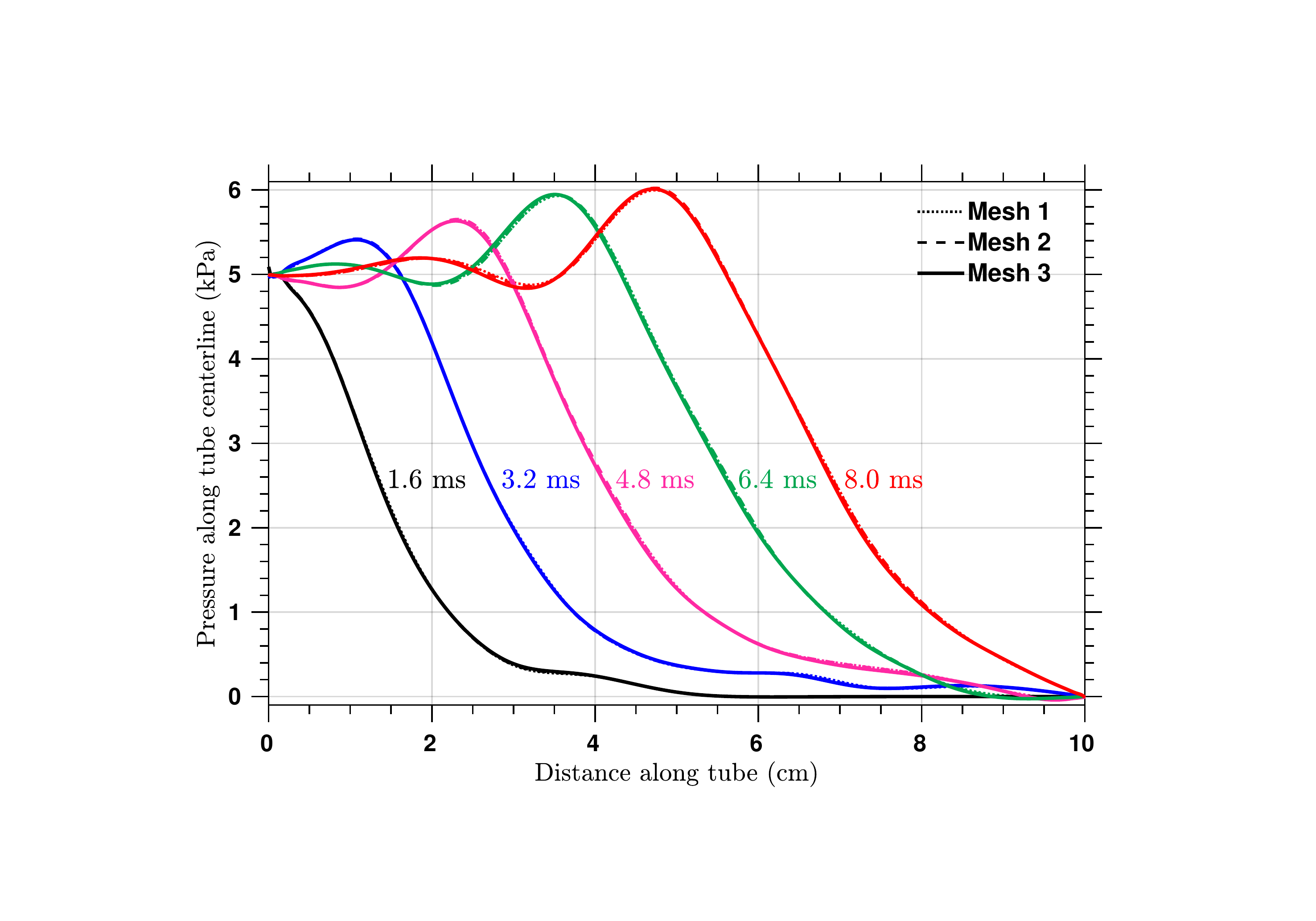}\\
(a) & (b) \\
\includegraphics[angle=0, trim=80 80 130 110, clip=true, scale = 0.32]{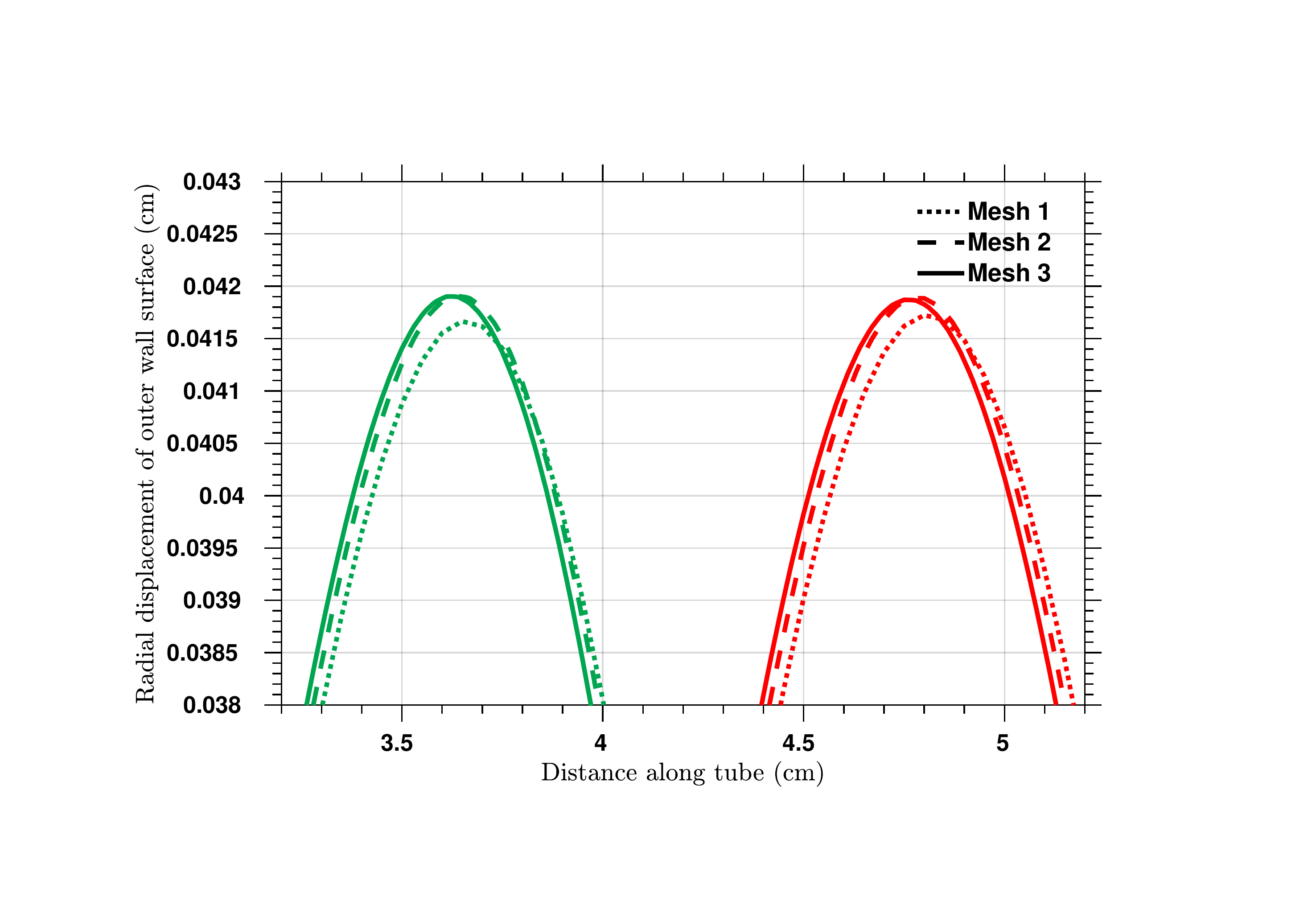} &
\includegraphics[angle=0, trim=80 80 130 110, clip=true, scale = 0.32]{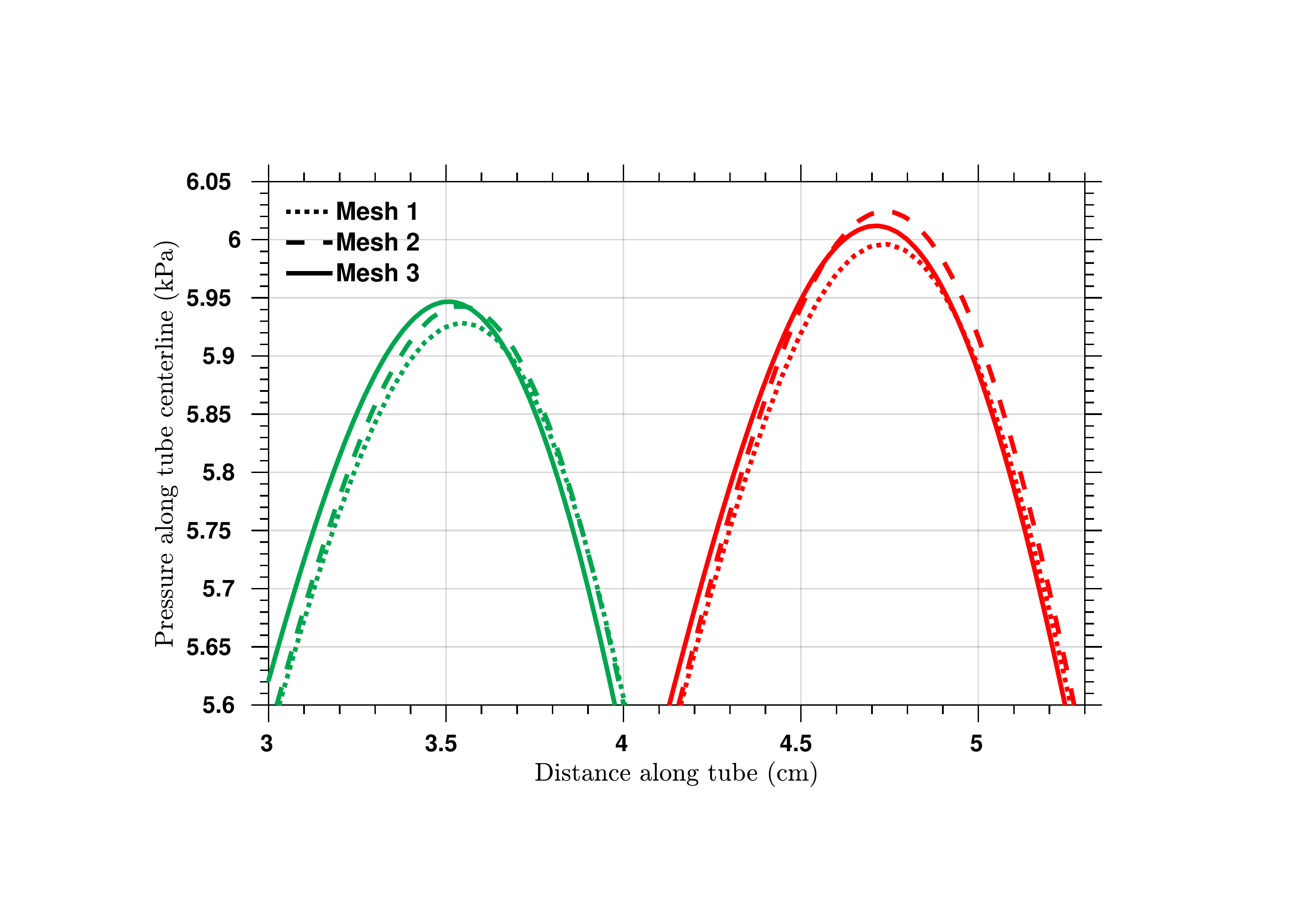}\\
(c) & (d)
\end{tabular}
\end{center}
\caption{The Greenshields-Weller benchmark problem: (a) Radial displacement of the outer wall surface; (b) Pressure along the tube centerline; (c) Detailed view of the radial displacement; (d) Detailed view of the pressure.}
\label{fig:greenshield_weller_disp_pressure_wave}
\end{figure}

Before investigating the solver performance, we first studied the wave propagation phenomenon with three different spatiotemporal resolutions (see Table \ref{table:gw_mesh}). Figure \ref{fig:greenshield_weller_disp_pressure_wave} depicts the radial displacement of the outer wall and the pressure along the centerline. We may observe from the snapshots that the pressure and radial displacement calculated from the three meshes are almost indistinguishable. Adopting the same definition of the pressure wavefront and wavelength given as in \cite{Greenshields2005,Liu2018,Passerini2013}, we may derive the wave velocity and frequency from the solution. From the result of the finest mesh (i.e., Mesh 3), we obtained a wave speed of $878$ cm/s and wave frequency $308.1$ Hz. The analytical solutions of the wave speed and frequency can be derived as functions of the pipe geometry and material properties, under proper assumptions \cite{Greenshields2005,Stuckenbruck1985}. The analytical solutions lead to an estimate of the wave speed to be $877$ cm/s and the wave frequency to be $308$ Hz. The numerical results agreed well with the theoretical estimates, which partly justifies the proposed FSI framework. Compared with the previous results \cite{Liu2018}, the discontinuity of fluid-solid interface pressure is the major difference in the formulation, which could explain the superior results obtained here.

\begin{table}
\small
\begin{center}
\tabcolsep=0.19cm
\renewcommand{\arraystretch}{1.2}
\begin{tabular}{c c c c}
\hline
Mesh &  No. nodes  & No. elements & $\Delta t$ [s] \\
\hline
1 & $9.73 \times 10^5$ & $5.71 \times 10^6$ & $2\times10^{-7}$ \\
2 & $2.29 \times 10^6$ & $1.35 \times 10^7$ & $1\times10^{-7}$ \\
3 & $4.02 \times 10^6$ & $2.38 \times 10^7$ & $5\times10^{-8}$ \\
\hline
\end{tabular}
\end{center}
\caption{Greenshields-Weller benchmark: spatial mesh and time step sizes }
\label{table:gw_mesh}
\end{table}

\subsubsection{Fixed-size scalability}
\label{sec:gw-strong-scaling}
In this part, we examined the parallel scalability of the proposed FSI solver technology by a fixed-size problem. We used Mesh 3 listed in Table \ref{table:gw_mesh} as well as Mesh 4, which contains $4.84 \times 10^7$ elements and $8.15 \times 10^6$ vertices. Both are anisotropic with boundary layer elements. The time step size was fixed to be $1.0 \times 10^{-7}$ s, and the problem was integrated up to $1.0 \times 10^{-6}$ s. The fixed-size scaling performance for SCR and SIMPLE preconditioners are reported in Tables \ref{table:strong_scaling based on SCR on Tai-Yi}, \ref{table:strong_scaling based on SIMPLE on Tai-Yi},  
 \ref{table:strong_scaling based on SCR on BJ}, and \ref{table:strong_scaling based on SIMPLE on BJ} respectively. The results demonstrate that the proposed preconditioning techniques exhibit a nearly optimal fixed-size scaling property for a rather wide range of processor counts.

\begin{table}[H]
\small
\begin{center}
\tabcolsep=0.19cm
\renewcommand{\arraystretch}{1.3}
\begin{tabular}{P{1.0cm} P{2.0cm} P{2.0cm} P{2.0cm} P{2.0cm} P{2.5cm} P{1.5cm} }
\hline
Proc. & $T_{A_1}$ (sec.) & $T_{A_2}$ (sec.) & $T_{L_1}$ (sec.) & $T_{L_2}$ (sec.) &$T_{\mathrm{total}}$ (sec.) & Efficiency   \\
\hline
2 & $7.08\times 10^{3}$ & $1.37\times 10^2$ & $2.78 \times 10^4$ & $5.25\times 10^3$ &$4.03\times 10^4$ & $-$ \\
4 & $3.58\times 10^{3}$ & $6.82\times 10^1$ & $1.38 \times 10^4$ & $2.57\times 10^3$ &$2.00\times 10^4$ & $100.66\%$ \\
8 & $1.84\times 10^{3}$ & $3.48\times 10^1$ & $6.95 \times 10^3$ & $1.34\times 10^3$ &$1.02\times 10^4$ & $99.09\%$ \\
16 & $8.76\times 10^{2}$ & $1.65\times 10^1$ & $3.57 \times 10^3$ & $6.71\times 10^2$ &$5.14\times 10^3$ & $98.02\%$ \\
20 & $7.13\times 10^{2}$ & $1.35\times 10^1$ & $2.98 \times 10^3$ & $5.39\times 10^2$ &$4.25\times 10^3$ & $94.86\%$ \\
32 & $4.55\times 10^{2}$ & $8.48\times 10^0$ & $1.88 \times 10^3$ & $3.56\times 10^2$ &$2.70\times 10^3$ & $93.34\%$ \\
40 & $3.82\times 10^{2}$ & $6.93\times 10^0$ & $1.59 \times 10^3$ & $3.51\times 10^2$ &$2.33\times 10^3$ & $86.33\%$ \\
80 & $2.48\times 10^{2}$ & $4.12\times 10^0$ & $9.57 \times 10^2$ & $2.05\times 10^2$ &$1.42\times 10^3$ & $71.13\%$ \\
\hline
\end{tabular}
\end{center}
\caption{The fixed-size scaling performance tested on Tai-Yi supercomputer using the SCR preconditioner and Mesh 3. In the table, $T_{A_1}$ and $T_{L_1}$ represent the timings of the tangent matrix assembly and linear system solve associated with the equation \eqref{eq:predictor-multi-corrector-linear-system}, respectively; $T_{A_2}$ and $T_{L_2}$ represent the timings of the right-hand side assembly and linear system solve associated with the mesh motion equations, respectively. The efficiency is computed based on the total runtime of the time integration $T_\mathrm{total}$.}
\label{table:strong_scaling based on SCR on Tai-Yi}
\end{table}

\begin{table}[htbp]
\small
\begin{center}
\tabcolsep=0.19cm
\renewcommand{\arraystretch}{1.3}
\begin{tabular}{P{1.0cm} P{2.0cm} P{2.0cm} P{2.0cm} P{2.0cm} P{2.5cm} P{1.5cm} }
\hline
Proc. & $T_{A_1}$ (sec.) & $T_{A_2}$ (sec.) & $T_{L1}$ (sec.) & $T_{L_2}$ (sec.) &$T_{\mathrm{total}}$ (sec.) & Efficiency   \\
\hline
2 & $6.60\times 10^{3}$ & $1.30\times 10^2$ & $1.84 \times 10^4$ & $4.94\times 10^3$ &$3.01\times 10^4$ & $-$ \\
4 & $3.35\times 10^{3}$ & $6.48\times 10^1$ & $9.26 \times 10^3$ & $2.42\times 10^3$ &$1.51\times 10^4$ & $99.47\%$ \\
8 & $1.71\times 10^{3}$ & $3.33\times 10^1$ & $4.72 \times 10^3$ & $1.27\times 10^3$ &$7.74\times 10^3$ & $97.03\%$ \\
16 & $8.79\times 10^{2}$ & $1.67\times 10^1$ & $2.43 \times 10^3$ & $6.68\times 10^2$ &$4.00\times 10^3$ & $93.98\%$ \\
20 & $7.15\times 10^{2}$ & $1.34\times 10^1$ & $1.98 \times 10^3$ & $5.37\times 10^2$ &$3.25\times 10^3$ & $92.59\%$ \\
32 & $4.57\times 10^{2}$ & $8.66\times 10^0$ & $1.28 \times 10^3$ & $3.55\times 10^2$ &$2.11\times 10^3$ & $89.23\%$ \\
40 & $3.92\times 10^{2}$ & $6.93\times 10^0$ & $1.04 \times 10^3$ & $3.53\times 10^2$ &$1.80\times 10^3$ & $83.53\%$ \\
80 & $2.40\times 10^{2}$ & $4.04\times 10^0$ & $6.22 \times 10^2$ & $2.03\times 10^2$ &$1.07\times 10^3$ & $70.09\%$ \\
\hline
\end{tabular}
\end{center}
\caption{The fixed-size scaling performance tested on Tai-Yi supercomputer using the SIMPLE preconditioner and Mesh 3. In the table, $T_{A_1}$ and $T_{L_1}$ represent the timings of the tangent matrix assembly and linear system solve associated with the equation \eqref{eq:predictor-multi-corrector-linear-system}, respectively; $T_{A_2}$ and $T_{L_2}$ represent the timings of the right-hand side assembly and linear system solve associated with the mesh motion equations, respectively. The efficiency is computed based on the total runtime of the time integration $T_\mathrm{total}$.}
\label{table:strong_scaling based on SIMPLE on Tai-Yi}
\end{table}

\begin{table}[htbp]
\small
\begin{center}
\tabcolsep=0.19cm
\renewcommand{\arraystretch}{1.3}
\begin{tabular}{P{1.0cm} P{2.0cm} P{2.0cm} P{2.0cm} P{2.0cm} P{2.5cm} P{1.5cm} }
\hline
Proc. & $T_{A_1}$ (sec.) & $T_{A_2}$ (sec.) & $T_{L_1}$ (sec.) & $T_{L_2}$ (sec.) &$T_{\mathrm{total}}$ (sec.) & Efficiency   \\
\hline
96 & $3.22\times 10^{2}$ & $5.79\times 10^0$ & $2.05 \times 10^3$ & $2.79\times 10^2$ &$2.66\times 10^3$ & $-$ \\
192 & $1.68\times 10^{2}$ & $3.18\times 10^0$ & $1.02 \times 10^4$ & $1.36\times 10^3$ &$1.33\times 10^4$ & $99.94\%$ \\
384 & $8.58\times 10^{1}$ & $1.64\times 10^0$ & $5.39 \times 10^2$ & $6.95\times 10^1$ &$6.97\times 10^2$ & $95.41\%$ \\
768 & $4.35\times 10^{1}$ & $8.33\times 10^{-1}$ & $2.85 \times 10^2$ & $3.90\times 10^1$ &$3.71\times 10^2$ & $89.76\%$ \\
1536 & $2.40\times 10^{1}$ & $4.69\times 10^{-1}$ & $1.59 \times 10^2$ & $3.62\times 10^1$ &$2.21\times 10^2$ & $75.17\%$ \\
3072 & $1.23\times 10^{1}$ & $2.64\times 10^{-1}$ & $9.51 \times 10^1$ & $2.49\times 10^1$ &$1.35\times 10^2$ & $61.77\%$ \\
6144 & $6.47\times 10^0$ & $1.72\times 10^{-1}$ & $7.08 \times 10^1$ & $2.54\times 10^1$ &$1.05\times 10^2$ & $39.54\%$ \\
\hline
\end{tabular}
\end{center}
\caption{The fixed-size scaling performance tested on BSCC-T6 supercomputer using the SCR preconditioner and Mesh 4. In the table, $T_{A_1}$ and $T_{L_1}$ represent the timings of the tangent matrix assembly and linear system solve associated with the equation \eqref{eq:predictor-multi-corrector-linear-system}, respectively; $T_{A_2}$ and $T_{L_2}$ represent the timings of the right-hand side assembly and linear system solve associated with the mesh motion equations, respectively. The efficiency is computed based on the total runtime of the time integration $T_\mathrm{total}$.}
\label{table:strong_scaling based on SCR on BJ}
\end{table}

\begin{table}[htbp]
\small
\begin{center}
\tabcolsep=0.19cm
\renewcommand{\arraystretch}{1.3}
\begin{tabular}{P{1.0cm} P{2.0cm} P{2.0cm} P{2.0cm} P{2.0cm} P{2.5cm} P{1.5cm} }
\hline
Proc. & $T_{A_1}$ (sec.) & $T_{A_2}$ (sec.) & $T_{L_1}$ (sec.) & $T_{L_2}$ (sec.) &$T_{\mathrm{total}}$ (sec.) & Efficiency   \\
\hline
96 & $3.23\times 10^{2}$ & $5.64\times 10^0$ & $9.33 \times 10^2$ & $2.77\times 10^2$ &$1.54\times 10^3$ & $-$ \\
192 & $1.66\times 10^{2}$ & $3.12\times 10^0$ & $4.73 \times 10^2$ & $1.36\times 10^2$ &$7.82\times 10^2$ & $98.64\%$ \\
384 & $9.09\times 10^{1}$ & $1.69\times 10^0$ & $2.64 \times 10^2$ & $7.37\times 10^1$ &$4.32\times 10^2$ & $89.14\%$ \\
768 & $4.75\times 10^{1}$ & $8.98\times 10^{-1}$ & $1.46 \times 10^2$ & $4.14\times 10^1$ &$2.38\times 10^2$ & $81.05\%$ \\
1536 & $2.35\times 10^{1}$ & $5.07\times 10^{-1}$ & $1.53 \times 10^2$ & $2.58\times 10^1$ &$2.05\times 10^2$ & $47.01\%$ \\
3072 & $1.33\times 10^{1}$ & $2.86\times 10^{-1}$ & $6.59 \times 10^1$ & $2.81\times 10^1$ &$1.10\times 10^2$ & $43.98\%$ \\
6144 & $6.59\times 10^0$ & $1.27\times 10^{-1}$ & $5.97 \times 10^1$ & $2.86\times 10^1$ &$9.75\times 10^1$ & $24.72\%$ \\
\hline
\end{tabular}
\end{center}
\caption{The fixed-size scaling performance tested on BSCC-T6 supercomputer using the SIMPLE preconditioner and Mesh 4. In the table, $T_{A_1}$ and $T_{L_1}$ represent the timings of the tangent matrix assembly and linear system solve associated with the equation \eqref{eq:predictor-multi-corrector-linear-system}, respectively; $T_{A_2}$ and $T_{L_2}$ represent the timings of the right-hand side assembly and linear system solve associated with the mesh motion equations, respectively. The efficiency is computed based on the total runtime of the time integration $T_\mathrm{total}$.}
\label{table:strong_scaling based on SIMPLE on BJ}
\end{table}

\begin{figure}[htbp]
\begin{center}
\includegraphics[angle=0, trim=170 80 160 110, clip=true, scale = 0.3]{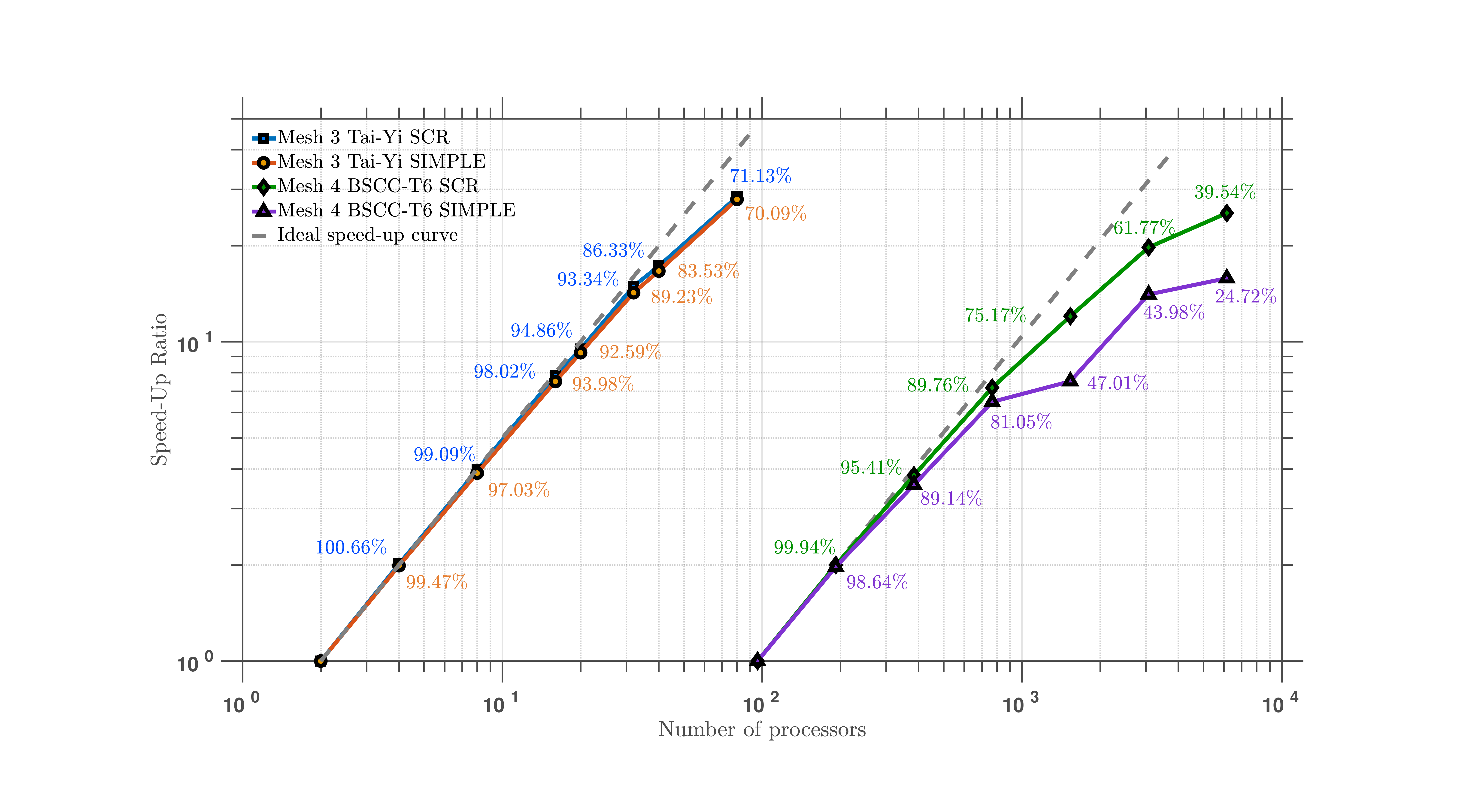} 
\end{center}
\caption{Fixed-size scalability of our solution strategy. Annotated efficiency rates are computed from the total runtime of the time integration.}
\label{fig:FSI_strong_scaling}
\end{figure}

\subsubsection{Isogranular scalability}
In this test, we investigated the isogranular scalability of the proposed FSI solution strategies. Four sets of anisotropic unstructured meshes with boundary layer elements were generated. We partitioned the meshes in such a way that each processor is assigned approximately $4.86 \times 10^4$ finite element nodes, and the number of processors increases from $2$ to $16$. The time step size was fixed to be $1.0 \times 10^{-7}$ s, and the problem was integrated up to $1.0 \times 10^{-6}$ s. This test was performed on the BSCC-T6 supercomputer, and the test results are summarized in Tables \ref{table:weak_scaling_scr} and \ref{table:weak_scaling_simple}. In both tables, the solvers at the intermediate level of the preconditioner, the mesh solver, and the assembly of \eqref{eq:predictor-multi-corrector-linear-system} are also reported. Since we only tested the preconditioning technique for solving \eqref{eq:predictor-multi-corrector-linear-system}, the mesh solver and the assembly procedure are identical. We observe that, with an eight-fold increase of the problem size, the averaged solution time for the discrete Laplacian operator increases by about 38\%, and the average time spent in the assembly of the linear system increases by approximately 14\%. Comparing the time spent on solving the block system \eqref{eq:predictor-multi-corrector-linear-system}, we see that the SCR preconditioner is more computationally expensive than the SIMPLE preconditioner. The averaged time spent on solving the block matrices $\boldsymbol{\mathrm A}$ and $\boldsymbol{\mathrm S}$ are also reported, from which we may identify that the additional cost of the SCR preconditioner primarily comes from the invocation of the inner level solver. The scaling results are comparable with the isogranular scaling results tested on a three-dimensional CFD problem reported in \cite{Elman2008}, suggesting the FSI solver enjoys a desirable isogranular scalability property.

\begin{table}[htbp]
\small
\begin{center}
\tabcolsep=0.19cm
\renewcommand{\arraystretch}{1.3}
\begin{tabular}{@{\extracolsep{4pt}}P{1.5cm} P{0.8cm} P{0.8cm} P{1.5cm} P{1.5cm} P{1.5cm} P{0.8cm} P{1.5cm} P{1.5cm} @{}}
\hline
No. nodes & Proc. & $\bar{n}_{L}$ & $\bar{T}_{L}$ (sec.) & $\bar{T}_{A}$ (sec.) & $\bar{T}_{S}$ (sec.) & $\bar{n}_{M}$ & $\bar{T}_{M}$ (sec.) & $\bar{T}_{AS}$ (sec.) \\
\hline
$9.73\times 10^{5}$ & 2 & 2.0 & $6.17 \times 10^1$ & $3.22 \times 10^0$ & $2.19 \times 10^1$ & 28.6 & $1.65 \times 10^1$ & $4.00 \times 10^1$ \\
$1.94\times 10^{6}$ & 4 & 2.1 & $6.71 \times 10^1$ & $3.23 \times 10^0$ & $2.36 \times 10^1$ & 30.6 & $1.84 \times 10^1$ & $4.38 \times 10^1$ \\
$3.89\times 10^{6}$ & 8 & 2.1 & $8.01 \times 10^1$ & $3.28 \times 10^0$ & $2.88 \times 10^1$ & 33.0 & $2.14 \times 10^1$ & $4.37 \times 10^1$ \\
$7.79\times 10^{6}$ & 16 & 2.3 & $1.10 \times 10^2$ & $3.48 \times 10^0$ & $3.81 \times 10^1$ & 35.0 & $2.28 \times 10^1$ & $4.57 \times 10^1$ \\
\hline
\end{tabular}
\end{center}
\caption{Comparison of the averaged iteration counts and CPU time in seconds using the SCR preconditioner. The values of $\bar{n}_{L}$ and $\bar{n}_{M}$ represent the averaged number of iterations for solving \eqref{eq:predictor-multi-corrector-linear-system} and the mesh motion equations, respectively. The values of $\bar{T}_{L}$, $\bar{T}_{A}$, $\bar{T}_{S}$, and $\bar{T}_{M}$ represent the averaged time for solving \eqref{eq:predictor-multi-corrector-linear-system}, the block matrix $\boldsymbol{\mathrm A}$, the Schur complement $\boldsymbol{\mathrm S}$,  and the mesh motion equations, respectively. The value of $\bar{T}_{AS}$ denotes the average timings for the assembly of the linear system \eqref{eq:predictor-multi-corrector-linear-system}.}
\label{table:weak_scaling_scr}
\end{table}

\begin{table}[htbp]
\small
\begin{center}
\tabcolsep=0.19cm
\renewcommand{\arraystretch}{1.3}
\begin{tabular}{@{\extracolsep{4pt}}P{1.5cm} P{0.8cm} P{0.8cm} P{1.5cm} P{1.5cm} P{1.5cm} P{0.8cm} P{1.5cm} P{1.5cm} @{}}
\hline
No. nodes & Proc. & $\bar{n}_{L}$ & $\bar{T}_{L}$ (sec.) & $\bar{T}_{A}$ (sec.) & $\bar{T}_{S}$ (sec.) & $\bar{n}_{M}$ & $\bar{T}_{M}$ (sec.) & $\bar{T}_{AS}$ (sec.) \\
\hline
$9.73\times 10^{5}$ & 2 & 6.4 & $6.10 \times 10^1$ & $2.93 \times 10^0$ & $2.91 \times 10^0$  & 28.6 & $1.66 \times 10^1$ & $4.03 \times 10^1$ \\
$1.94\times 10^{6}$ & 4 & 6.4 & $6.37 \times 10^1$ & $2.96 \times 10^0$ & $3.25 \times 10^0$ & 30.6 & $1.84 \times 10^1$ & $4.29 \times 10^1$ \\
$3.89\times 10^{6}$ & 8 & 6.4 & $7.03 \times 10^1$ & $3.05 \times 10^0$ & $3.99 \times 10^0$ & 33.0 & $2.13 \times 10^1$  & $4.36 \times 10^1$ \\
$7.79\times 10^{6}$ & 16 & 6.6 & $8.54 \times 10^1$  & $3.27 \times 10^0$ & $5.52 \times 10^0$ & 35.0 & $2.28 \times 10^1$ & $4.55 \times 10^1$ \\
\hline
\end{tabular}
\end{center}
\caption{Comparison of the averaged iteration counts and CPU time in seconds using the SIMPLE preconditioner. The values of $\bar{n}_{L}$ and $\bar{n}_{M}$ represent the averaged number of iterations for solving \eqref{eq:predictor-multi-corrector-linear-system} and the mesh motion equations, respectively. The values of $\bar{T}_{L}$, $\bar{T}_{A}$, $\bar{T}_{S}$, and $\bar{T}_{M}$ represent the averaged time for solving \eqref{eq:predictor-multi-corrector-linear-system}, the block matrix $\boldsymbol{\mathrm A}$, the Schur complement $\boldsymbol{\mathrm S}$,  and the mesh motion equations, respectively. The value of $\bar{T}_{AS}$ denotes the average timings for the assembly of the linear system \eqref{eq:predictor-multi-corrector-linear-system}.}
\label{table:weak_scaling_simple}
\end{table}

\subsubsection{Robustness with respect to material properties}
Besides scalability, we examined the robustness of the solver technology with varying material properties. We used Mesh 3 with 80 processors on the Tai-Yi supercomputer. The fluid properties were fixed, and the solid density varied from $10^{-2}$ to $10^2$ and Young's modulus $E$ varied from $10^6$ to $10^8$. We observe from Table \ref{table:gw_robustness} that both preconditioners are insensitive to Young's modulus. For heavier solids, both nonlinear and linear solvers converge with less number of iterations. For the SIMPLE preconditioner, the averaged time for solving the linear system \eqref{eq:predictor-multi-corrector-linear-system} mildly decreases with the increase of the solid density, while for the SCR preconditioner the trend is not clear. Still, the averaged time for solving the linear system remains within a rather narrow range, suggesting the overall solver technology is robust concerning material properties.

\begin{table}[htbp]
\small
\begin{center}
\tabcolsep=0.19cm
\renewcommand{\arraystretch}{1.3}
\begin{tabular}{@{\extracolsep{4pt}}P{1.5cm} P{1.5cm} P{1.0cm} P{1.0cm} P{1.5cm} P{1.0cm} P{1.0cm} P{1.5cm} @{}} \\
\hline
\multirow{2}{*}{$\rho_0^s$} & \multirow{2}{*}{$E$} & \multicolumn{3}{c}{SCR} & \multicolumn{3}{c}{SIMPLE} \\
\cline{3-5} \cline{6-8}
& & $\bar{l}$ & $\bar{n}_L$ & $\bar{T}_L$ & $\bar{l}$ & $\bar{n}$ & $\bar{T}_L$ \\
\hline
$1.0 \times 10^{-2}$ & $1.0 \times 10^7$ & $1.9$ & $2.4$ & $2.05 \times 10^1$ & $1.9$ & $7.6$ & $1.87 \times 10^1$ \\
$1.0 \times 10^{-1}$ & $1.0 \times 10^7$ & $1.9$ & $2.4$ & $2.10 \times 10^1$ & $1.9$ & $7.3$ & $1.82 \times 10^1$ \\
$1.0 \times 10^{0}$ & $1.0 \times 10^7$ & $1.8$ & $2.4$ & $2.58 \times 10^1$ & $1.8$ & $6.6$ & $1.70 \times 10^1$ \\
$1.0 \times 10^{1}$ & $1.0 \times 10^7$ & $1.6$ & $2.6$ & $3.14 \times 10^1$ & $1.7$ & $6.4$ & $1.63 \times 10^1$ \\
$1.0 \times 10^{2}$ & $1.0 \times 10^7$ & $1.0$ & $2.2$ & $2.24 \times 10^1$ & $1.0$ & $6.0$ & $1.43 \times 10^1$ \\
\hline
$1.0 \times 10^{0}$ & $1.0 \times 10^6$ & $1.8$ & $2.4$ & $2.55 \times 10^1$ & $1.8$ & $6.6$ & $1.70 \times 10^1$ \\
$1.0 \times 10^{0}$ & $1.0 \times 10^8$ & $1.8$ & $2.4$ & $2.53 \times 10^1$ & $1.8$ & $6.7$ & $1.72 \times 10^1$ \\
\hline
\end{tabular}
\end{center}
\caption{Comparison of the solver performance with varying material properties. The values of $\bar{l}$, $\bar{n}_L$, and $\bar{T}_L$ represent the averaged number of multi-correctors in the nonlinear solver, the averaged number of the Krylov iterations for solving \eqref{eq:predictor-multi-corrector-linear-system}, and the averaged time for solving \eqref{eq:predictor-multi-corrector-linear-system}, respectively.}
\label{table:gw_robustness}
\end{table}

\subsection{Patient-specific vascular FSI}
\begin{figure}[htbp]
\begin{center}
\includegraphics[angle=0, trim=0 102 70 0, clip=true, scale = 0.35]{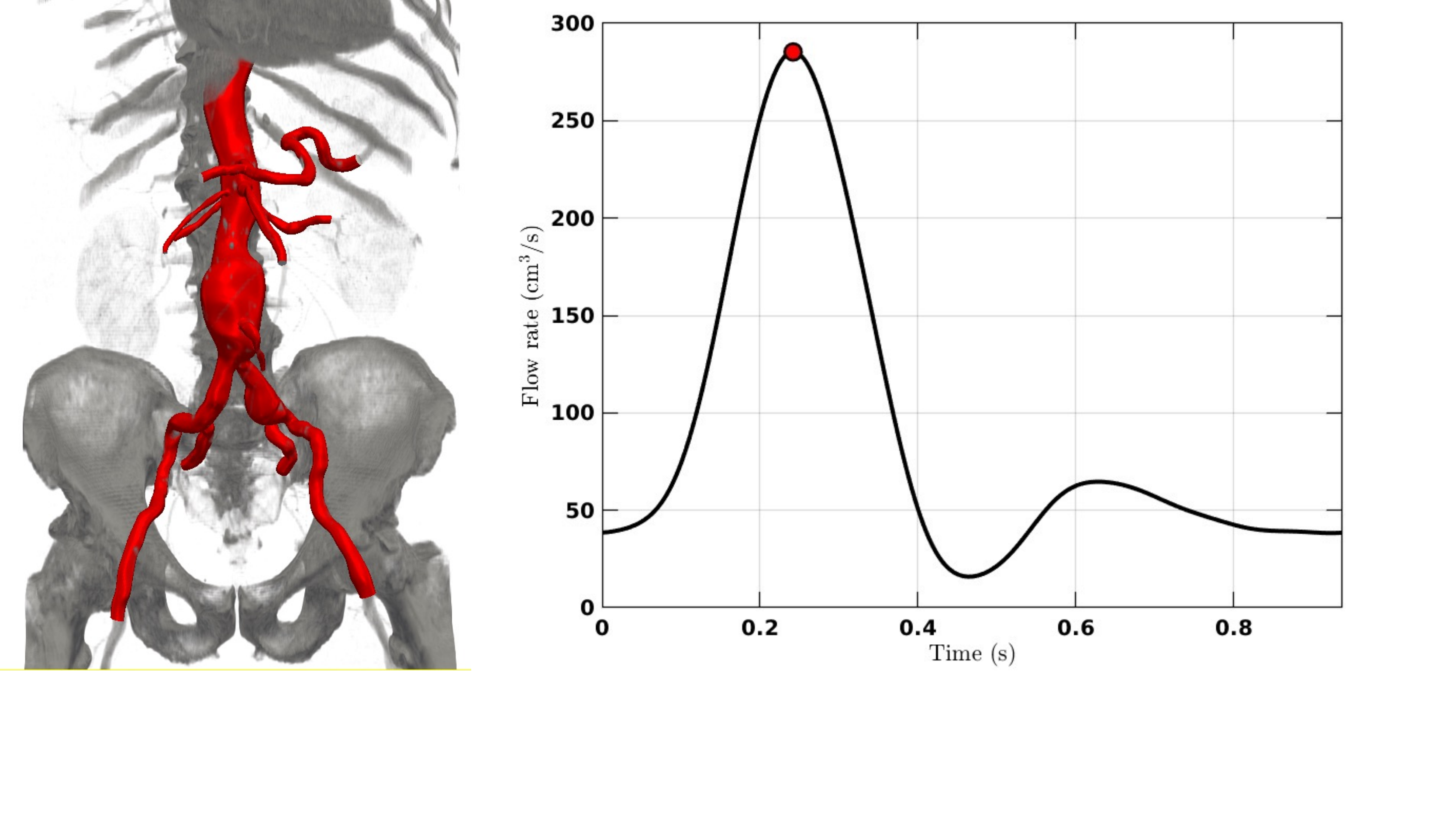} 
\end{center}
\caption{Volume rendering of the medical imaging data (left), and the volumetric flow waveform prescribed on the inlet surface (right). The red dot represents the peak systole.}
\label{fig:FSI_AAA_model}
\end{figure}

\begin{figure}[htbp]
\begin{center}
\includegraphics[angle=0, trim=0 0 130 10, clip=true, scale = 0.5]{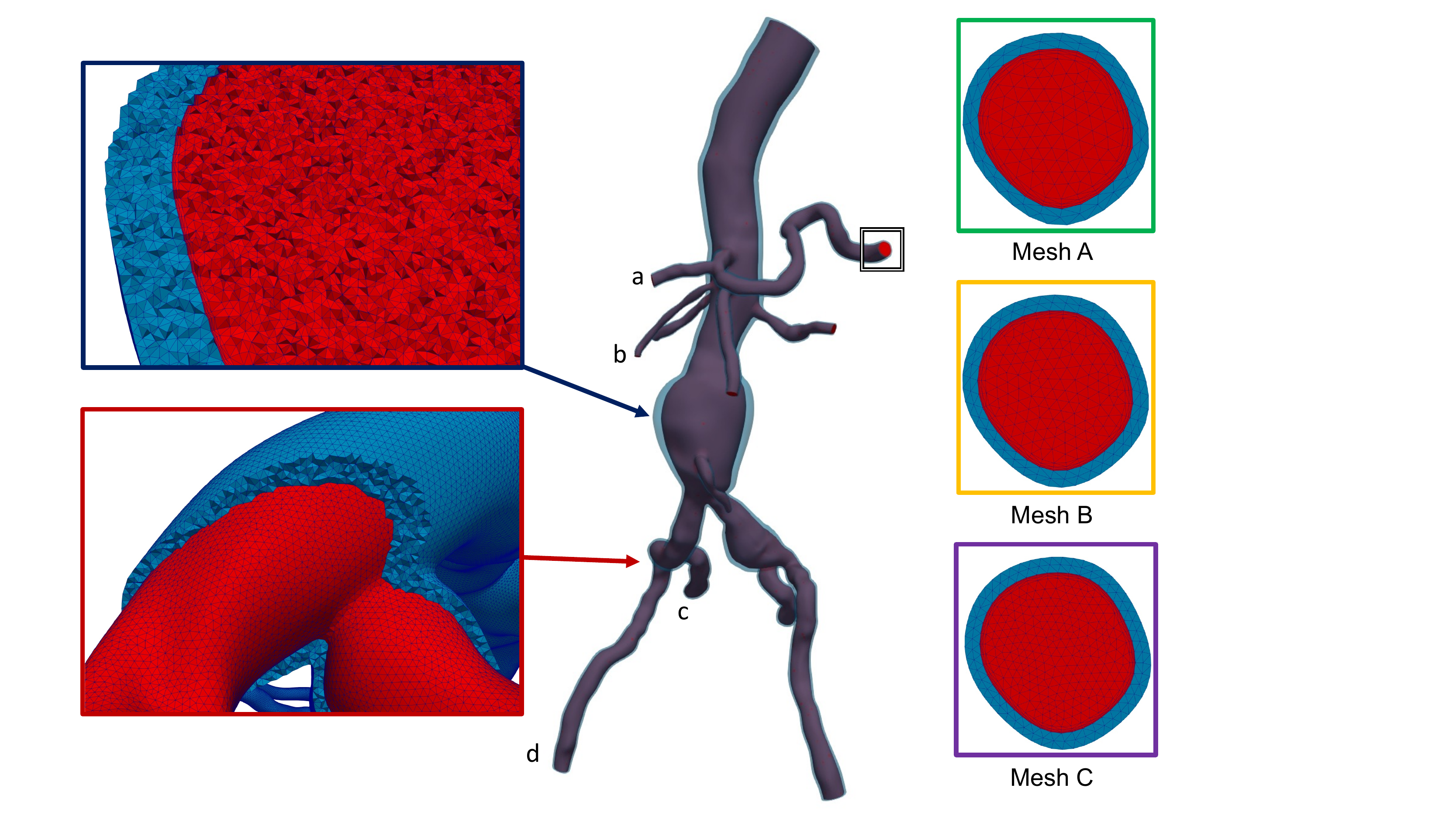} 
\end{center}
\caption{Illustration of the meshes used for the model. The volumetric mesh of Mesh C is shown in the middle. The three meshes of the splenic artery outlet surface are depicted on the right.}
\label{fig:FSI_AAA_mesh}
\end{figure}

\begin{figure}[htbp]
\begin{center}
\begin{tabular}{cc}
\includegraphics[angle=0, trim=20 0 0 20, clip=true, scale = 0.22]{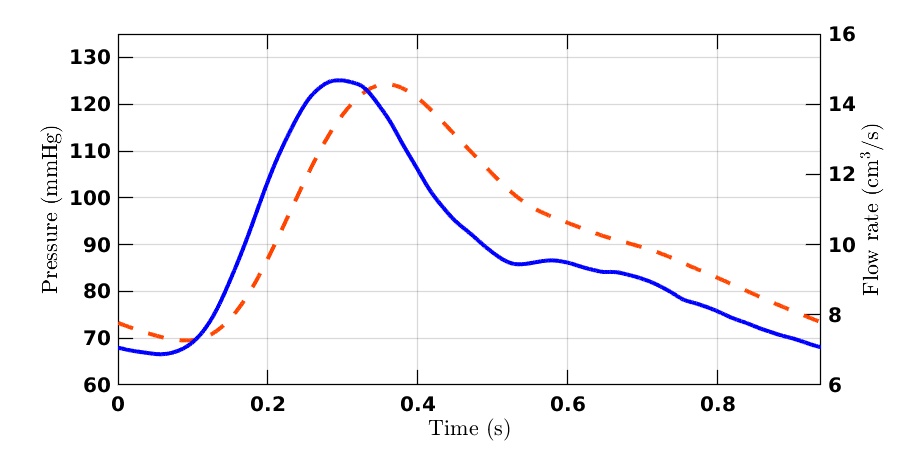} &
\includegraphics[angle=0, trim=20 0 0 20, clip=true, scale = 0.22]{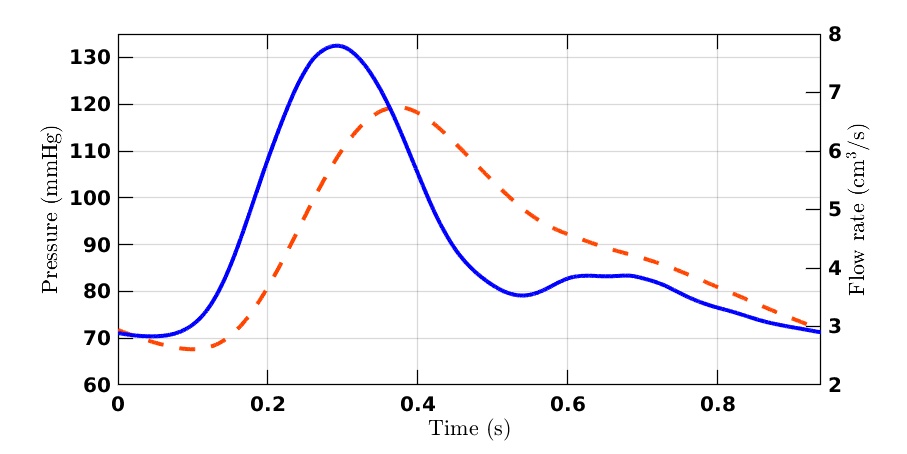} \\
(a) & (b) \\
\includegraphics[angle=0, trim=20 0 0 20, clip=true, scale = 0.22]{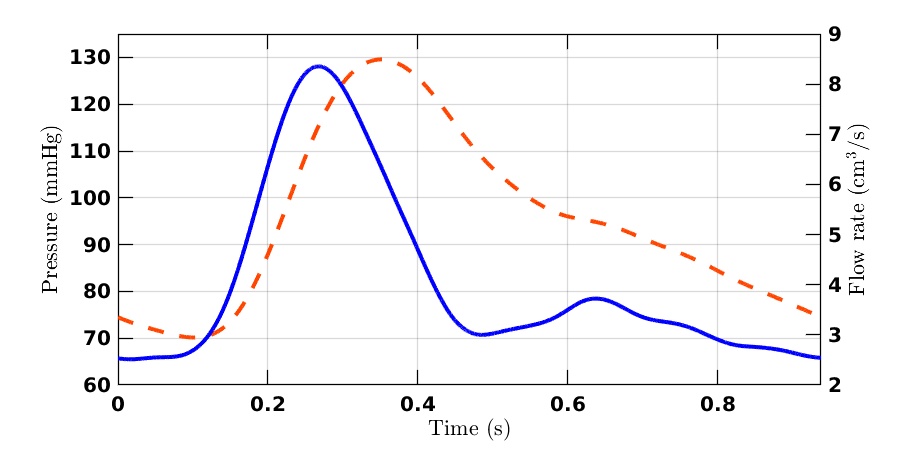} &
\includegraphics[angle=0, trim=20 0 0 20, clip=true, scale = 0.22]{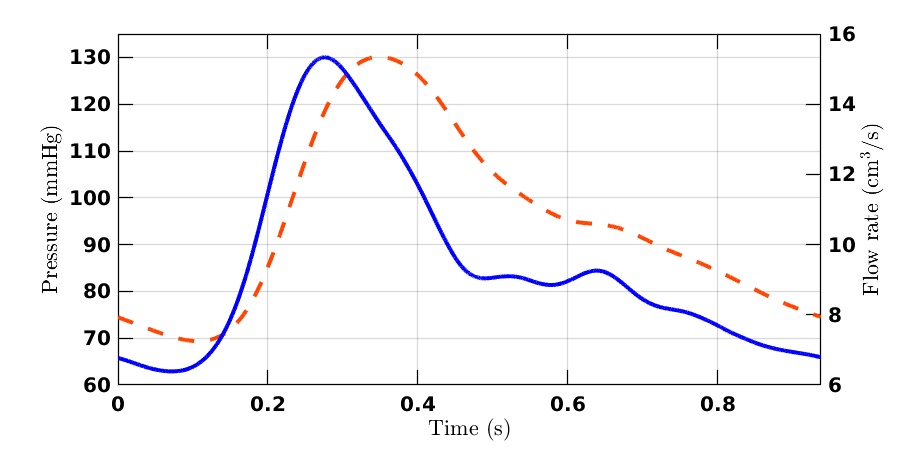} \\
(c) & (d) \\
\multicolumn{2}{c}{ \includegraphics[angle=0, trim=329 23 300 410, clip=true, scale = 0.4]{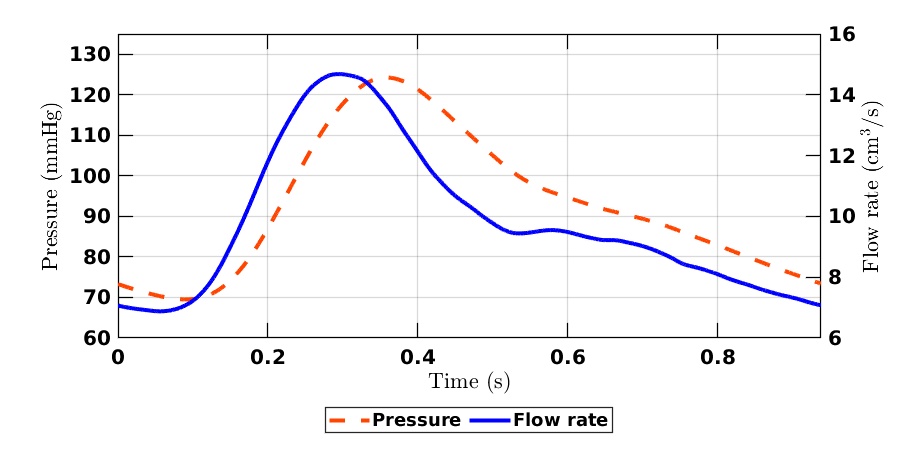} }
\end{tabular}
\end{center}
\caption{The flow and pressure waveforms on four outlets of the model computed based on Mesh C. The locations of the outlets are depicted in Figure \ref{fig:FSI_AAA_mesh}.}
\label{fig:FSI_AAA_pressure_flow_outlets}
\end{figure}

\begin{figure}[htbp]
\begin{center}
\includegraphics[angle=0, trim=0 0 260 0, clip=true, scale = 0.5]{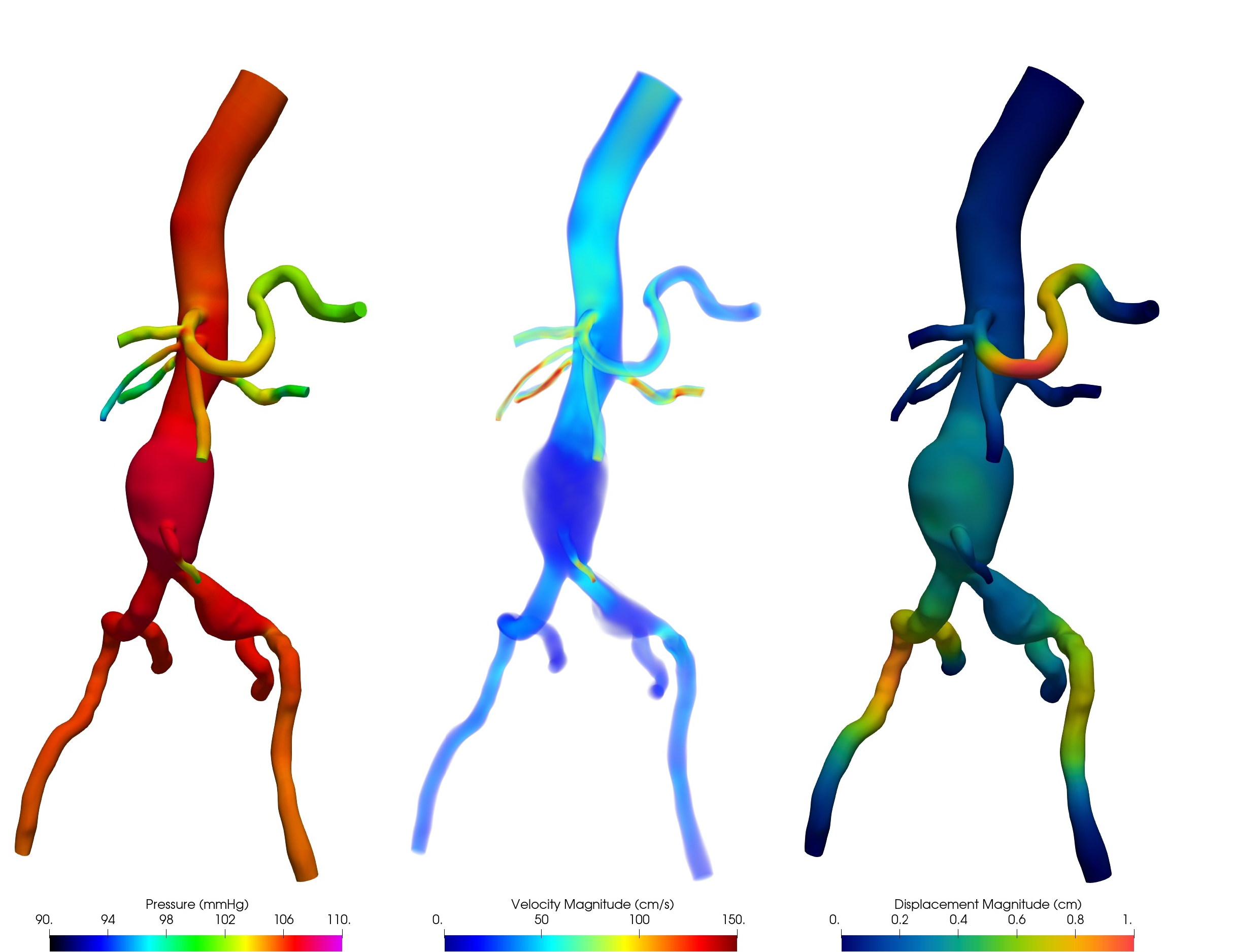} 
\end{center}
\caption{The pressure, velocity, and wall displacement computed based on Mesh C near the peak systole.}
\label{fig:FSI_AAA_pres_velo_disp_snapshot}
\end{figure}

\begin{figure}[htbp]
\begin{center}
\includegraphics[angle=0, trim=0 0 130 10, clip=true, scale = 0.5]{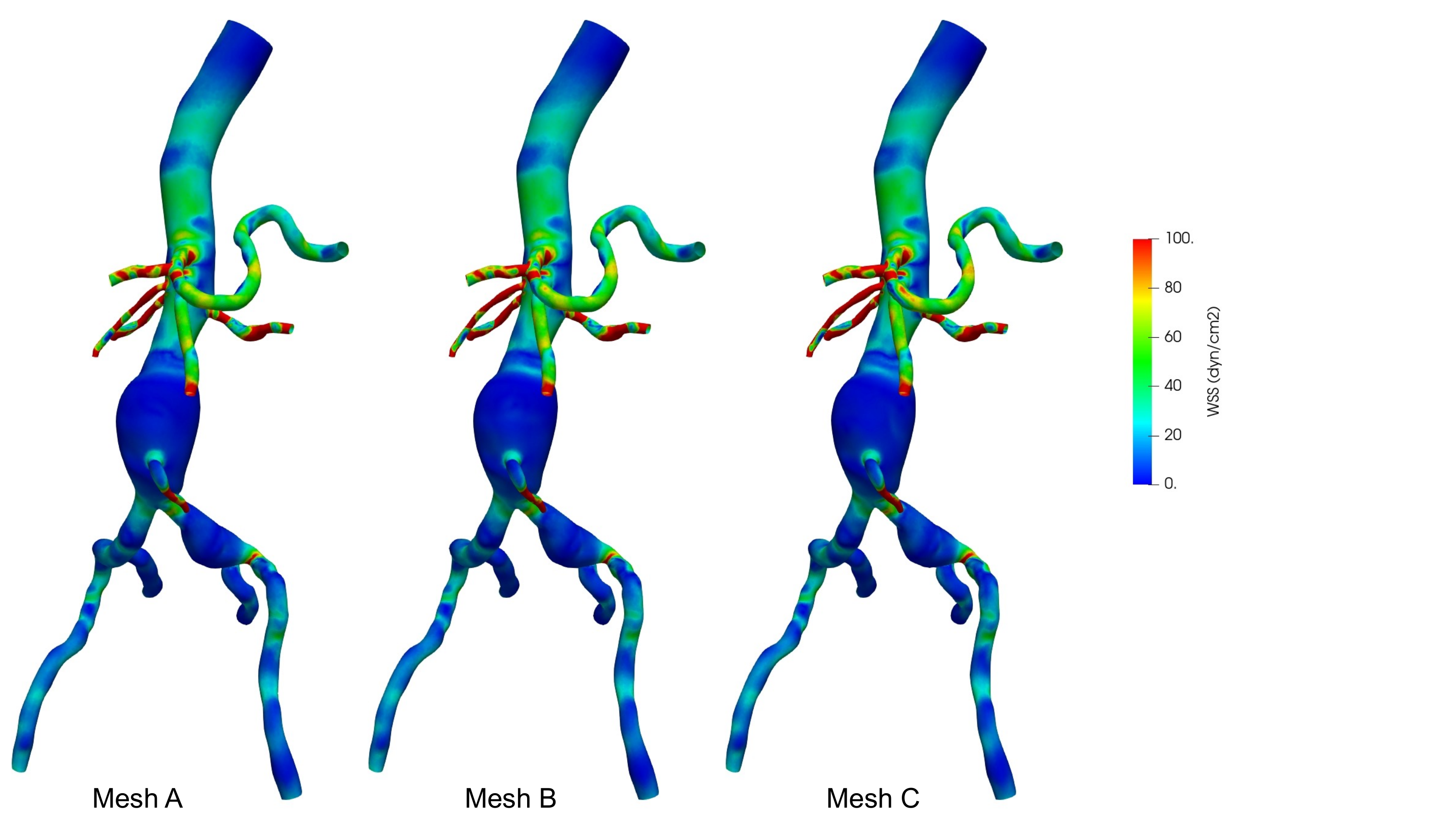} 
\end{center}
\caption{The wall shear stress computed by the three meshes near the peak systole.}
\label{fig:FSI_AAA_WSS_convergence}
\end{figure}

In the second problem, we considered a subject-specific model of a diseased abdominal aorta with a fusiform aneurysm, which is constructed from the CT angiogram of a 75-year-old male. The image data was acquired from the Vascular Model Repository \cite{VMR}. In specific, this abdominal aorta aneurysm (AAA) model extends from the supraceliac aorta, ends after the common iliac artery bifurcation, and contains 11 outlets in total (See Figure \ref{fig:FSI_AAA_model}). We generated unstructured finite element meshes using the pipeline proposed in Section \ref{sec:geometric_modeling_mesh_generation}, which contain three boundary layers at a thickness gradation factor of $0.6$. In the meantime, we strived to have at least two layers of elements across the vessel wall thickness (see Figure \ref{fig:FSI_AAA_mesh}). The details of the meshes are listed in Table \ref{table:aaa_mesh}. The meshes created here are arguably the most anatomically detailed and geometrically sophisticated FSI mesh for AAA analysis to date. Previous studies either consider only the region near the aneurysm \cite{Scotti2005,Scotti2007} or use thin-walled assumptions \cite{Lan2022c}, let alone the vast majority of studies adopt the rigid-wall assumption \cite{Achille2014,Les2010}. The material properties used in the FSI simulations are summarized in Table \ref{table:aaa_material_properties}. The inflow was prescribed based on a parabolic profile scaled by a representative flow waveform for the supraceliac aortic flow (see Figure \ref{fig:FSI_AAA_model}), with the cardiac period being $T_p = 0.938$ seconds. Since our model construction started from the supraceliac aorta, the parabolic flow profile is a reasonable assumption. On the outlets, three-element RCR models were coupled to replicate the resistance and capacitance effects of the downstream vasculature. The RCR model parameters were tuned to achieve target flow splits as well as inlet systolic and diastolic pressures \cite{Lan2022c}. Simulations were performed with all three meshes following the initialization procedure outlined in Sec. \ref{subsec:prestressing}. For Meshes A and B, the time step size was chosen to be $T_p/2500$, while for Mesh C, the time step size was fixed to be $T_p/3200$. Three cardiac cycles were simulated. With the help of the initialization procedure, cycle-to-cycle periodic solutions were achieved in the second cycle, and the results from the last cycle is presented here. The simulations of this model problem were performed on the BSCC-T6 supercomputer. The simulated pressure and flow waveform on four representative outlets are presented in Figure \ref{fig:FSI_AAA_pressure_flow_outlets}, and the snapshots of the pressure, velocity, and wall displacement near the peak systole are depicted in Figure \ref{fig:FSI_AAA_pres_velo_disp_snapshot}. The wall shear stress near the peak systole computed by the three meshes is illustrated in Figure \ref{fig:FSI_AAA_WSS_convergence}, from which it can be concluded that the calculated wall shear stress is mesh-independent.

\begin{table}[htbp]
\small
\begin{center}
\renewcommand{\arraystretch}{1.0}
\begin{tabular}{l l}
\hline
Solid properties \\
Young's modulus & $E = 1.00\times{10^{7}}$ dyn/cm$^{2}$ \\
Poisson's ratio  & $\nu = 0.5$ \\
Reference density & $\rho_{0}^{s}=1.00$ g/cm$^{3}$ \\[0.5em]
Fluid properties \\
Dynamic viscosity & $\bar{\mu}= 4.00 \times 10^{-2}$ poise  \\
Density & $\rho_{0}^{f}=1.06$ g/cm$^{3}$ \\
\hline
\end{tabular}
\end{center}
\caption{Material parameters used in the AAA vascular FSI model. }
\label{table:aaa_material_properties}
\end{table}

\begin{table}[htbp]
\small
\begin{center}
\tabcolsep=0.19cm
\renewcommand{\arraystretch}{1.2}
\begin{tabular}{c c c}
\hline
Mesh &  No. nodes  & No. elements \\
\hline
A & $1.33 \times 10^6$ & $8.09 \times 10^6$ \\
B & $2.66 \times 10^6$ & $1.63 \times 10^7$ \\
C & $5.31 \times 10^6$ & $3.29 \times 10^7$ \\
\hline
\end{tabular}
\end{center}
\caption{Patient-specific vascular FSI: spatial mesh details.}
\label{table:aaa_mesh}
\end{table}

\subsubsection{Fixed-size scalability}
Using the patient-specific AAA model, we examined the proposed FSI solution strategy under the parallel setting. We considered the three meshes with the time step size fixed to be $1.0 \times 10^{-4}$ and integrated for $10$ time steps. It can be observed that there is a region where superlinear speedup is achieved for all six scaling curves, which is likely due to the increase of cache available to the parallel execution \cite{Ristov2016}. For mesh A, the parallel efficiency maintains at around 90 \% and 97 \% for the SCR and SIMPLE preconditioners, respectively, after a ten-fold increase of processors. For the two finer meshes, the efficiency drops gradually to around 70 \% and 80 \% for the SCR and SIMPLE preconditioners, respectively. The parallel scalability achieved here is comparable to those obtained from single physics problems \cite{Liu2019,Liu2020}. This demonstrates that our proposed solver technology for the vascular FSI problem is indeed well-parallelized and scalable on supercomputers.

\begin{figure}[htbp]
\begin{center}
\includegraphics[angle=0, trim=170 80 160 110, clip=true, scale = 0.25]{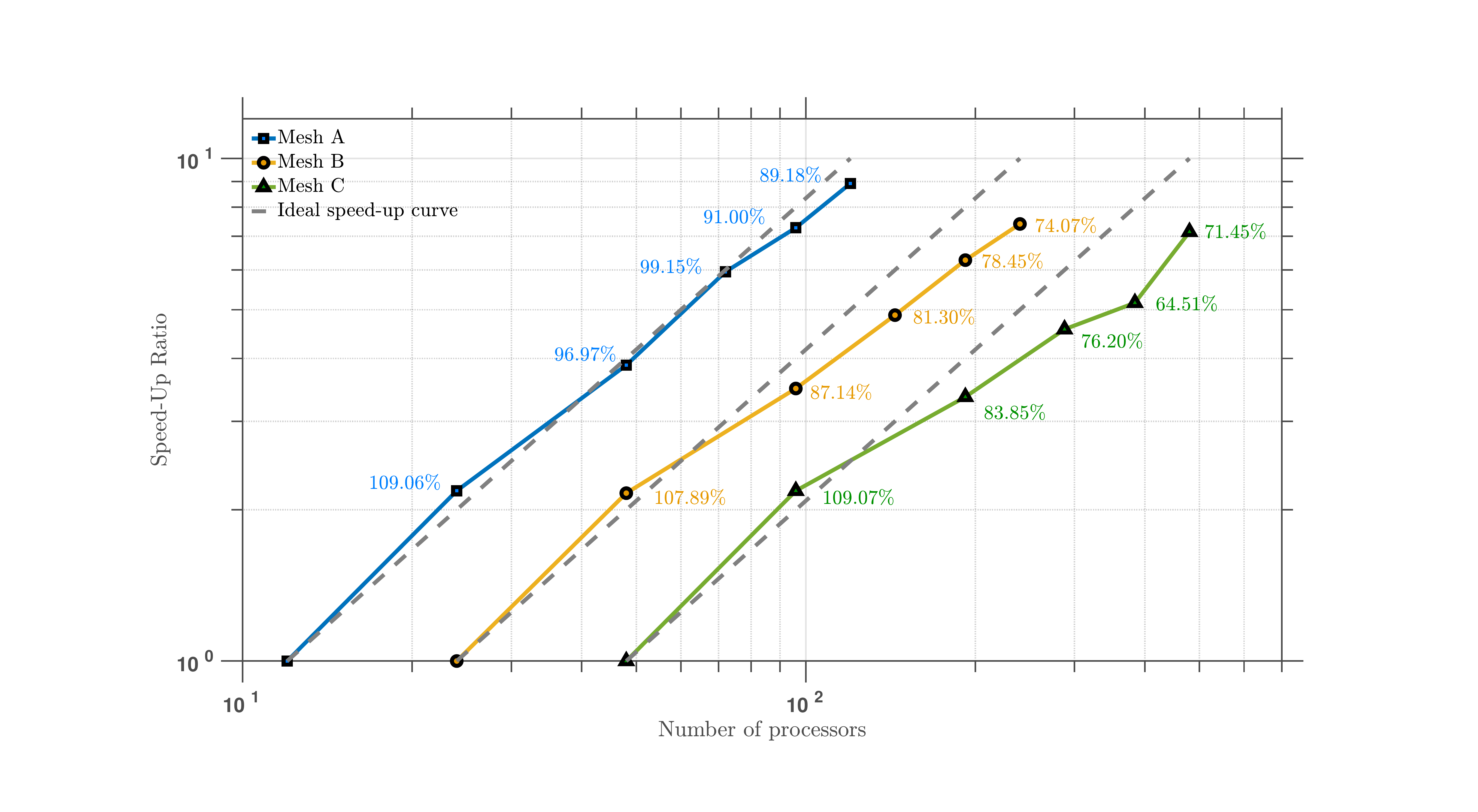} \\
\includegraphics[angle=0, trim=170 80 160 110, clip=true, scale = 0.25]{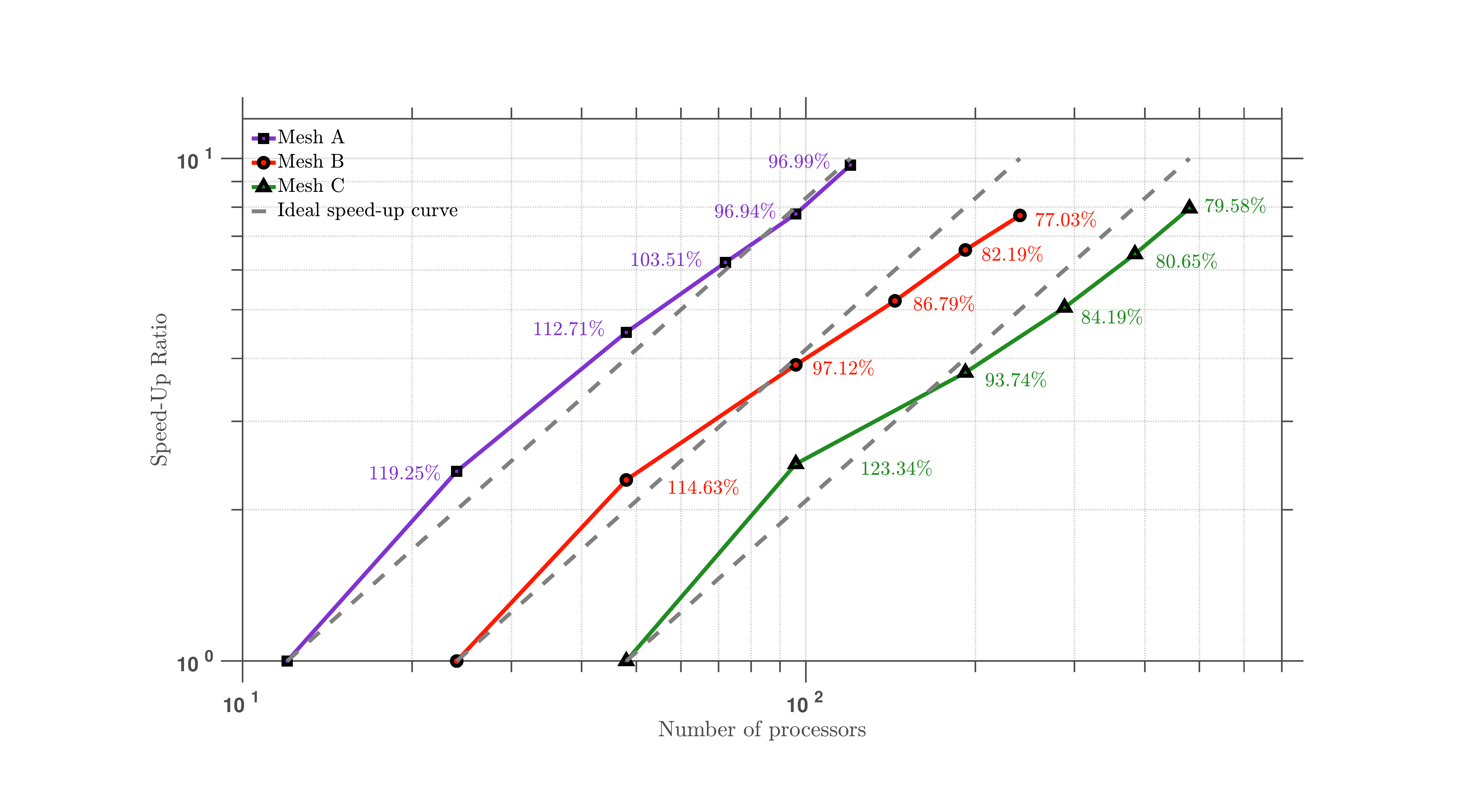} 
\end{center}
\caption{Fixed-size scalability of our solution strategy using the SCR (top) and SIMPLE (bottom) preconditioners for the AAA vascular FSI problem. Annotated efficiency rates are computed from the total runtime of the time integration}
\label{fig:FSI_AAA_strong_scaling}
\end{figure}

\subsubsection{Robustness with respect to the time step size}
In the last, we studied the solver's performance with varying time step sizes to examine the solver's capability in handling implicit time integrations. We used Mesh C with 480 processors, and the collected statistics of solver performance are listed in Table \ref{table:aaa_dt_robustness}. For the two smaller time steps, the averaged time for solving the linear system \eqref{eq:predictor-multi-corrector-linear-system} remains at about the same level. For the largest $\Delta t$, the SCR preconditioner still can drive the residual below the prescribed tolerance with around $5$ Krylov iterations. The SIMPLE preconditioner, on the other hand, failed to converge within the prescribed maximum number of iterations. With the increase of $\Delta t$, the SIMPLE preconditioner clearly demanded more Krylov iterations, while the iteration of the SCR preconditioner grew mildly. Under the current solver setting and the largest time step size, the SCR preconditioner needed more than 200 seconds for solving the linear system \eqref{eq:predictor-multi-corrector-linear-system}, which is still too cost-prohibitive. The efficiency of the SCR preconditioner can be further improved by properly tuning the inner solver, noticing that there is a lot of leeway in the choice of inner solver options and tolerances \cite{Liu2019a,Liu2020}.

\begin{table}[htbp]
\small
\begin{center}
\tabcolsep=0.19cm
\renewcommand{\arraystretch}{1.3}
\begin{tabular}{@{\extracolsep{4pt}}P{2.0cm} P{1.0cm} P{1.5cm} P{1.0cm} P{1.5cm} @{}} \\
\hline
\multirow{2}{*}{$\Delta t$} & \multicolumn{2}{c}{SCR} & \multicolumn{2}{c}{SIMPLE} \\
\cline{2-3} \cline{4-5}
& $\bar{n}_L$ & $\bar{T}_L$ & $\bar{n}$ & $\bar{T}_L$ \\
\hline
$1.0 \times 10^{-5}$ & $2.1$ & $4.69 \times 10^0$ & $6.3$ & $3.93 \times 10^0$ \\
$1.0 \times 10^{-4}$ & $3.1$ & $2.63 \times 10^0$ & $44.1$ & $2.80 \times 10^0$ \\
$1.0 \times 10^{-3}$ & $5.1$ & $2.33 \times 10^2$ & NC & NC \\
\hline
\end{tabular}
\end{center}
\caption{Comparison of the solver performance with varying time step sizes. The values of $\bar{n}_L$, and $\bar{T}_L$ represent the averaged number of the Krylov iterations and the averaged time for solving \eqref{eq:predictor-multi-corrector-linear-system}, respectively. The symbol NC stands for non-convergence as the linear solver reaches the maximum number of iterations with the residual norm greater than the prescribed relative and absolute tolerances.}
\label{table:aaa_dt_robustness}
\end{table}

\section{Conclusion}
\label{sec:conclusion}
We presented a comprehensive suite of computational FSI modeling techniques for vascular problems in this work. Adopting the generalized-$\alpha$ scheme allows a fully implicit integration of the FSI system with desirable numerical properties. In particular, the classical generalized-$\alpha$ scheme is rectified to ensure second-order accuracy \cite{Liu2021a}. The long-lasting overshoot phenomenon is resolved theoretically. Combining with the unified continuum formulation \cite{Liu2018}, the generalized-$\alpha$ scheme achieves optimal high-frequency dissipation in both fluid and solid subproblems, circumventing another issue plaguing the conventional FSI formulation. Thanks to the unified continuum formulation, the fully discrete problem presents a uniform mathematical structure in both subproblems, rendering the parallelization of the algorithm convenient and straightforward. We systematically designed a suite of solver strategies by exploiting the block structure. Factorization is performed to segregate the kinematics from the mechanical balance equations in the multi-corrector iterations. The decoupled kinematics field is conveniently addressed by the AMG method because it is governed by a simple elliptic operator. The discrete balance equations exhibit a two-by-two block structure, and we invoked the Schur complement reduction strategy to further untangle the physics within the Krylov iterations. Sparse approximation of the Schur complement and AMG method were utilized to furnish the preconditioner design, and the well-known SIMPLE preconditioner can be conveniently recovered as a special case of the SCR procedure. Numerical results demonstrated that the solver technology enjoys appealing parallel scalability and robustness properties. These attributes reinforce the conclusion that the unified continuum formulation is advantageous in general FSI problems.

In addition to the FSI computation techniques, we also developed a suite of techniques for subject-specific vascular modeling. First, a pipeline is proposed for generating vascular FSI meshes from medical images. Different from most existing strategies, we avoided manipulating directly on the luminal wall surface mesh to generate the volumetric wall mesh. Instead, we chose to start with the segmentation of the exterior wall surfaces from medical image data. In doing so, we are able to leverage CAD and polygon mesh processing tools for surface fairing and editing. With high-quality lumen and wall geometric models constructed, mesh generation algorithms are invoked to ensure that the mesh is $C^0$-continuous across the fluid-solid interface and is endowed with boundary layers in the fluid subdomain. Second, inspired by the prior work \cite{Hsu2011}, a prestress generation procedure has been devised for the proposed FSI formulation to ensure physiologically realistic simulations. This technique is indispensable for simulations in systemic circulations. We want to mention that the development of the prestress technique inspires us to release the pressure continuity across the fluid-solid interface, which results in a better dynamic coupling condition in the discrete FSI formulation.

Future work includes performing verification and validation studies of the proposed FSI framework against analytic solutions and experimental data. Some initial studies have been performed using a membrane wall model \cite{Lan2022b,Lan2022}. It is also intriguing to evaluate the impact of the prestress generated from the proposed algorithm, as the algorithm cannot guarantee the uniqueness of the prestress. Generation of a more refined arterial wall model with intima, media, and adventitia layers embedded with different families of fibers from subject-specific imaging data will also be pursued.

\section*{Acknowledgments}
This work is supported by the National Natural Science Foundation of China [Grant Number 12172160], Southern University of Science and Technology [Grant Number Y01326127], and the Guangdong-Hong Kong-Macao Joint Laboratory for Data-Driven Fluid Mechanics and Engineering Applications [Grant Number 2020B1212030001]. Computational resources are provided by the Center for Computational Science and Engineering at the Southern University of Science and Technology and the Beijing Super Cloud Computing Center.

\appendix

\section{An analysis of the no-overshoot property of the first-order generalized-$\alpha$ scheme}
\label{sec:overshoot-analysis}
We consider a single-degree-of-freedom (SDOF) model problem in the following form,
\begin{align}
\label{app:SDOF-model-problem}
\ddot{d} + \omega^2 d = 0,
\end{align}
where $d$ is the Fourier coefficient and $\omega$ is the undamped frequency of vibration. We consider two generalized-$\alpha$ schemes applied to this problem. The first-order generalized-$\alpha$ scheme \cite{Jansen2000} for this equation reads as follows,
\begin{align*}
\begin{bmatrix}
1 & 0 \\
0 & 1
\end{bmatrix}
\begin{Bmatrix}
 \dot{d}_{n+\alpha_m} \\
 \dot{v}_{n+\alpha_m}
\end{Bmatrix}
+
\begin{bmatrix}
0 & - 1 \\
\omega^2 & 0
\end{bmatrix}
\begin{Bmatrix}
 d_{n+\alpha_f} \\
 v_{n+\alpha_f}
\end{Bmatrix}
=
\begin{Bmatrix}
0 \\
0
\end{Bmatrix},
\end{align*}
where
\begin{align*}
& \dot{d}_{n+\alpha_m} = \alpha_m \dot{d}_{n+1} + ( 1 -\alpha_m ) \dot{d}_n, \quad &&
\dot{v}_{n+\alpha_m} = \alpha_m \dot{v}_{n+1} + ( 1 -\alpha_m ) \dot{v}_n, \displaybreak[2] \\
& d_{n+\alpha_f} = \alpha_f d_{n+1} + ( 1 -\alpha_f ) d_n, \quad && v_{n+\alpha_f} = \alpha_f v_{n+1} + ( 1 -\alpha_f ) v_n, \\
& d_{n+1} = d_n + \Delta t [ \gamma \dot{d}_{n+1} + ( 1 - \gamma ) \dot{d}_n ], \quad &&
v_{n+1} = v_n + \Delta t [ \gamma \dot{v}_{n+1} + ( 1 - \gamma ) \dot{v}_n ].
\end{align*}
The approximations of displacement, displacement time derivative, velocity and velocity time derivative at time step $t_{n+1}$ can be derived from the equations above, that is,
\begin{align}
\label{app:1st-gen-alpha-disp-begin}
& d_{n+1} =  \frac{1}{D_1}\left( \left( \alpha_m^2 + ( \alpha_f^2 - \alpha_f ) \gamma^2 \Omega^2 \right) d_n + \alpha_m \gamma \Delta t v_n  + \alpha_m ( \alpha_m - \gamma ) \Delta t \dot{d}_n + ( \alpha_m - \gamma) \alpha_f \gamma \Delta t^2 \dot{v}_n \right), \\
& \dot{d}_{n+1} = \frac{1}{\gamma \Delta t} ( d_{n+1} - d_n ) + \frac{\gamma - 1}{\gamma} \dot{d}_n, \\
& v_{n+1} = \frac{\alpha_m}{\alpha_f \gamma \Delta t} ( d_{n+1} - d_n ) + \frac{\gamma - \alpha_m}{\gamma \alpha_f} \dot{d}_n + \frac{\alpha_f - 1}{\alpha_f} v_n, \displaybreak[2] \\
& \dot{v}_{n+1} = \frac{\alpha_m}{\alpha_f \gamma^2 \Delta t^2} ( d_{n+1} - d_n ) - \frac{1}{\alpha_f \gamma \Delta t} v_n + \frac{\gamma - 1}{\gamma} \dot{v}_n + \frac{\gamma - \alpha_m}{\alpha_f \gamma^2 \Delta t} \dot{d}_n,
\end{align}
where $D_1 := \alpha_m^2 + \alpha_f^2 \gamma^2 \Omega^2$, and $\Omega := \omega \Delta t$.

The second-order generalized-$\alpha$ scheme \cite{Chung1993} for the same model problem can be written as
\begin{align*}
\ddot{d}_{n + \alpha_m} + \omega^2 d_{n + \alpha_f} = 0,
\end{align*}
in which,
\begin{align*}
& \ddot{d}_{n + \alpha_m} = \alpha_m \ddot{d}_{n+1} + ( 1 - \alpha_m ) \ddot{d}_n, && d_{n + \alpha_f} = \alpha_f d_{n+1} + ( 1 - \alpha_f ) d_n, \displaybreak[2] \\
& \dot{d}_{n+1} = \dot{d}_n + \Delta t [ ( 1 - \gamma ) \ddot{d}_n + \gamma \ddot{d}_{n+1} ], && d_{n+1} = d_n + \Delta t \dot{d}_n + \Delta t^2 [ ( \frac{1}{2} - \beta ) \ddot{d}_n + \beta \ddot{d}_{n+1} ].
\end{align*}
Consequently, the displacement and its derivatives at time step $t_{n+1}$ can be written as 
\begin{align}
& d_{n+1} = \frac{1}{D_2} \left( \left( \alpha_m + \alpha_f \beta \Omega^2 - \beta \Omega^2 \right) d_n + \alpha_m \Delta t \dot{d}_n + ( \frac{\alpha_m}{2} - \beta ) \Delta t^2 \ddot{d}_n \right), \displaybreak[2] \\
& \dot{d}_{n+1} = \frac{\gamma}{\Delta t \beta} ( d_{n+1} - d_n ) + ( 1 - \frac{\gamma}{\beta} ) \dot{d}_n + \Delta t ( 1 - \frac{\gamma}{2\beta} ) \ddot{d}_n, \displaybreak[2] \\
\label{app:2nd-gen-alpha-disp-ddot-end}
& \ddot{d}_{n+1} = \frac{1}{\Delta t^2 \beta} ( d_{n+1} - d_n ) - \frac{1}{\Delta t \beta} \dot{d}_n - \frac{1 - 2 \beta}{2 \beta} \ddot{d}_n,
\end{align}
wherein $D_2 := \alpha_m + \alpha_f \beta \Omega^2$.

The numerical solutions of the SDOF problem \eqref{app:SDOF-model-problem} based on the two generalized-$\alpha$ schemes can be cast into the following two forms, respectively,
\begin{align}
\label{app:SDOF-iteration-form}
\bm X_{n+1} = \bm A_1 \bm X_{n}, \qquad  \bm Y_{n+1} = \bm A_2 \bm Y_{n}, \qquad \forall n \in \{ 0, 1, \cdots, N_{\mathrm{ts}}-1 \},
\end{align}
where $N_{\mathrm{ts}}$ is the number of time steps, $\bm X_n := \{ d_n, \, \Delta t v_n, \, \Delta t \dot{d}_n, \, \Delta t^2 \dot{v}_n \}^T$ for the first-order generalized-$\alpha$ scheme, $\bm Y_n := \{ d_n, \, \Delta t \dot{d}_n, \, \Delta t^2 \ddot{d}_n \}^T$ for the second-order generalized-$\alpha$ scheme, and $\bm A_1$ and $\bm A_2$ are the amplification matrices with sizes $4 \times 4$ and $3 \times 3$, respectively. To gain insights into the performance of the algorithms, we derive the explicit forms of the amplification matrices from \eqref{app:1st-gen-alpha-disp-begin}-\eqref{app:2nd-gen-alpha-disp-ddot-end},
\begin{align*}
& \bm A_1 = \frac{1}{D_1} 
\begin{bmatrix}
\alpha_m^2 + D_f \gamma^2 \Omega^2 & \alpha_m \gamma & \alpha_m ( \alpha_m - \gamma ) & \gamma \alpha_f ( \alpha_m - \gamma ) \\[1.5mm]
- \gamma \Omega^2 \alpha_m & \gamma^2 \Omega^2 D_f + \alpha_m^2 & \gamma \Omega^2 \alpha_f ( \gamma - \alpha_m ) & \alpha_m ( \alpha_m - \gamma ) \\[1.5mm]
- \gamma \Omega^2 \alpha_f & \alpha_m & ( \gamma^2 - \gamma ) \Omega^2 \alpha_f^2 + D_m & \alpha_f ( \alpha_m - \gamma ) \\[1.5mm]
- \Omega^2 \alpha_m & - \gamma \Omega^2 \alpha_f & \Omega^2 \alpha_f ( \gamma - \alpha_m ) & ( \gamma^2 - \gamma ) \Omega^2 \alpha_f^2 + D_m
\end{bmatrix},\\
& \bm A_2 = \frac{1}{D_2}
\begin{bmatrix}
D_2 - \beta \Omega^2 & \alpha_m & \frac{\alpha_m}{2} - \beta \\[1.5mm]
- \gamma \Omega^2 & D_2 - \gamma \alpha_f \Omega^2 & D_2 - \gamma (\frac{\alpha_f}{2}\Omega^2 + 1) \\[1.5mm]
- \Omega^2 & - \alpha_f \Omega^2 & D_2 - ( \frac{\alpha_f }{2}\Omega^2 + 1 )
\end{bmatrix},
\end{align*}
where $D_f := \alpha_f^2 - \alpha_f$ and $D_m := \alpha_m^2 - \alpha_m$. The overshoot behavior can be obtained by analyzing the results at the first time step. Given the initial conditions $d(0) = d_0$ and $\dot{d}(0) = v_0$, and from the SDOF model problem we get the compatible initial conditions for $v(t)$ and $\dot{v}(t)$ as $v(0) = \dot{d_0}$, and $\dot{v}(0) = -\omega^2 d_0$. The solutions of the two generalized-$\alpha$ schemes in the first time step can be calculated from \eqref{app:SDOF-iteration-form} with $n=0$. We are only interested in the situation when the time step is large enough since both algorithms are convergent and thus will not suffer from the overshoot phenomenon when $\Omega \rightarrow 0$. The behavior of the solutions when $\Omega \rightarrow \infty$ in the two schemes is summarized as follows. For the first-order generalized-$\alpha$ scheme, we have 
\begin{align*}
& d_1 = ( 1 - \frac{\alpha_m}{\alpha_f \gamma} ) d_0, \qquad \dot{d_1} = \frac{\gamma - 1}{\gamma} \dot{d_0}, \qquad v_1 = ( 1 - \frac{\alpha_m}{\alpha_f \gamma} ) \dot{d}_0, \qquad \dot{v}_1 = \frac{1 - \gamma}{\gamma} \ddot{d}_0,
\end{align*}
and for the second-order generalized-$\alpha$ scheme, we have
\begin{align*}
d_1 = \frac{\alpha_f - 1}{\alpha_f} d_0, \qquad \dot{d}_1 = ( 1 - \frac{\gamma}{\beta} ) \dot{d}_0 + ( \frac{\gamma}{2 \beta} - 1 ) \omega \Omega d_0, \qquad \ddot{d}_1 = ( 1 - \frac{1}{2 \beta} ) \ddot{d}_0.
\end{align*}
From the expression of $\dot{d}_1$ above, the second-order generalized-$\alpha$ scheme tends to overshoot the velocity equation linearly with respect to $\Omega$. And the response of $\dot{d}_1$ is exactly the same as that in the HHT-$\alpha$ scheme \cite[p.~538]{Hughes1987}. In contrast, all kinematic fields of the first-order generalized-$\alpha$ scheme are independent of the frequency $\Omega$, and this scheme thus does not exhibit unbounded error in its first step. In other words, the first-order generalized-$\alpha$ scheme is evidently superior to its second-order counterpart in maintaining a good quality of the dynamic behavior. This analysis result confirms the numerical observation made in \cite{Kadapa2017}.

\section{Schur complement reduction preconditioner}
\label{sec:schur_complement_reduction_PC}
We state the definition of the SCR preconditioner $\boldsymbol{\mathcal P}_{\textup{SCR}}$ in terms of the following algorithm.
\begin{algorithm}[H]
\caption{The calculation of $\bm y = \boldsymbol{\mathcal P}_{\textup{SCR}}^{-1} \bm r$ with $\bm r = [ \bm r_{\bm v}; \bm r_p]^T$ and $\bm y = [ \bm y_{\bm v}; \bm y_p]^T$.}
\label{algorithm:exact_block_factorization}
\begin{algorithmic}[1]
\State \texttt{Solve the linear system} 
\begin{align*}
\boldsymbol{\mathrm A} \hat{\bm y}_{\bm v} = \bm r_{\bm v}
\end{align*}
\texttt{by GMRES preconditioned by $\boldsymbol{\mathrm P}_{\mathrm A}$, with $\delta_{\mathrm A}$ and $\mathrm n^{\textup{max}}_{\mathrm A}$ prescribed.}
\State \texttt{Update the mass residual by $\bm r_{p} \gets  \bm r_{p} - \boldsymbol{\mathrm C} \hat{\bm r}_{\bm v}$.}
\State \texttt{Solve the linear system} 
\begin{align*}
\boldsymbol{\mathrm S} \bm y_p = \bm r_{p} 
\end{align*}
\texttt{by GMRES preconditioned by $\boldsymbol{\mathrm P}_{\mathrm S}$, with $\delta_{\mathrm S}$ and $\mathrm n^{\textup{max}}_{\mathrm S}$ prescribed.}
\State \texttt{Update the momentum residual by $\bm r_{\bm v} \gets \bm r_{\bm v} - \boldsymbol{\mathrm B} \bm y_{p}$.}
\State \texttt{Solve the linear system}
\begin{align*}
\boldsymbol{\mathrm A} \bm y_{\bm v} = \bm r_{\bm v}
\end{align*}
\texttt{by GMRES preconditioned by $\boldsymbol{\mathrm P}_{\mathrm A}$, with $\delta_{\mathrm A}$ and $\mathrm n^{\textup{max}}_{\mathrm A}$ prescribed.}
\end{algorithmic}
\end{algorithm}
In the third step of Algorithm \ref{algorithm:exact_block_factorization}, one needs to solve the linear system associated with the Schur complement $\boldsymbol{\mathrm S}$, whose definition is given by $\boldsymbol{\mathrm D} - \boldsymbol{\mathrm C} \boldsymbol{\mathrm A}^{-1} \boldsymbol{\mathrm B}$. The equation associated with $\boldsymbol{\mathrm S}$ can be solved in a matrix-free manner with the action of $\boldsymbol{\mathrm S}$ on a vector given by the following algorithm.
\begin{algorithm}[H]
\caption{The matrix-free algorithm of $\boldsymbol{\mathrm S}$ acting on a vector $\bm x_p$.}
\label{algorithm:matrix_free_mat_vec_for_S}
\begin{algorithmic}[1]
\State \texttt{Compute the matrix-vector multiplication $\bar{\bm x}_p \gets \boldsymbol{\mathrm B} \bm x_p$.}
\State \texttt{Invoke GMRES preconditioned by $\boldsymbol{\mathrm P}_{\mathrm I}$ to solve the linear system} 
\begin{align*}
\boldsymbol{\mathrm A} \tilde{\bm x}_p = \bar{\bm x}_p
\end{align*}
\texttt{for $\tilde{\bm x}_p$, with $\delta_{\mathrm I}$ and $\mathrm n^{\textup{max}}_{\mathrm I}$ prescribed.}
\State \texttt{Compute the matrix-vector multiplication $\bar{\bm x}_p \gets \boldsymbol{\mathrm C} \tilde{\bm x}_p$.} \algorithmiccomment{The vector $\bar{\bm x}_p$ is reused.}
\State \texttt{Compute the matrix-vector multiplication $\hat{\bm x}_p \gets \boldsymbol{\mathrm D} \bm x_p$.}
\State \Return $\hat{\bm x}_p - \bar{\bm x}_p$.
\end{algorithmic}
\end{algorithm}

\bibliographystyle{plain}
\bibliography{fsi}

\end{document}